\documentclass[12pt]{article}

\usepackage[utf8]{inputenc}
\usepackage{textcomp}
\usepackage{chetnew}
\usepackage{tikz}
\usepackage{amssymb,amsfonts,mathrsfs,dsfont,yfonts,bbm}\usetikzlibrary{calc}
\usepackage{xcolor}
\usepackage{braket}

\newcommand{\CJ}{\mathcal{J}}
\newcommand{\DV}{\mathds{V}}

\newcommand{\CW}{\mathcal{W}}
\newcommand{\CL}{\mathcal{L}}
\newcommand{\CH}{\mathcal{H}}

\newcommand{\CB}{\mathcal{B}}
\newcommand{\CC}{\mathcal{C}}
\newcommand{\CO}{\mathcal{O}}
\newcommand{\CT}{\mathcal{T}}
\newcommand{\CI}{\mathcal{I}}
\newcommand{\CN}{\mathcal{N}}
\newcommand{\CS}{\mathcal{S}}
\newcommand{\CM}{\mathcal{M}}
\newcommand{\CP}{\mathcal{P}}

\newcommand*{\boxcoloro}{orange}
\makeatletter
\newcommand{\boxedo}[1]{\textcolor{\boxcoloro}{%
\tikz[baseline={([yshift=-1ex]current bounding box.center)}] \node [rectangle, minimum width=1ex,rounded corners,draw] {\normalcolor\m@th$\displaystyle#1$};}}
 \makeatother
\newcommand*{\boxcolorr}{red}
\makeatletter
\newcommand{\boxedr}[1]{\textcolor{\boxcolorr}{%
\tikz[baseline={([yshift=-1ex]current bounding box.center)}] \node [rectangle, minimum width=1ex,rounded corners,draw] {\normalcolor\m@th$\displaystyle#1$};}}
 \makeatother
\newcommand*{\boxcolorb}{blue}
\makeatletter
\newcommand{\boxedb}[1]{\textcolor{\boxcolorb}{%
\tikz[baseline={([yshift=-1ex]current bounding box.center)}] \node [rectangle, minimum width=1ex,rounded corners,draw] {\normalcolor\m@th$\displaystyle#1$};}}
 \makeatother
\newcommand*{\boxcolorg}{green}
\makeatletter
\newcommand{\boxedg}[1]{\textcolor{\boxcolorg}{%
\tikz[baseline={([yshift=-1ex]current bounding box.center)}] \node [rectangle, minimum width=1ex,rounded corners,draw] {\normalcolor\m@th$\displaystyle#1$};}}
 \makeatother
 \newcommand*{\boxcolorp}{purple}
\makeatletter
\newcommand{\boxedp}[1]{\textcolor{\boxcolorp}{%
\tikz[baseline={([yshift=-1ex]current bounding box.center)}] \node [rectangle, minimum width=1ex,rounded corners,draw] {\normalcolor\m@th$\displaystyle#1$};}}
 \makeatother
  \newcommand*{\boxcolorc}{cyan}
\makeatletter
\newcommand{\boxedc}[1]{\textcolor{\boxcolorc}{%
\tikz[baseline={([yshift=-1ex]current bounding box.center)}] \node [rectangle, minimum width=1ex,rounded corners,draw] {\normalcolor\m@th$\displaystyle#1$};}}
 \makeatother
  \newcommand*{\boxcolory}{yellow}
\makeatletter
\newcommand{\boxedy}[1]{\textcolor{\boxcolory}{%
\tikz[baseline={([yshift=-1ex]current bounding box.center)}] \node [rectangle, minimum width=1ex,rounded corners,draw] {\normalcolor\m@th$\displaystyle#1$};}}
 \makeatother
\newcommand{\be}{\begin{equation}}
\newcommand{\ee}{\end{equation}}
\newcommand{\bea}{\begin{eqnarray}}
\newcommand{\eea}{\end{eqnarray}}
\newcommand{\DZ}{\mathds{Z}}
\usepackage{amsmath}	% required for `\align' (yatex added)

\title{Qudit Stabilizer Codes, CFTs, and \\[2mm] Topological Surfaces}

\author{Matthew Buican$^{1}$ and Rajath Radhakrishnan$^{2}$}

\affiliation{\smallskip ${}^{1}$CTP and Department of Physics and Astronomy\\
Queen Mary University of London, London E1 4NS, UK\\
${}^{2}$International Centre for Theoretical Physics, Strada Costiera 11, Trieste 34151, Italy}

\abstract{We study general maps from the space of rational CFTs with a fixed chiral algebra and associated Chern-Simons (CS) theories to the space of qudit stabilizer codes with a fixed generalized Pauli group. We consider certain natural constraints on such a map and show that the map can be described as a graph homomorphism from an orbifold graph, which captures the orbifold structure of CFTs, to a code graph, which captures the structure of self-dual stabilizer codes. By studying explicit examples, we show that this graph homomorphism cannot always be a graph embedding. However, we construct a physically motivated map from universal orbifold subgraphs of CFTs to operators in a generalized Pauli group. We show that this map results in a self-dual stabilizer code if and only if the surface operators in the bulk CS theories corresponding to the CFTs in question are self-dual. For CFTs admitting a stabilizer code description, we show that the full abelianized generalized Pauli group can be obtained from twisted sectors of certain 0-form symmetries of the CFT. Finally, we connect our construction with SymTFTs, and we argue that many equivalences between codes that arise in our setup correspond to equivalence classes of bulk topological surfaces under fusion with invertible surfaces.}

\begin{document}
\setcounter{tocdepth}{2}
\maketitle
\toc

\bigskip

\section{Introduction}

Any physical system designed for controlled manipulation of quantum information must deal with errors. Therefore, quantum error-correcting codes (or quantum codes for short) play a crucial role in quantum computation. Quantum error correction involves encoding information in a ``code subspace'' of a Hilbert space. The errors acting on this subspace are detected using syndrome measurements and then corrected using appropriate unitary operations. Stabilizer codes are an important and well-studied class of quantum codes, where the code subspace is determined by an abelian group called the \lq\lq stabilizer group" \cite{Gottesman:1997qd,Gottesman:1998se}. We start with a set of $n$ qudits with product Hilbert space
\be
\CH:=\CH_{1} \otimes \dots \otimes \CH_{n}~,
\ee
where the complex dimensions of the Hilbert space factors are (potentially different) integers $d_i\ge2$ for each $\CH_i$.\footnote{The case of $d_i=2$ is the standard qubit. In what follows, we will sometimes refer to $d_i=3$ as a \lq\lq qutrit" and to $d_i=4$ as a \lq\lq quadit."} The generalized Pauli operators, $\{X,Z\}$, acting on each Hilbert space together form the generalized Pauli group, $\CP_n$ (see Sec. \ref{sec:quditcodes} for details). The code subspace, $\CC\subset\CH$, is the $+1$ eigenspace of a non-trivial abelian subgroup, $\CS<\CP_n$, called the \lq\lq stabilizer group."\footnote{Note that restricting to any other common eigenvalue instead will not lead to a subgroup of $\CP_n$.} The subgroup of $\CP_{n}$ built from the set of elements which commute with $\CS$ but are not in $\CS$, acts on the code subspace as \lq\lq logical" operators. The subgroup of elements in $\CP_n$ that commute with at least one element of $\CS$ are the errors that can be corrected. For reasons we will explain below, we will be particularly interested in self-dual codes, which are codes with a 1-dimensional code subspace (we briefly discuss non-self-dual codes in Appendix \ref{ap:non-self-dual}).

Stabilizer codes are arguably one of the most well-studied classes of quantum codes. In fact, the impact of stabilizer codes goes beyond quantum computation. For example, they play a crucial role in bulk reconstruction in AdS/CFT \cite{Almheiri:2014lwa,Pastawski:2015qua}. The Toric code \cite{Kitaev:1997wr} and, more generally, Levin-Wen/Turaev Viro lattice models are stabilizer codes \cite{Koenig:2010uua}, whose low-energy limits are described by Chern-Simons theories. Moreover, exploration of stabilizer codes lead to the discovery of fractons \cite{Haah:2011drr}.  

\begin{figure}[h!]
    \centering

\tikzset{every picture/.style={line width=0.75pt}} %set default line width to 0.75pt        

\begin{tikzpicture}[x=0.75pt,y=0.75pt,yscale=-1,xscale=1]

%Shape: Parallelogram [id:dp04467542223368459] 
\draw   (444.58,203.85) -- (382.59,235) -- (381.04,83.79) -- (443.03,52.64) -- cycle ;
%Shape: Parallelogram [id:dp3372825549064301] 
\draw  [color={rgb, 255:red, 245; green, 166; blue, 35 }  ,draw opacity=1 ][fill={rgb, 255:red, 248; green, 198; blue, 155 }  ,fill opacity=1 ] (353.82,203.85) -- (291.83,235) -- (290.28,83.79) -- (352.27,52.64) -- cycle ;
%Shape: Parallelogram [id:dp7882685111742219] 
\draw   (265.95,203.85) -- (203.95,235) -- (202.4,83.79) -- (264.4,52.64) -- cycle ;
%Straight Lines [id:da08633876367548532] 
\draw [color={rgb, 255:red, 139; green, 6; blue, 24 }  ,draw opacity=1 ]   (234.17,143.82) -- (290.38,144.08) ;
%Straight Lines [id:da7942076943577906] 
\draw  [dash pattern={on 4.5pt off 4.5pt}]  (290.38,144.08) -- (322.05,143.82) ;
%Straight Lines [id:da7885048110044185] 
\draw [color={rgb, 255:red, 139; green, 6; blue, 24 }  ,draw opacity=1 ]   (322.05,143.82) -- (381.14,144.08) ;
%Straight Lines [id:da7076530663801579] 
\draw [color={rgb, 255:red, 139; green, 6; blue, 24 }  ,draw opacity=1 ] [dash pattern={on 4.5pt off 4.5pt}]  (381.14,144.08) -- (412.81,143.82) ;
%Shape: Ellipse [id:dp6838728927850737] 
\draw  [color={rgb, 255:red, 139; green, 6; blue, 24 }  ,draw opacity=1 ][fill={rgb, 255:red, 139; green, 6; blue, 24 }  ,fill opacity=1 ] (316.29,143.82) .. controls (316.29,142.25) and (317.58,140.98) .. (319.17,140.98) .. controls (320.76,140.98) and (322.05,142.25) .. (322.05,143.82) .. controls (322.05,145.39) and (320.76,146.66) .. (319.17,146.66) .. controls (317.58,146.66) and (316.29,145.39) .. (316.29,143.82) -- cycle ;
%Shape: Ellipse [id:dp18115810421979095] 
\draw  [color={rgb, 255:red, 139; green, 6; blue, 24 }  ,draw opacity=1 ][fill={rgb, 255:red, 139; green, 6; blue, 24 }  ,fill opacity=1 ] (228.41,143.82) .. controls (228.41,142.25) and (229.7,140.98) .. (231.29,140.98) .. controls (232.88,140.98) and (234.17,142.25) .. (234.17,143.82) .. controls (234.17,145.39) and (232.88,146.66) .. (231.29,146.66) .. controls (229.7,146.66) and (228.41,145.39) .. (228.41,143.82) -- cycle ;
%Shape: Ellipse [id:dp5608346266372702] 
\draw  [fill={rgb, 255:red, 139; green, 6; blue, 24 }  ,fill opacity=1 ] (412.81,143.82) .. controls (412.81,142.25) and (414.1,140.98) .. (415.69,140.98) .. controls (417.28,140.98) and (418.57,142.25) .. (418.57,143.82) .. controls (418.57,145.39) and (417.28,146.66) .. (415.69,146.66) .. controls (414.1,146.66) and (412.81,145.39) .. (412.81,143.82) -- cycle ;
%Straight Lines [id:da6157708076318841] 
\draw    (227.69,264.5) -- (408.58,263.51) ;
\draw [shift={(410.58,263.5)}, rotate = 179.69] [color={rgb, 255:red, 0; green, 0; blue, 0 }  ][line width=0.75]    (10.93,-3.29) .. controls (6.95,-1.4) and (3.31,-0.3) .. (0,0) .. controls (3.31,0.3) and (6.95,1.4) .. (10.93,3.29)   ;
\draw   (235.42,267.68) .. controls (232.16,266.01) and (228.88,265.02) .. (225.58,264.7) .. controls (228.9,264.36) and (232.22,263.33) .. (235.58,261.65) ;

% Text Node
\draw (215,118.4) node [anchor=north west][inner sep=0.75pt]    {$p$};
% Text Node
\draw (411,117.4) node [anchor=north west][inner sep=0.75pt]    {$q$};
% Text Node
\draw (290,81.4) node [anchor=north west][inner sep=0.75pt]    {$S$};
% Text Node
\draw (207.4,82.19) node [anchor=north west][inner sep=0.75pt]    {$\Sigma _{1}$};
% Text Node
\draw (386.04,83.19) node [anchor=north west][inner sep=0.75pt]    {$\Sigma _{2}$};
% Text Node
\draw (313,269.4) node [anchor=north west][inner sep=0.75pt]    {$I$};

\end{tikzpicture}
    \caption{The pairing of 2d CFT left and right movers on $\Sigma_1$ and $\Sigma_2$ respectively is specified by an abelian CS theory on $X=\Sigma \times I$ (the $\Sigma_i$ are isomorphic to $\Sigma$) with the surface operator, $S$, inserted in the bulk. The left-mover, $p$, can be paired with the right-mover, $q$, if and only if the corresponding Wilson line in the bulk TQFT can form a junction on $S$.}
    \label{fig:TQFT-CFT intro}
\end{figure}
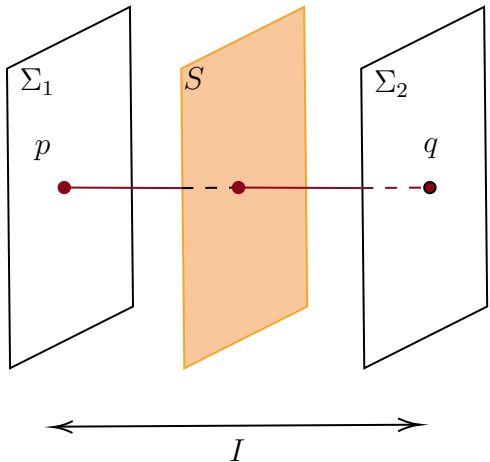

Recent studies in conformal field theory have revealed that Narain CFTs are closely related to quantum stabilizer codes \cite{Dymarsky:2020bps,dymarsky2020quantum,buican2021quantum} (see also \cite{Harvey:2020jvu,Dymarsky:2021xfc,Henriksson:2022dnu,Kawabata:2023rlt,Kawabata:2023iss,Kawabata:2023usr,Alam:2023qac,Furuta:2023xwl,Kawabata:2023nlt}). The description of Narain CFTs in terms of stabilizer codes has been used to identify CFTs maximizing the spectral gap \cite{Angelinos:2022umf}, in understanding fake modular invariants \cite{Dymarsky:2021xfc}, and in the description of holographic theories \cite{Dymarsky:2020pzc,Aharony:2023zit}. In particular, in \cite{buican2021quantum}, the authors of this paper, along with Anatoly Dymarsky, showed that, starting with certain RCFTs, one can naturally construct a huge class of associated qubit stabilizer codes (including all of those in \cite{Dymarsky:2020bps,dymarsky2020quantum}). Since these RCFTs are closely related with corresponding bulk Chern-Simons (CS) theories via the 2d/3d correspondence \cite{Kapustin:2010if} described in Fig. \ref{fig:TQFT-CFT intro}, quantum codes arising in this context are also intimately connected with CS theories \cite{buican2021quantum}. In particular, the quantum code construction in \cite{buican2021quantum} captures the following QFT data:
\begin{itemize}
\item The spectrum of CFT primary operators, their fusion group, and the corresponding 1-form symmetry of the associated bulk CS theories
\item Certain 0-form symmetries of the RCFT and corresponding twisted-sector states (see also \cite{Buican:2023bzl} for comments on non-invertible bulk CS 0-form symmetries arising in the construction of \cite{buican2021quantum})
\item A way to understand orbifolding (i.e., gauging the RCFT 0-form symmetries in the previous bullet) at the level of quantum codes
\end{itemize}

In this work, we extend the analysis in \cite{buican2021quantum} and study the properties of more general maps from the ``space of RCFTs and associated CS theories" to the ``space of stabilizer codes". More precisely, we will study a general chiral algebra, $\DV$, associated with an Abelian RCFT. An Abelian RCFT is defined by the property that its chiral primaries form an abelian group, $K$, under fusion. Given $\DV$ and the associated chiral primaries, an RCFT is determined by a consistent pairing of the left and right movers. As alluded to above, all consistent pairings can be elegantly understood through a bulk 3d CS theory \cite{Kapustin:2010if}. Indeed, the representations of $\DV$ label Wilson lines in a bulk abelian CS theory, which we will denote as $\CI$. The group, $K$, is the 1-form symmetry group of $\CI$ (see also the discussion in \cite{buican2021quantum}). The consistent pairings of the left and right movers are in one-to-one correspondence with surface operators in $\CI$ (see Fig. \ref{fig:TQFT-CFT intro}).

Therefore, in order to classify all RCFTs with chiral algebra $\DV$, we have to classify the surface operators in $\CI$. All such surface operators can be obtained from higher-gauging a 1-form symmetry group $Q\le K$ on a 2-manifold with discrete torsion, $[\sigma] \in H^2(Q,U(1))$ \cite{Roumpedakis:2022aik}. We will denote the resulting surface operator as $S(Q,[\sigma])$. In all CS theories, we have the trivial surface operator, and so the corresponding ``Cardy case" RCFT with charge-conjugation modular invariant is also universal. We will denote this CFT as $\CT$, and, by construction, the corresponding Abelian fusion group of untwisted primaries is $K$. More generally, the CFT corresponding to $S(Q,[\sigma])$ will be denoted as $\CT/(Q,[\sigma])$. In the 2d CFT, higher-gauging the 1-form symmetry of the bulk TQFT corresponds to gauging a 0-form symmetry (or, in more traditional language, to orbifolding). Indeed, the CFT, $\CT/(Q,[\sigma])$, is obtained from gauging a 0-form symmetry, $Q$, of the CFT, $\CT$, with discrete torsion $[\sigma]$ \cite{fuchs2002tft,fuchs2004tft,Frohlich:2009gb}. The untwisted primaries of this CFT form an abelian group, $K_{(Q,[\sigma])}$, under fusion.\footnote{Note that $K_{(Q,[\sigma])}$ is not always isomorphic to $K$.} 

The above discussion implies there are a finite number of RCFTs with a given chiral algebra, $\mathds{V}$. Therefore, the ``the space of abelian RCFTs" for a given $\mathds{V}$ is a finite set. We will study maps from this set of RCFTs to quantum stabilizer codes constructed from a system of qudits. More precisely, given the group, $K$, associated with the chiral algebra, $\DV$, consider the decomposition
\be
K \cong \DZ_{n_1} \times \dots \times \DZ_{n_k}~.
\ee
We choose the system of qudits to have the smallest Hilbert space that can faithfully realize this symmetry (we relax this condition in Appendix \ref{ap:non-self-dual}). In other words, we take
\be
\CH_{\DV}:=\CH_{1} \otimes \dots \otimes \CH_{k}~,
\ee
with factors of dimensions $n_1,\cdots,n_k$ respectively. We will consider maps from the finite set of abelian RCFTs with chiral algebra $\DV$ to the finite number of stabilizer codes constructed from the Pauli group, $\CP_{\DV}$, acting on $\CH_{\DV}$
\be
\mu: \{\text{ abelian RCFTs with chiral algebra $\DV$ } \} \to \{\text{ stabilizer codes } \subset \CP_{\DV} \}~. 
\ee

Let $\CS_{(Q,[\sigma])}$ and $\CC_{(Q,[\sigma])}$ be the stabilizer group and the code subspace associated with $\CT/(Q,[\sigma])$ respectively. What are some properties that $\mu$ should satisfy? One fundamental property of a CFT is the \textit{state-operator correspondence}. We would like this relation to be reproduced by the quantum code. Indeed, there is a coarse-grained state-operator correspondence at the level of quantum codes. Consider the stabilizer group, $\CS$: the code subspace, $\CC$, is uniquely fixed as the $+1$ eigenspace of $\CS$. On the other hand, given $\CC$, $\CS$ is uniquely fixed as the maximal subgroup of the generalized Pauli group that acts trivially on $\CC$. From this discussion, it is natural to require that the CFT to quantum code map satisfies
\vspace{0.2cm}

\noindent
{\it Untwisted primaries (and descendants) of $\CT/(Q,[\sigma])$  $\leftrightarrow$  Stabilizer group $\CS_{(Q,[\sigma])}$}~,

\vspace{0.1cm}

\noindent
{\it Corresponding states of $\CT/(Q,[\sigma])$ $\leftrightarrow$ Code subspace $\CC_{(Q,[\sigma])}$}~.

\vspace{0.2cm}
\noindent
One subtlety of this discussion is that the CFT state-operator correspondence is one-to-one, while the stabilizer group / code subspace correspondence is generally not (e.g., we can have $\dim_{\mathbb{C}}(\CC)=1$ with arbitrarily large $|\CS|$). Therefore, we allow some coarse graining in the action of operators on states and define the CFT code space to be the space closed under the action of the primaries (and their descendants). In other words, stabilizers in the RCFT and CS theory correspond to local operators and bulk Wilson lines which can form a junction on the surface operator as in Fig. \ref{fig:TQFT-CFT intro}, respectively. 

The stabilizer group is a finite abelian group. What aspect of the operators of the RCFT should the stabilizer group capture? Even though $\CT/(Q,[\sigma])$ has an infinite number of operators, it has a finite number of conformal families and corresponding primary operators. These primaries (and their descendants) form the abelian group, $K_{(Q,[\sigma])}$, under fusion. Therefore, it is natural to require that 
\be\label{Cond1}
K_{(Q,[\sigma])} \cong \CS_{(Q,[\sigma])}~,
\ee
as groups. Imposing this requirement then forces the quantum code to be self-dual. Since the abelian RCFTs for a given chiral algebra are all related to each other under gauging 0-form symmetries, the stabilizer codes associated with these RCFTs must also be related to each other. More precisely, note that gauging finite symmetries of an RCFT is an invertible operation.

In particular, the set of primary operators of $\CT$ invariant under the 0-form symmetry, $Q$, remain as local operators of the CFT, $\CT/(Q,[\sigma])$, obtained after orbifolding. In this sense, two CFTs can share a non-trivial number of local operators. These ``shared primaries" are captured by the group $K_{\CT/(Q_1,[\sigma_1])} \cap K_{\CT/(Q_2,[\sigma_2])}$. Since these groups are mapped to the stabilizer groups of the corresponding quantum codes, it is natural to require that 
\be\label{Cond2}
K_{(Q_1,[\sigma_1])} \cap K_{(Q_2,[\sigma_2])} \cong  \CS_{(Q_1,[\sigma_1])} \cap \CS_{(Q_2,[\sigma_2])}~.
\ee

In this paper, we will show that these constraints can be written as a graph homomorphism
\be
\mu: \Gamma_{\DV} \to \Gamma_{\CP_{\DV}}~,
\ee
where $\Gamma_{\DV}$ is a colored graph constructed using the orbifold structure of abelian RCFTs with chiral algebra $\DV$ called the \lq\lq orbifold graph," and $\Gamma_{\CP_{\DV}}$ is a graph constructed from the stabilizer groups in $\CP_{\DV}$ called the \lq\lq code graph."

We study the conditions that should be satisfied for such a graph homomorphism to exist. A natural question that arises is:
\vspace{0.2cm}

\noindent
{\it Do all CFTs with a given chiral algebra, $\DV$, admit a map to stabilizer codes?}
\vspace{0.2cm}

\noindent In other words, is the map $\mu$ an embedding of the graph $\Gamma_{\DV}$ in the graph $\Gamma_{\CP_{\DV}}$? We will argue the answer is \lq\lq no," by showing that the consistency conditions \eqref{Cond1} and \eqref{Cond2} cannot be simultaneously satisfied for a particular class of chiral algebras related to 3d discrete gauge theories with gauge group $\mathbb{Z}_{2^r}$ and $r>1$ (of the \lq\lq prime" CS theories,\footnote{These are theories that cannot be factorized into product CS theories closed separately under fusion and with trivial mutual braiding.} these examples have a particularly rich set of allowed surface operators).

More generally, given the impossibility of finding a universal embedding, $\Gamma_{\DV}\hookrightarrow\Gamma_{\CP_{\DV}}$, satisfying the properties in \eqref{Cond1} and \eqref{Cond2}, we ask if there is a $\mu$ defined for universal non-trivial subgraphs, $\Gamma\subset\Gamma_{\DV}$, that is an embedding in $\Gamma_{\CP_{\DV}}$.\footnote{In Appendix \ref{ap:non-self-dual} we take an orthogonal approach and relax conditions \eqref{Cond1} and \eqref{Cond2}. We then show that a universal embedding, $\Gamma_{\DV}\hookrightarrow\Gamma_{\CP_{\DV}}$, does exist in this case.} Using the classification of Abelian CS theories, we show that such $\Gamma$'s and a corresponding $\mu$ exist. Moreover, we argue that there is a particularly interesting physical choice for $\mu$ and these $\Gamma$'s such that our construction is well-defined whenever we consider CFTs with pairings of left and right movers determined by self-dual surface operators in the corresponding CS theories.\footnote{These are operators that are insensitive to the orientation of the 2-manifold on which they are defined.} In other words, self-duality of the code (which follows from the condition in \eqref{Cond1}) is equivalent to self-duality of the CS surface operator. 

\vspace{0.2cm}
{
\noindent
\it
Self-dual surface operators of bulk CS theory $\xrightarrow{\hspace{0.3cm} \mu \hspace{0.3cm}}$ Self-dual stabilizer codes.
}
\vspace{0.2cm}

\noindent
This construction naturally generalises the CFT to qubit codes map constructed in \cite{buican2021quantum}, where we only considered $Q$ with order-two elements.\footnote{In particular, the stabilizer code condition on the absence of 1-form anomaly of bulk CS $Q$-lines in \cite{buican2021quantum} is replaced by the self-duality of the surface operator defining the RCFT in question. Although we will not pursue it further in this paper, it is possible to show that the self-duality of the surface operators we will construct is related to the absence of a certain (mixed) 1-form 't Hooft anomaly in the bulk.}

Much of the above discussion can be reformulated and several new aspects of the CFT/CS to code map more naturally clarified by folding the diagram in Fig. \ref{fig:TQFT-CFT intro} as in Fig. \ref{folded}. This reformulation has several important properties:
\begin{figure}
    \centering

\tikzset{every picture/.style={line width=0.75pt}} %set default line width to 0.75pt        

\begin{tikzpicture}[x=0.75pt,y=0.75pt,yscale=-1,xscale=1]

%Shape: Parallelogram [id:dp04467542223368459] 
\draw   (264.58,205.85) -- (202.59,237) -- (201.04,85.79) -- (263.03,54.64) -- cycle ;
%Shape: Parallelogram [id:dp3372825549064301] 
\draw  [color={rgb, 255:red, 0; green, 0; blue, 0 }  ,draw opacity=1 ][fill={rgb, 255:red, 74; green, 74; blue, 74 }  ,fill opacity=0.3 ] (144.82,205.85) -- (82.83,237) -- (81.28,85.79) -- (143.27,54.64) -- cycle ;
%Straight Lines [id:da7885048110044185] 
\draw [color={rgb, 255:red, 139; green, 6; blue, 24 }  ,draw opacity=1 ]   (113.05,145.82) -- (201.14,146.08) ;
%Straight Lines [id:da7076530663801579] 
\draw [color={rgb, 255:red, 139; green, 6; blue, 24 }  ,draw opacity=1 ] [dash pattern={on 4.5pt off 4.5pt}]  (201.14,146.08) -- (232.81,145.82) ;
%Shape: Ellipse [id:dp6838728927850737] 
\draw  [color={rgb, 255:red, 139; green, 6; blue, 24 }  ,draw opacity=1 ][fill={rgb, 255:red, 139; green, 6; blue, 24 }  ,fill opacity=1 ] (107.29,145.82) .. controls (107.29,144.25) and (108.58,142.98) .. (110.17,142.98) .. controls (111.76,142.98) and (113.05,144.25) .. (113.05,145.82) .. controls (113.05,147.39) and (111.76,148.66) .. (110.17,148.66) .. controls (108.58,148.66) and (107.29,147.39) .. (107.29,145.82) -- cycle ;
%Shape: Ellipse [id:dp5608346266372702] 
\draw  [fill={rgb, 255:red, 139; green, 6; blue, 24 }  ,fill opacity=1 ] (232.81,145.82) .. controls (232.81,144.25) and (234.1,142.98) .. (235.69,142.98) .. controls (237.28,142.98) and (238.57,144.25) .. (238.57,145.82) .. controls (238.57,147.39) and (237.28,148.66) .. (235.69,148.66) .. controls (234.1,148.66) and (232.81,147.39) .. (232.81,145.82) -- cycle ;
%Shape: Parallelogram [id:dp08465439621193405] 
\draw   (580.58,204.85) -- (518.59,236) -- (517.04,84.79) -- (579.03,53.64) -- cycle ;
%Shape: Parallelogram [id:dp872694981867432] 
\draw  [color={rgb, 255:red, 0; green, 0; blue, 0 }  ,draw opacity=1 ][fill={rgb, 255:red, 74; green, 74; blue, 74 }  ,fill opacity=0.38 ] (460.82,204.85) -- (398.83,236) -- (397.28,84.79) -- (459.27,53.64) -- cycle ;
%Straight Lines [id:da4566002574468726] 
\draw [color={rgb, 255:red, 139; green, 6; blue, 24 }  ,draw opacity=1 ] [dash pattern={on 4.5pt off 4.5pt}]  (517,135) -- (548.67,134.75) ;
%Shape: Ellipse [id:dp7114840307184083] 
\draw  [color={rgb, 255:red, 139; green, 6; blue, 24 }  ,draw opacity=1 ][fill={rgb, 255:red, 139; green, 6; blue, 24 }  ,fill opacity=1 ] (446.05,134.82) .. controls (446.05,133.25) and (447.34,131.98) .. (448.93,131.98) .. controls (450.52,131.98) and (451.81,133.25) .. (451.81,134.82) .. controls (451.81,136.39) and (450.52,137.66) .. (448.93,137.66) .. controls (447.34,137.66) and (446.05,136.39) .. (446.05,134.82) -- cycle ;
%Shape: Ellipse [id:dp8307151774248447] 
\draw  [fill={rgb, 255:red, 139; green, 6; blue, 24 }  ,fill opacity=1 ] (545.79,134.75) .. controls (545.79,133.18) and (547.08,131.9) .. (548.67,131.9) .. controls (550.26,131.9) and (551.55,133.18) .. (551.55,134.75) .. controls (551.55,136.31) and (550.26,137.59) .. (548.67,137.59) .. controls (547.08,137.59) and (545.79,136.31) .. (545.79,134.75) -- cycle ;
%Straight Lines [id:da974823539980397] 
\draw [color={rgb, 255:red, 139; green, 6; blue, 24 }  ,draw opacity=1 ]   (474,160) -- (517,160) ;
%Shape: Ellipse [id:dp4207428867175196] 
\draw  [color={rgb, 255:red, 139; green, 6; blue, 24 }  ,draw opacity=1 ][fill={rgb, 255:red, 139; green, 6; blue, 24 }  ,fill opacity=1 ] (430.24,160) .. controls (430.24,158.43) and (431.53,157.16) .. (433.12,157.16) .. controls (434.71,157.16) and (436,158.43) .. (436,160) .. controls (436,161.57) and (434.71,162.84) .. (433.12,162.84) .. controls (431.53,162.84) and (430.24,161.57) .. (430.24,160) -- cycle ;
%Shape: Ellipse [id:dp7331664324546432] 
\draw  [color={rgb, 255:red, 139; green, 6; blue, 24 }  ,draw opacity=1 ][fill={rgb, 255:red, 139; green, 6; blue, 24 }  ,fill opacity=1 ] (547.29,159.82) .. controls (547.29,158.25) and (548.58,156.98) .. (550.17,156.98) .. controls (551.76,156.98) and (553.05,158.25) .. (553.05,159.82) .. controls (553.05,161.39) and (551.76,162.66) .. (550.17,162.66) .. controls (548.58,162.66) and (547.29,161.39) .. (547.29,159.82) -- cycle ;
%Straight Lines [id:da5658596148834714] 
\draw [color={rgb, 255:red, 139; green, 6; blue, 24 }  ,draw opacity=1 ] [dash pattern={on 4.5pt off 4.5pt}]  (517,160) -- (531.99,159.91) -- (547.29,159.82) ;
%Straight Lines [id:da8195484317063146] 
\draw [color={rgb, 255:red, 139; green, 6; blue, 24 }  ,draw opacity=1 ]   (432,68) -- (433.12,157.16) ;
%Curve Lines [id:da584046343475869] 
\draw [color={rgb, 255:red, 139; green, 6; blue, 24 }  ,draw opacity=1 ]   (495,157) .. controls (496,152) and (501,138) .. (517,135) ;
%Curve Lines [id:da8803215217984794] 
\draw [color={rgb, 255:red, 139; green, 6; blue, 24 }  ,draw opacity=1 ]   (448.93,134.82) .. controls (472,162) and (482,178) .. (493,163) ;
%Straight Lines [id:da4608651358059902] 
\draw [color={rgb, 255:red, 139; green, 6; blue, 24 }  ,draw opacity=1 ]   (436,160) -- (467,160) ;

% Text Node
\draw (155,122.4) node [anchor=north west][inner sep=0.75pt]    {$( q,\overline{p})$};
% Text Node
\draw (83,83.4) node [anchor=north west][inner sep=0.75pt]    {$\mathcal{B}$};
% Text Node
\draw (206.04,85.19) node [anchor=north west][inner sep=0.75pt]    {$\Sigma $};
% Text Node
\draw (476,116.4) node [anchor=north west][inner sep=0.75pt]    {$( q,\overline{p})$};
% Text Node
\draw (399,82.4) node [anchor=north west][inner sep=0.75pt]    {$\mathcal{B}$};
% Text Node
\draw (522.04,84.19) node [anchor=north west][inner sep=0.75pt]    {$\Sigma $};
% Text Node
\draw (438,163.4) node [anchor=north west][inner sep=0.75pt]    {$( r,\overline{s})$};

\end{tikzpicture}
    \caption{Left: Folding the diagram in Fig. \ref{fig:TQFT-CFT intro} turns the surface operator, $S$, into a gapped boundary, $\mathcal{B}$. Right: The non-trivial monodromy between local and twisted sector operators in the CFT is captured by the non-trivial braiding of bulk line operators $(q,\bar p)$ and $(r,\bar s)$.}
    \label{folded}
\end{figure}
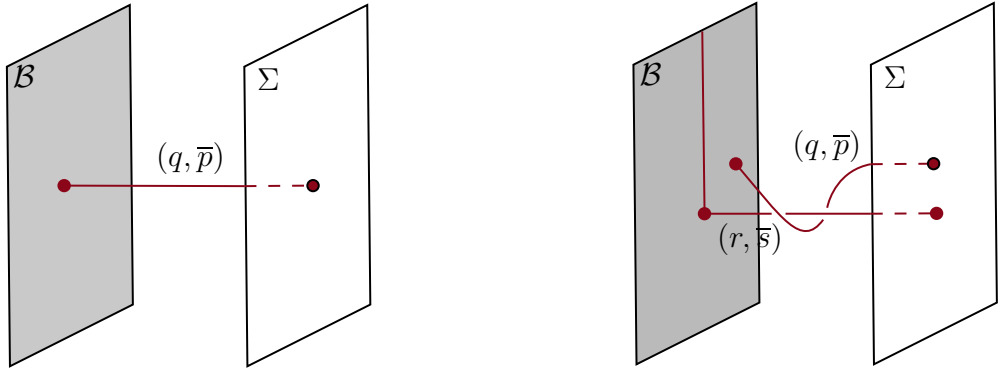

\begin{itemize}
\item Gapped boundaries are specified by Lagrangian subgroups \cite{Kapustin:2010if}. As we will review below, abelianized Pauli groups have symplectic vector spaces associated with them. In this context, self-dual stabilizer codes correspond to Lagrangian vector subspaces with respect to this symplectic form. Therefore, it is natural that the self-duality property of the quantum code can be related to the Lagrangian property of $K_{(Q,[\sigma])}$ (see Sec. \ref{sec:gapped boundaries} for details).\footnote{Relations between systems that do not have gauge fields and CS theory are known to occur in other settings. For example, see the interesting relation between the theory of a 3d $\CN=4$ hypermultiplet and $U(1|1)$ CS theory \cite{Mikhaylov:2015nsa} that arises via B-twisting (the twisting gives rise to a vector field).}
\item In the setup of Fig. \ref{folded}, stabilizers correspond to Wilson lines that stretch between the boundary supporting the CFT and end on the gapped boundary. On the other hand, errors are related to Wilson lines that form a non-trivial junction on the gapped boundary. The resulting junction and line living on the gapped boundary encode the non-trivial mutual monodromy of lines corresponding to stabilizers and lines corresponding to errors. 
\item The construction in Fig. \ref{folded} naturally gives us a SymTFT \cite{Freed:2012bs,Freed:2022qnc,Kaidi:2022cpf,Kaidi:2023maf} for the Abelian 0-form symmetry, $K_{(Q,[\sigma])}$, of $\CT/(Q,[\sigma])$. In particular, different gapped boundary conditions in Fig. \ref{fig:TQFT-CFT intro} correspond to different vertices in $\Gamma_{\mathds{V}}$ and hence to different orbifold theories. Transitions between different gapped boundaries can be obtained by considering fusion of condensation surfaces constructed via higher gauging with these boundaries.
\item By gauging some of the CS 0-form symmetries implemented by surfaces in the previous bullet, we can obtain SymTFTs with non-invertible lines \cite{Kaidi:2022cpf}. In certain cases, bringing these lines to the boundary gives rise to Tambara-Yamigami (TY) categories that describe dualities in the RCFTs we study. More generally, gauging invertible 0-form symmetries trivializes the corresponding surfaces and leads to new non-invertible lines. This way of thinking suggests that theories specified by bulk surfaces that are related by fusion with invertible surfaces should give rise to equivalent quantum codes and that the non-invertible lines that arise in the 0-form gauging play the role of elements of the Clifford group. Indeed, we show that this is the case in large classes of theories.
\end{itemize}

\begin{table}[h!]
\centering
 \begin{tabular}{||c | c||} 
 \hline
 Abelian RCFT & Qudit quantum code \\ [0.5ex] 
 \hline\hline
 Primaries forming Abelian group $K$ & System of qudits with Hilbert space $\CH_{\DV}$. \\
Group $K_{(Q,[\sigma])}$ of primary operators & Stabilizer group $\CS_{(Q,[\sigma])}$ \\
Twisted-sector operators & Error operators
\\ 
CFT Hilbert space & Code subspace $\CC_{(Q,[\sigma])}$\\
Twisted-sector states & States in $\CH_{\DV}$ but not in $\CC_{(Q,[\sigma])}$ \\
[1ex] 
 \hline
 \end{tabular}
 \caption{The Abelian RCFT to qudit code map.}
 \label{tb:CFTCodemap}
\end{table}

\begin{table}[h!]
\centering
 \begin{tabular}{||c | c||} 
 \hline
 Folded Abelian CS theory & Qudit quantum code \\ [0.5ex] 
 \hline\hline
Invertible lines & System of qudits with Hilbert space $\CH_{\DV}$. \\
Lines ending on gapped boundary & Stabilizer group $\CS_{(Q,[\sigma])}$ \\
Lines forming junction on gapped boundary & Error operators
\\ 
Fusion of $S$ with an invertible surface & Certain code equivalences\\
[1ex] 
 \hline
 \end{tabular}
 \caption{The Abelian CS to qudit code map.}
 \label{tb:CSCodemap}
\end{table}

To summarize the results described here, we include table \ref{tb:CFTCodemap} for RCFT / code relations and table \ref{tb:CSCodemap} for CS / code relations. 

The structure of the paper is as follows. In Sec. \ref{sec:qudit codes and CFT-Code map} we start with a review of qudit stabilizer codes and define a colored graph of stabilizer groups $\Gamma_{\CP}$ (i.e., the code graph). We then study the structure of Abelian RCFTs with a particular chiral algebra, $\DV$, and define the orbifold graph, $\Gamma_{\DV}$. We end this section by studying the graph homomorphism, $\mu$, from $\Gamma_{\DV}$ to $\Gamma_{\CP_{\DV}}$ in various explicit examples. We comment on the fact that, for the chiral algebras whose representations form the $E_{2^r}$ Modular Tensor Category (MTC), $\mu$ cannot be an embedding. In Sec. \ref{sec:abelian RCFTS to qudit codes} we define a map from RCFTs to generalized Pauli group elements and show that the it results in a stabilizer code precisely when the surface operator of the bulk Chern-Simons theory corresponding to the CFT is self-dual. In Sec. \ref{sec:generalized pauli group} we study the symmetries of Abelian RCFTs and map the twisted-sector operators to error operators in the quantum code. Finally, in Sec. \ref{sec:gapped boundaries} we discuss the relationship between gapped boundaries and self-dual quantum codes. We also show that certain gapped interfaces correspond to non-self-dual quantum codes. In addition, we study bulk 0-form gauging from the SymTFT point of view to argue that code equivalences in large classes of theories are related to fusion of the defining bulk surfaces with invertible surfaces.  We conclude with some comments and future directions.

We also include three appendices. In appendix \ref{Ap:1}, we prove the impossibility of universally embedding the graph $\Gamma_{\DV}$ in $\Gamma_{\CP_{\DV}}$ by considering the example of Rep$(\DV)=E_{2^r}$ MTCs with $r>1$. In Appendix \ref{ap:comparing with previous work}, we compare the map $\mu$ studied in this paper to our previous work \cite{buican2021quantum}. Finally, in Appendix \ref{ap:non-self-dual}, we discuss a map from abelian RCFTs to non-self-dual quantum codes. 

\smallskip
\noindent
{\bf Note added:} This work is based on notes written by the authors in summer of 2022. In particular, the discussion of folded CS theories and the relation between Lagrangian subgroups and self-dual quantum codes in Sec. \ref{sec:gapped boundaries} was presented by the authors during the workshop \href{https://nms.kcl.ac.uk/gerard.watts/DefSym22/}{Defects and Symmetry 2022} (e.g., see p. 32 of the publicly available slides \href{https://nms.kcl.ac.uk/gerard.watts/DefSym22/Radhakrishnan_From%20Top%20Defects%20To%20QECCs%20Final.pdf}{here}). See also the paper \cite{Barbar:2023ncl} which relates Lagrangian subgroups in abelian CS theory to self-dual additive codes. We thank the authors of \cite{Barbar:2023ncl} for sending us their draft and for encouraging us to publish our old notes.

\section{Qudit stabilizer codes and the CFT-Code map}
In this subsection we begin with a brief review of qudit stabilizer codes before discussing general aspects of the CFT-code map. In particular, in the later parts of this section, we define the orbifold and code graphs and show there is a natural homomorphism between the two. We conclude the section with some examples and argue that the CFT-code map cannot always be an embedding of the orbifold graph into the code graph.

\label{sec:qudit codes and CFT-Code map}

\subsection{Qudit stabilizer codes}

\label{sec:quditcodes}

In this section, we study quantum stabilizer codes acting on systems of qudits of varying dimensions. We start with a quick review of stabilizer codes for qudits of a fixed dimension (e.g., see \cite{Gottesman:1998se,Farinholt_2014,Haah_2017}).

To that end, consider a qudit with a $d$-dimensional Hilbert space, $\CH$. Let us define $X$ and $Z$ operators acting on $\CH$ as follows
\be
X \ket{j}= \ket{j+1 \text{ mod } d}~,~ Z \ket{j}= z^j \ket{j}~,
\ee
where $0\leq j <d$, and $z$ is a primitive $d^{\rm th}$ root of unity. Clearly, the operators $X$ and $Z$ have order $d$ and satisfy
\be
ZX= z XZ~.
\ee

Next, let $\hat z$ be a primitive $d^{\text{th}}$ root of unity when $d$ is odd and a primitive $2d^{\text{th}}$ root of unity when $d$ is even. Let us define the generalized Pauli group, $\CP_n$, which acts on $n$ qudits, to have elements of the form
\be
G(\vec \alpha,\vec \beta):= \hat z^{\lambda} ~ X^{\alpha_1} \otimes \cdots \otimes X^{\alpha_n} \circ Z^{\beta_1} \otimes ... \otimes Z^{\beta_n}=\hat z^{\lambda}X^{\alpha_1}Z^{\beta_1}\otimes\cdots\otimes X^{\alpha_n}Z^{\beta_n} ~,
\ee
determined by the pair of vectors $\vec{\alpha}, \vec \beta \in \DZ_{d}^{n}$ with $\lambda \in \DZ_{d}$ if $d$ is odd, and $\lambda \in \DZ_{2d}$ if $d$ is even. In order to simplify notation, we will write a general Pauli group element (modulo an overall root of unity) as $X^{\vec \alpha} \circ Z^{\vec \beta}$.
We then find the commutation relation
\be
\label{Paulicommutconst}
G(\vec \alpha^{(1)},\vec \beta^{(1)}) G(\vec \alpha^{(2)},\vec \beta^{(2)})= z^{\beta^{(1)}\cdot \alpha^{(2)}-\alpha^{(1)}\cdot \beta^{(2)}} ~G(\vec \alpha^{(2)},\vec \beta^{(2)}) G(\vec \alpha^{(1)},\vec \beta^{(1)})~.
\ee
Clearly, $\CP_n$ is a non-Abelian group . 

Let us consider the abelianized generalized Pauli group, $V_n:=\CP_n/\langle \hat z\rangle$, where $\langle \hat z \rangle$ is the cyclic subgroup generated by $\hat z$ (i.e., the center of $\CP_n$). This group is isomorphic to $\DZ_{d}^{2n}$ \cite{Farinholt_2014, Haah_2017}. We can define a symplectic inner product on $V_n$ given by\footnote{When $d$ is the power of  a prime number, then $V_n$ is in fact a sympelctic vector space over the finite field $\mathds{F}_d$. More generally, it is a symplectic module \cite{Farinholt_2014}.}
\be
\label{eq:symplectic}
\omega((\vec \alpha^{(1)},\vec \beta^{(1)}), (\vec \alpha^{(2)},\vec \beta^{(2)})):=\beta^{(1)}\cdot \alpha^{(2)}-\alpha^{(1)}\cdot \beta^{(2)}~.
\ee

A stabilizer group is an isotropic abelian subgroup, $\CS<\CP_n$. In the case of a self-dual stabilizer group, it is a Lagrangian subgroup (here $\dim_{\mathbb{C}}(\CC)=1$). The subspace, $\CC\subset\CH$, on which information is encoded is defined as the $+1$ eigenspace of $\CS$. The logical operators that act non-trivially on $\CC$ are the elements of $\CP_n$ which commute with $\CS$ but are not in $\CS$. The errors that this code can correct are the elements of $\CP_n$ which do not commute with at least one element of $\CS$.  

In the discussion below, we will need to discuss a product of qudit codes of different dimensions. Therefore, let us consider a system of qudits with Hilbert space 
\be
\CH:= \CH_{1}^{\otimes n_1} \otimes \dots \otimes \CH_{k}^{\otimes n_k},
\ee
and dimensions $d_1,\cdots,d_k$, respectively. The total number of qudits (including those of all dimensions) is $n=n_1+\cdots+ n_k$. Now, consider the Pauli group 
\be
\label{eq:genPauli}
\CP:= \CP_{n_1} \times \dots \times \CP_{n_k}~,
\ee
where $\CP_{n_i}$ acts on the $n_i$ qudits with Hilbert space $\CH_i^{\otimes n_i}$ and dimension $d_i$. For example, suppose we have a product of $n_1$ qudits with dim$(\CH_1)=d_1$ and $n_2$ qudits with dim$(\CH_1)=d_2$. An element of the generalized Pauli group (modulo overall factors of roots of unity) will be denoted as follows
\be
(X_{(d_1)}^{\vec \alpha} \circ Z_{(d_1)}^{\vec \beta}) \otimes (X_{(d_2)}^{\vec \gamma} \circ Z_{(d_2)}^{\vec \delta})=X_{(d_1)}^{\vec \alpha} \otimes X_{(d_2)}^{\vec \gamma} \circ Z_{(d_1)}^{\vec \beta} \otimes Z_{(d_2)}^{\vec \delta}~,
\ee
where $\vec \alpha, \vec \beta$ are length $n_1$ vectors, while $\vec \gamma, \vec \delta$ are length $n_2$ vectors. The subscripts on the $X$ and $Z$ matrices specify the dimension of the qudit Hilbert spaces they act on. 

Of particular importance for us are self-dual stabilizer codes (although we will consider more general codes in Appendix \ref{ap:non-self-dual}). These are codes for which the code subspace is 1-dimensional. In this case, the stabilizer group has order $|\CS|=d_1^{n_1}\dots d_k^{n_k}$. To understand this statement, consider the projector 
\be
\Omega_{\CC}:=\frac{1}{|\CS|} \sum_{s\in \CS} s~,
\ee
onto the code subspace. The dimension of $\CC$ is the trace of this operator, and, noting that both $X$ and $Z$ are traceless, we find
\be
\text{dim}_{\mathbb{C}}(\CC)={\rm Tr}\ \Omega_{\CC}= \frac{\text{dim}(\CH)}{|\CS|}= \frac{d_1^{n_1}\dots d_k^{n_k}}{|\CS|}~.
\ee
Therefore, when the stabilizer group has order $d_1^n\dots d_k^{n_k}$, the code subspace is 1-dimensional. In this case, $\CS$ defines a Lagrangian subgroup of $V$. There are clearly no non-trivial logical operators. In particular, all elements in the generalized Pauli group which are not in the stabilizer group are errors that act on the code subspace. The resulting code is an error-detection code. An error operator acting on the code subspace necessarily takes a state outside the code subspace. Then, by definition, there is at least one element in the stabilizer group $\CS$ for which the eigenvalue of the this state is not equal to $1$. Therefore, $\CS$ can be used to detect errors. 

Given a set of qudits of various dimensions and the generalized Pauli group, $\CP$, defined as in \eqref{eq:genPauli}, we can construct a colored graph of self-dual stabilizer codes, $\Gamma_{\CP}$, called the \lq\lq code graph" as follows:
\begin{itemize}
    \item {\it Vertices} are labeled by self-dual stabilizer groups, $\CS_i<\CP$.
    \item If $|\CS_i \cap \CS_j|>1$ (for $i\ne j$), then a non-trivial {\it edge} is labelled by the group, $\CS_i \cap \CS_j$, between the vertices labelled by $\CS_i$ and $\CS_j$.
\end{itemize}

Unitary operators which leave the generalized Pauli group invariant under conjugation are called Clifford operators. These define an equivalence relation on the set of stabilizer codes and the codes in an equivalence class are called equivalent codes. We will study Clifford operations in detail in Sec. \ref{CodeEqSec} where we study dualities between CFTs.

\subsection{General structure of a CFT-Code map}

In the previous subsection, we studied qudit stabilizer codes and defined the graph, $\Gamma_{\CP}$, of self-dual stabilizer codes. In this section, we will study the structure of abelian RCFTs sharing a fixed chiral algebra, $\DV$, and define the \lq\lq orbifold graph," $\Gamma_{\DV}$. We will then show that the map from CFTs to quantum codes satisfying some natural constraints is a graph homomorphism from $\Gamma_{\DV}$ to $\Gamma_{\CP}$. 

\subsubsection{Abelian RCFTs and the Orbifold Graph}

\begin{figure}[h!]
    \centering

\tikzset{every picture/.style={line width=0.75pt}} %set default line width to 0.75pt        

\begin{tikzpicture}[x=0.75pt,y=0.75pt,yscale=-1,xscale=1]

%Shape: Parallelogram [id:dp04467542223368459] 
\draw   (444.58,203.85) -- (382.59,235) -- (381.04,83.79) -- (443.03,52.64) -- cycle ;
%Shape: Parallelogram [id:dp3372825549064301] 
\draw  [color={rgb, 255:red, 245; green, 166; blue, 35 }  ,draw opacity=1 ][fill={rgb, 255:red, 248; green, 198; blue, 155 }  ,fill opacity=1 ] (353.82,203.85) -- (291.83,235) -- (290.28,83.79) -- (352.27,52.64) -- cycle ;
%Shape: Parallelogram [id:dp7882685111742219] 
\draw   (265.95,203.85) -- (203.95,235) -- (202.4,83.79) -- (264.4,52.64) -- cycle ;
%Straight Lines [id:da08633876367548532] 
\draw [color={rgb, 255:red, 139; green, 6; blue, 24 }  ,draw opacity=1 ]   (234.17,143.82) -- (290.38,144.08) ;
%Straight Lines [id:da7942076943577906] 
\draw  [dash pattern={on 4.5pt off 4.5pt}]  (290.38,144.08) -- (322.05,143.82) ;
%Straight Lines [id:da7885048110044185] 
\draw [color={rgb, 255:red, 139; green, 6; blue, 24 }  ,draw opacity=1 ]   (322.05,143.82) -- (381.14,144.08) ;
%Straight Lines [id:da7076530663801579] 
\draw [color={rgb, 255:red, 139; green, 6; blue, 24 }  ,draw opacity=1 ] [dash pattern={on 4.5pt off 4.5pt}]  (381.14,144.08) -- (412.81,143.82) ;
%Shape: Ellipse [id:dp6838728927850737] 
\draw  [color={rgb, 255:red, 139; green, 6; blue, 24 }  ,draw opacity=1 ][fill={rgb, 255:red, 139; green, 6; blue, 24 }  ,fill opacity=1 ] (316.29,143.82) .. controls (316.29,142.25) and (317.58,140.98) .. (319.17,140.98) .. controls (320.76,140.98) and (322.05,142.25) .. (322.05,143.82) .. controls (322.05,145.39) and (320.76,146.66) .. (319.17,146.66) .. controls (317.58,146.66) and (316.29,145.39) .. (316.29,143.82) -- cycle ;
%Shape: Ellipse [id:dp18115810421979095] 
\draw  [color={rgb, 255:red, 139; green, 6; blue, 24 }  ,draw opacity=1 ][fill={rgb, 255:red, 139; green, 6; blue, 24 }  ,fill opacity=1 ] (228.41,143.82) .. controls (228.41,142.25) and (229.7,140.98) .. (231.29,140.98) .. controls (232.88,140.98) and (234.17,142.25) .. (234.17,143.82) .. controls (234.17,145.39) and (232.88,146.66) .. (231.29,146.66) .. controls (229.7,146.66) and (228.41,145.39) .. (228.41,143.82) -- cycle ;
%Shape: Ellipse [id:dp5608346266372702] 
\draw  [fill={rgb, 255:red, 139; green, 6; blue, 24 }  ,fill opacity=1 ] (412.81,143.82) .. controls (412.81,142.25) and (414.1,140.98) .. (415.69,140.98) .. controls (417.28,140.98) and (418.57,142.25) .. (418.57,143.82) .. controls (418.57,145.39) and (417.28,146.66) .. (415.69,146.66) .. controls (414.1,146.66) and (412.81,145.39) .. (412.81,143.82) -- cycle ;
%Straight Lines [id:da6157708076318841] 
\draw    (227.69,264.5) -- (408.58,263.51) ;
\draw [shift={(410.58,263.5)}, rotate = 179.69] [color={rgb, 255:red, 0; green, 0; blue, 0 }  ][line width=0.75]    (10.93,-3.29) .. controls (6.95,-1.4) and (3.31,-0.3) .. (0,0) .. controls (3.31,0.3) and (6.95,1.4) .. (10.93,3.29)   ;
\draw   (235.42,267.68) .. controls (232.16,266.01) and (228.88,265.02) .. (225.58,264.7) .. controls (228.9,264.36) and (232.22,263.33) .. (235.58,261.65) ;

% Text Node
\draw (215,118.4) node [anchor=north west][inner sep=0.75pt]    {$\vec{g} +\vec{p}$};
% Text Node
\draw (411,117.4) node [anchor=north west][inner sep=0.75pt]    {$\vec{p}$};
% Text Node
\draw (290,81.4) node [anchor=north west][inner sep=0.75pt]    {$S( Q,[ \sigma ])$};
% Text Node
\draw (207.4,82.19) node [anchor=north west][inner sep=0.75pt]    {$\Sigma _{1}$};
% Text Node
\draw (386.04,83.19) node [anchor=north west][inner sep=0.75pt]    {$\Sigma _{2}$};
% Text Node
\draw (313,269.4) node [anchor=north west][inner sep=0.75pt]    {$I$};

\end{tikzpicture}
    \caption{The pairing of 2d CFT left and right movers on $\Sigma_1$ and $\Sigma_2$ of $\CT/(Q,[\sigma])$ is specified by an abelian CS theory on $X=\Sigma \times I$, with the surface operator, $S(Q,[\sigma])$, inserted in the bulk.}
    \label{fig:TQFT-CFT non-trivial surface}
\end{figure}
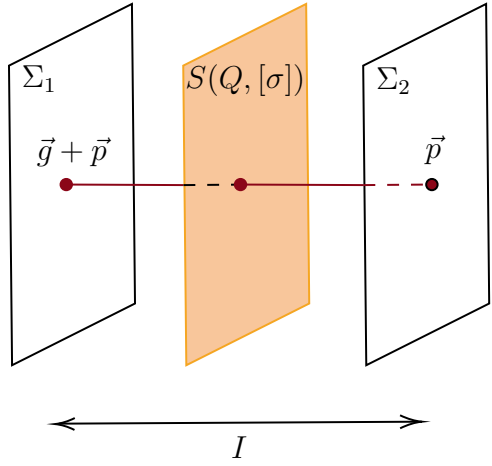

Consider CFTs with a fixed chiral algebra, $\mathds{V}$. The chiral primaries of these CFTs are labelled by elements of the set, Rep$(\mathds{V})$, of irreducible representations of $\mathds{V}$. We will assume that the group formed by the chiral primaries (and their descendants) under fusion is an abelian group, $K$. Any such theory is an orbifold of the \lq\lq Cardy case" RCFT, $\CT$, for $\mathds{V}$ (e.g., see the discussion in \cite{fuchs2002tft,fuchs2004tft,Frohlich:2009gb}). This latter RCFT has $T^2$ partition function
\begin{equation}\label{ccZ}
Z_{\CT}(q)=\sum_{\vec p}\chi_{\vec p}(q)\bar\chi_{\overline{\vec p}}(\bar q)~, \ \ \vec p+\overline{\vec p}=\vec 0~, \ \ N_{\overline{\vec p}},N_{\vec p},N_{\vec 0}\in{\rm Rep}(\mathds{V})~.
\end{equation}
In \eqref{ccZ}, Rep$(\DV)$ is a Modular Tensor Category (MTC) whose objects are in one-to-one correspondence with the (genuine) line operators of a bulk Abelian Chern-Simons theory, $\CI$.\footnote{We will return to non-genuine lines (i.e, lines attached to surfaces) later when discussing code equivalence and CS theory.} The set of such theories have been completely classified \cite{wall}. Indeed, they can be written as direct products of the following \lq\lq prime" factors
\begin{eqnarray}\label{MTCclass}
A_{2^r}&\sim&\mathbb{Z}_{2^r}~,\ A_{q^r}\sim\mathbb{Z}_{p^r}~,\ B_{2^r}\sim\mathbb{Z}_{2^r}~, B_{q^r}\sim\mathbb{Z}_{q^r}~,\ C_{2^r}\sim\mathbb{Z}_{2^r}~, \cr D_{2^r}&\sim&\mathbb{Z}_{2^r}\ , E_{2^r}\sim\mathbb{Z}_{2^r}\times\mathbb{Z}_{2^r},\ \ \, F_{2^r}\sim\mathbb{Z}_{2^r}\times\mathbb{Z}_{2^r}~,
\end{eqnarray}
where the labels on the lefthand sides of \eqref{MTCclass} denote CS theories\footnote{Note these groups are {\it not} related to the Lie groups denoted by the same letters. Instead, we are using the nomenclature in \cite{wang2020and}.} with Wilson line fusion rules given by the abelian groups on the righthand sides, and $q>2$ is a prime number. Therefore, we should think of a vector label, $\vec p$, denoting an element of ${\rm Rep}(\mathds{V})$ as valued in the following product group 
\begin{eqnarray}\label{pval}
\vec{p}\in\prod_r\Big(\mathbb{Z}_{2^r}^{n_{A_{2^r}}}\times\mathbb{Z}_{2^r}^{n_{B_{2^r}}}\times\mathbb{Z}_{2^r}^{n_{C_{2^r}}}\times\mathbb{Z}_{2^r}^{n_{D_{2^r}}}&\times& \Big[\mathbb{Z}_{2^r}\times\mathbb{Z}_{2^r}\Big]^{n_{E_{2^r}}}\times\Big[\mathbb{Z}_{2^r}\times\mathbb{Z}_{2^r}\Big]^{n_{F_{2^r}}}\cr&\times&\prod_q\Big[\mathbb{Z}_{q^r}^{n_{A_{q^r}}}\times\mathbb{Z}_{q^r}^{n_{B_{q^r}}}\Big]\Big):=K~,
\end{eqnarray}
where $n_{X}$ is the number of independent factors of CS theory $X$. The CFT, $\CT$, with charge-conjugation partition function corresponds to the trivial surface operator of the bulk CS theory. This is the surface operator that acts trivially on the Wilson lines \cite{Kapustin:2010if,Fuchs:2012dt} (see Fig. \ref{fig:TQFT-CFT non-trivial surface} with trivial $S(Q,[\sigma])$).

We can construct all other RCFTs sharing chiral algebra $\mathds{V}$ from $\CT$ by orbifolding $\CT$ by a non-anomalous 0-form symmetry subgroup, $Q\le K$. In other words, the topological defects (Verlinde lines in this case) implementing $Q$ must have trivial $F$ symbols in $H^3(Q,U(1))$. More explicitly, $F$ can be written as
\be
\label{eq:Fmatrixdef}
F(\vec g, \vec h,\vec k)= \prod_i \bigg\{
	\begin{array}{ll}
		1 & \mbox{   if } h_i + k_i < n_i\\
		\theta(e_i)^{g_in_i}       & \mbox{  if } h_i + k_i \geq n_i ~,
	\end{array}
\ee
where $e_i$ is a basis for the cyclic factors in \eqref{pval}, and $\vec g= \sum_i g_i e_i$. In this equation, $n_i$ is the order of the $i^{\text{th}}$ cyclic factor, and $\theta_{\vec p}:={\rm exp}(2\pi ih_{\vec p})$, where $h_{\vec p}$ is the holomorphic scaling dimension of an operator in representation $\vec p$. If $Q$ is non-anomalous, $F$ is a 3-coboundary
\be
\label{eq:anomaly vanishing conditions}
F(\vec h_1,\vec h_2,\vec h_3)= \frac{\tau(\vec h_2, \vec h_3) \tau(\vec h_1,\vec h_2 + \vec h_3) }{\tau(\vec h_1 + \vec h_2, \vec h_3)   \tau(\vec h_1, \vec h_2) }   ~ \forall \vec h_1, \vec h_2, \vec h_3 \in Q~,
\ee
with 2-cochain, $\tau$. In fact, there is a short-cut that allows one to efficiently check whether $Q$ is non-anomalous. Indeed, $Q$ is non-anomalous if and only if $\theta_{\vec h}^{O_{\vec h}}=1$ for all $\vec h \in Q$, and $O_{\vec h}$ is the order of $\vec h$ in $Q$ \cite{fuchs2004tft}.

However, to calculate the partition function of the CFT after orbifolding, we need to make a choice of the $\tau$ solving \eqref{eq:anomaly vanishing conditions}. The orbifold torus partition function is then
\begin{equation}
\label{HZ}
Z_{\CT/(Q,[\sigma])}=\sum_{\vec g\in Q}\sum_{\vec p\in B_{\vec g}}\chi_{\vec p}(q)\bar\chi_{\overline{\vec p+\vec g}}(\bar q)~,
\end{equation}
where $[\sigma]$ is an equivalence class in $H^2(Q,U(1))$ corresponding to the discrete torsion, and 
\be
\label{pconst2}
B_{\vec g}:=\left\{\vec p\ \Big|\ S_{\vec h,\vec  p} ~ \Xi(\vec h,\vec g)=1 ~,\ \forall \vec h \in Q\right\}~,
\ee
where we take\footnote{Note that our $S$ matrix is unnormalized. It differs from the unitary $S$ matrix by a normalization factor $\sqrt{|K|}$, where $|K|$ is the number of Wilson lines in the CS theory associated with our RCFT. \label{normS}} 
\bea
\label{SXiZ2k}
S_{\vec h,\vec p}:={\theta_{\vec h+\vec p}\over\theta_{\vec h}\theta_{\vec p}}~,~~~ \ \Xi(\vec g,\vec h):=R(\vec{h},\vec g) \frac{\tau(\vec h,\vec g)\sigma(\vec h,\vec g)}{\tau(\vec g, \vec h)\sigma(\vec g, \vec h)}~,\ \ ~.
\eea
and $R(\vec h, \vec g)$ can be written in terms of $\theta_{\vec g}$ as
\be
\label{eq:Rmatrixdef}
R(\vec h,\vec g)= \prod_i (\theta_{e_i})^{h_ig_i} \prod_{i<j} (S_{e_i,e_j})^{h_i g_j}~.
\ee
In this expression, the $e_i$ form a basis for the cyclic factors in \eqref{pval}, and $\vec g= \sum_i g_i e_i$. Note that both $R(\vec h, \vec g)$ and $\tau(\vec g, \vec h)$ depend on a choice of basis in Rep$(\mathds{V})$, but $\Xi(\vec g, \vec h)$ is basis independent. We denote $K_{(Q,[\sigma])}$ to be the group formed by the primaries of $\mathcal{T}/(Q,[\sigma])$ under fusion. Note that for the charge-conjugation RCFT, we have $K_{(\DZ_1,[1])}\cong K$.  However, more generally this is not necessarily the case.

In the language of the bulk TQFT, the group $Q$ is a 1-form symmetry group implemented by certain line operators. Orbifolding the CFT, $\CT$, by $Q$ corresponds to higher-gauging $Q$ on a surface to obtain a surface operator, $S(Q,[\sigma])$ \cite{Roumpedakis:2022aik}. The partition function of the CFT $\CT/(Q,[\sigma])$ is determined by the action of this surface operator on the Wilson lines as in Fig. \ref{fig:TQFT-CFT non-trivial surface} \cite{Kapustin:2010if,Fuchs:2012dt}. 

We are now ready to define the RCFT \lq\lq orbifold graph," $\Gamma_{\mathds{V}}$. In particular, for a given chiral algebra, $\Gamma_{\mathds{V}}$ is a colored graph defined as follows:
\begin{itemize}
    \item The {\it vertices} are labeled by the groups, $K_{(Q_i,[\sigma_i])}$, formed by the primaries of the CFT, $\CT/(Q_i,[\sigma_i])$, under fusion.
    \item If $|K_{(Q_i,[\sigma_i])}\cap K_{(Q_j,[\sigma_j])}|>1$ (for $i\ne j$), then a non-trivial {\it edge} between the vertices corresponding to $K_{(Q_i,[\sigma_i])}$ and $K_{(Q_j,[\sigma_j])}$ is labelled by the group $K_{(Q_i,[\sigma_i])}\cap K_{(Q_j,[\sigma_j])}$.
\end{itemize}
The orbifold graph captures the finite set of abelian RCFTs with a specified chiral algebra. In particular, the edges and coloring of the graph capture the orbifold structure of these RCFTs.\footnote{The orbifold graph is closely related to the orbifold groupoid studied in \cite{Gaiotto:2020iye}.} 

\subsubsection{From orbifold graphs to qudit stabilizer codes}
In previous sections, we studied qudit stabilizer codes and defined the code graph, $\Gamma_{\CP}$. We also studied abelian RCFTs with a given chiral algebra and defined the orbifold graph, $\Gamma_{\DV}$. In this section, we study general maps from CFTs to quantum codes which satisfy certain natural assumptions and explain how $\Gamma_{\CP}$ and $\Gamma_{\DV}$ are related.

Given a chiral algebra $\mathds{V}$, $K$ is an abelian group by assumption. Therefore, we have a decomposition
\be
K \cong \DZ_{p_1^{n_1}} \times \dots \times \DZ_{p_{k}^{n_k}}~,
\ee
into cyclic groups of prime power order. Consider a system of $n$ qudits with Hilbert space 
\be
\CH:=\CH_{1} \otimes \dots \otimes \CH_{n}~,
\ee
with dimensions $p_1^{n_1}, \cdots , p_{k}^{n_k}$ respectively. Next, consider the direct product of Pauli groups 
\be
\CP_{\DV}:= \CP_{p_1^{n_1}} \times \cdots \times \CP_{p_k^{n_k}}~,
\ee
acting on this Hilbert space. We want to associate a stabilizer code constructed from $\CP_{\DV}$ with an abelian RCFT, $\CT/(Q,[\sigma])$, having chiral algebra $\DV$. To that end, let $S_{(Q,[\sigma])}$ be the stabilizer group of this code. From the discussion in the introduction, recall that it is natural to require that
\be
\label{eq:CFTcode condition 1}
K_{(Q,[\sigma])} \cong \CS_{(Q,[\sigma])}~.
\ee
Note that the CFT, $\CT$, has $p_1^{n_1}\dots p_k^{n_k}$ primary operators. Moreover, since all orbifolds, $\CT/(Q,[\sigma])$, are obtained via an invertible gauging of a finite symmetry group, we find that $\CT/(Q,[\sigma])$ has $p_1^{n_1}\dots p_k^{n_k}$ primary operators as well. Therefore, we find that $|K_{(Q,[\sigma])} |=p_1^{n_1}\dots p_k^{n_k}$. Then, for \eqref{eq:CFTcode condition 1} to hold, we require that $|S_{(Q,[\sigma]}|=p_1^{n_1}\dots p_k^{n_k}$. In other words, $S_{(Q,[\sigma])}$ is a self-dual stabilizer code. As an additional constraint note that, in order for the quantum codes to capture the orbifold structure of the RCFTs, it is natural to require that
\be
\label{eq:CFTcode condition 2}
K_{(Q_1,[\sigma_1])} \cap K_{(Q_2,[\sigma_2])} \cong  \CS_{(Q_1,[\sigma_1])} \cap \CS_{(Q_2,[\sigma_2])}~.
\ee
Therefore, in terms of $\Gamma_{\DV}$ and $\Gamma_{\CP_{\DV}}$, we have a graph homomorphism
\be
\mu: \Gamma_{\DV} \to \Gamma_{\CP_{\DV}}~. 
\ee
In the next section, we study explicit examples of $\mu$ for certain choices of $\DV$. 

\subsection{Examples}

Consider a chiral algebra, $\DV$, satisfying ${\rm Rep}(\DV)\cong B_{q^r}$ as MTCs, where $q>2$ is prime (see \cite{wang2020and} for the full data of this MTC\footnote{This theory is a Galois conjugate of $SU(q^r)_1$.}). This MTC has fusion rules $\DZ_{q^r}$, and the topological spins are given by 
\be
\theta(p)=e^{\frac{2 \pi i p^2}{q^r}}~,
\ee
where $p \in \{0,1,...,q^r-1\}$. Consider the CFT $\CT$ with partition function given by
\be 
Z_{\CT}:= \sum_i \chi_{p} \bar \chi_{\bar p}~,
\ee
where $p \in \{0,1,...,q^r-1\}$. Other CFTs with the same chiral algebra can be constructed by orbifolding by non-anomalous chiral algebra-preserving symmetries. Since the $F$ symbols for the full category, $B_{q^r}$, are trivial, we can gauge any subgroup of $\DZ_{q^r}$. Subgroups of $\DZ_{q^r}$ are generated by $q^{r-t}$ for $t=0,...,r$.

Let us consider $q=3$ and the first few values for $r$ below. Since the fusion of chiral primaries form the group $\DZ_{3^r}$, we choose a single qudit with a Hilbert space of dimension $3^r$. 

\subsubsection{$r=1$}

In this case we have two CFTs: the charge-conjugation CFT, $\CT$, and the orbifold, $\CT/\DZ_3$. They have the partition functions
\bea
Z_{\CT}:= \chi _0 \bar{\chi }_{\bar{0}}+\chi _1 \bar{\chi }_{\bar{1}}+\chi _2 \bar{\chi }_{\bar{2}}~,\nonumber\\
Z_{\CT/\DZ_3}:= \chi _0 \bar{\chi }_{\bar{0}}+\chi _1 \bar{\chi }_{\bar{2}}+\chi _2 \bar{\chi }_{\bar{1}}~.
\eea
These CFTs are mapped to stabilizer groups in the Pauli group acting on the Hilbert space of a single qutrit. The two CFTs above can be depicted using the graph in Fig. \ref{fig:orbgraph1} with two vertices. 
\begin{figure}[h!]
\centering

\tikzset{every picture/.style={line width=0.75pt}} %set default line width to 0.75pt        

\begin{tikzpicture}[x=0.75pt,y=0.75pt,yscale=-1,xscale=1]

%Shape: Circle [id:dp48759178582002205] 
\draw  [fill={rgb, 255:red, 0; green, 0; blue, 0 }  ,fill opacity=1 ] (405,99.21) .. controls (405,97.44) and (406.44,96) .. (408.21,96) .. controls (409.98,96) and (411.42,97.44) .. (411.42,99.21) .. controls (411.42,100.98) and (409.98,102.42) .. (408.21,102.42) .. controls (406.44,102.42) and (405,100.98) .. (405,99.21) -- cycle ;
%Shape: Circle [id:dp6351142949072132] 
\draw  [fill={rgb, 255:red, 0; green, 0; blue, 0 }  ,fill opacity=1 ] (405,178.21) .. controls (405,176.44) and (406.44,175) .. (408.21,175) .. controls (409.98,175) and (411.42,176.44) .. (411.42,178.21) .. controls (411.42,179.98) and (409.98,181.42) .. (408.21,181.42) .. controls (406.44,181.42) and (405,179.98) .. (405,178.21) -- cycle ;
%Shape: Circle [id:dp8307635979760861] 
\draw  [fill={rgb, 255:red, 0; green, 0; blue, 0 }  ,fill opacity=1 ] (133,98.21) .. controls (133,96.44) and (134.44,95) .. (136.21,95) .. controls (137.98,95) and (139.42,96.44) .. (139.42,98.21) .. controls (139.42,99.98) and (137.98,101.42) .. (136.21,101.42) .. controls (134.44,101.42) and (133,99.98) .. (133,98.21) -- cycle ;
%Shape: Circle [id:dp3568358241006482] 
\draw  [fill={rgb, 255:red, 0; green, 0; blue, 0 }  ,fill opacity=1 ] (133,177.21) .. controls (133,175.44) and (134.44,174) .. (136.21,174) .. controls (137.98,174) and (139.42,175.44) .. (139.42,177.21) .. controls (139.42,178.98) and (137.98,180.42) .. (136.21,180.42) .. controls (134.44,180.42) and (133,178.98) .. (133,177.21) -- cycle ;
%Straight Lines [id:da21082779182767653] 
\draw    (319.5,142) -- (359,142) ;
\draw [shift={(361,142)}, rotate = 180] [color={rgb, 255:red, 0; green, 0; blue, 0 }  ][line width=0.75]    (10.93,-3.29) .. controls (6.95,-1.4) and (3.31,-0.3) .. (0,0) .. controls (3.31,0.3) and (6.95,1.4) .. (10.93,3.29)   ;

% Text Node
\draw (422,169.4) node [anchor=north west][inner sep=0.75pt]    {$\CS_{1} \ \cong \ \mathbb{Z}_{3} =\langle l\rangle $};
% Text Node
\draw (421,89.4) node [anchor=north west][inner sep=0.75pt]    {$\CS_{2} \ \cong \mathbb{Z}_{3} =\langle m\rangle $};
% Text Node
\draw (150,168.4) node [anchor=north west][inner sep=0.75pt]    {$\mathbb{Z}_{3} =\langle \chi _{1}\overline{\chi }_{\overline{1}} \rangle ~ (\mathcal{\CT})$};
% Text Node
\draw (149,88.4) node [anchor=north west][inner sep=0.75pt]    {$\mathbb{Z}_{3} =\langle \chi _{1}\overline{\chi }_{\overline{2}} \rangle ~ (\mathcal{T}/\DZ_3 )$};
% Text Node
\draw (332,116.4) node [anchor=north west][inner sep=0.75pt]    {$\mu $};

\end{tikzpicture}
\caption{Left: The orbifold graph of two CFTs that do not share any non-trivial primaries. The vertices are labelled by the isomorphism class of the group formed by primaries under fusion. Right: A graph of two stabilizer groups which do not share any non-trivial elements.}
\label{fig:orbgraph1}
\end{figure}
The generalized Pauli group elements $l$ and $m$ must be two different order $3$ elements so that distinct quantum codes are associated with the two CFTs. One consistent choice is
\be
l= X_{(3)} ~, \ \ \  m=Z_{(3)}~.
\ee

\subsubsection{$r=2$}

\begin{figure}[h!]
\centering
\tikzset{every picture/.style={line width=0.75pt}} %set default line width to 0.75pt        

\begin{tikzpicture}[x=0.75pt,y=0.75pt,yscale=-1,xscale=1]

%Straight Lines [id:da6943475046616141] 
\draw    (91.21,53.41) -- (91.21,145.64) ;
%Straight Lines [id:da21082779182767653] 
\draw    (335.5,151) -- (377,151) ;
\draw [shift={(379,151)}, rotate = 180] [color={rgb, 255:red, 0; green, 0; blue, 0 }  ][line width=0.75]    (10.93,-3.29) .. controls (6.95,-1.4) and (3.31,-0.3) .. (0,0) .. controls (3.31,0.3) and (6.95,1.4) .. (10.93,3.29)   ;
%Straight Lines [id:da7884297379136038] 
\draw    (91.21,149.98) -- (91.21,242.22) ;
%Straight Lines [id:da4812627386892334] 
\draw    (425.21,55.41) -- (425.21,147.64) ;
%Straight Lines [id:da7239970577170927] 
\draw    (425.21,151.98) -- (425.21,244.22) ;
%Shape: Circle [id:dp5593770605748442] 
\draw  [fill={rgb, 255:red, 0; green, 0; blue, 0 }  ,fill opacity=1 ] (88,56.61) .. controls (88,54.84) and (89.44,53.41) .. (91.21,53.41) .. controls (92.98,53.41) and (94.42,54.84) .. (94.42,56.61) .. controls (94.42,58.39) and (92.98,59.82) .. (91.21,59.82) .. controls (89.44,59.82) and (88,58.39) .. (88,56.61) -- cycle ;
%Shape: Circle [id:dp11600223962328615] 
\draw  [fill={rgb, 255:red, 0; green, 0; blue, 0 }  ,fill opacity=1 ] (88,242.22) .. controls (88,240.45) and (89.44,239.01) .. (91.21,239.01) .. controls (92.98,239.01) and (94.42,240.45) .. (94.42,242.22) .. controls (94.42,243.99) and (92.98,245.43) .. (91.21,245.43) .. controls (89.44,245.43) and (88,243.99) .. (88,242.22) -- cycle ;
%Shape: Circle [id:dp6469519138869866] 
\draw  [fill={rgb, 255:red, 0; green, 0; blue, 0 }  ,fill opacity=1 ] (88,148.85) .. controls (88,147.08) and (89.44,145.64) .. (91.21,145.64) .. controls (92.98,145.64) and (94.42,147.08) .. (94.42,148.85) .. controls (94.42,150.62) and (92.98,152.06) .. (91.21,152.06) .. controls (89.44,152.06) and (88,150.62) .. (88,148.85) -- cycle ;
%Shape: Circle [id:dp10678127815418592] 
\draw  [fill={rgb, 255:red, 0; green, 0; blue, 0 }  ,fill opacity=1 ] (422,247.43) .. controls (422,245.66) and (423.44,244.22) .. (425.21,244.22) .. controls (426.98,244.22) and (428.42,245.66) .. (428.42,247.43) .. controls (428.42,249.2) and (426.98,250.64) .. (425.21,250.64) .. controls (423.44,250.64) and (422,249.2) .. (422,247.43) -- cycle ;
%Shape: Circle [id:dp25708101694649865] 
\draw  [fill={rgb, 255:red, 0; green, 0; blue, 0 }  ,fill opacity=1 ] (422,151.98) .. controls (422,150.21) and (423.44,148.77) .. (425.21,148.77) .. controls (426.98,148.77) and (428.42,150.21) .. (428.42,151.98) .. controls (428.42,153.76) and (426.98,155.19) .. (425.21,155.19) .. controls (423.44,155.19) and (422,153.76) .. (422,151.98) -- cycle ;
%Shape: Circle [id:dp22435247442153128] 
\draw  [fill={rgb, 255:red, 0; green, 0; blue, 0 }  ,fill opacity=1 ] (422,58.61) .. controls (422,56.84) and (423.44,55.41) .. (425.21,55.41) .. controls (426.98,55.41) and (428.42,56.84) .. (428.42,58.61) .. controls (428.42,60.39) and (426.98,61.82) .. (425.21,61.82) .. controls (423.44,61.82) and (422,60.39) .. (422,58.61) -- cycle ;

% Text Node
\draw (105,143.8) node [anchor=north west][inner sep=0.75pt]    {$\mathbb{Z}_{3} \times \mathbb{Z}_{3} =\langle \chi _{3}\overline{\chi }_{\overline{3}} ,\chi _{3}\overline{\chi }_{\overline{6}} \rangle  ~ (\CT/\DZ_3)$};
% Text Node
\draw (104,39.6) node [anchor=north west][inner sep=0.75pt]    {$\mathbb{Z}_{9} =\langle \chi _{1}\overline{\chi }_{\overline{8}} \rangle ~ (\CT/\DZ_9)$};
% Text Node
\draw (105,92.97) node [anchor=north west][inner sep=0.75pt]    {$\mathbb{Z}_{3} =\langle \chi _{3}\overline{\chi }_{\overline{6}} \rangle $};
% Text Node
\draw (104,235.3) node [anchor=north west][inner sep=0.75pt]    {$\mathbb{Z}_{9} =\langle \chi _{1}\overline{\chi }_{\overline{1}} \rangle ~ (\CT)$};
% Text Node
\draw (104,190.82) node [anchor=north west][inner sep=0.75pt]    {$\mathbb{Z}_{3} =\langle \chi _{3}\overline{\chi }_{\overline{3}} \rangle $};
% Text Node
\draw (439,145.8) node [anchor=north west][inner sep=0.75pt]    {$\CS_{2} \ \cong \mathbb{Z}_{3} \times \mathbb{Z}_{3} =\langle l^{3} ,m^{3} \rangle $};
% Text Node
\draw (438,41.6) node [anchor=north west][inner sep=0.75pt]    {$\CS_{3} \ \cong \mathbb{Z}_{9} =\langle m\rangle $};
% Text Node
\draw (439,94.97) node [anchor=north west][inner sep=0.75pt]    {$\mathbb{Z}_{3} =\langle m^{3} \rangle $};
% Text Node
\draw (439,237.3) node [anchor=north west][inner sep=0.75pt]    {$\CS_{1} \cong \mathbb{Z}_{9} =\langle l\rangle $};
% Text Node
\draw (438,192.82) node [anchor=north west][inner sep=0.75pt]    {$\mathbb{Z}_{3} =\langle l^{3} \rangle $};
% Text Node
\draw (348,127.4) node [anchor=north west][inner sep=0.75pt]    {$\mu $};

\end{tikzpicture}
\caption{Left: The orbifold graph of three CFTs. The edges of the graph show the primaries shared by the CFTs. Right: The three CFTs are mapped to three stabilizer groups $S_1,S_2$ and $S_3$.}
\label{fig:orbgraph2}
\end{figure}
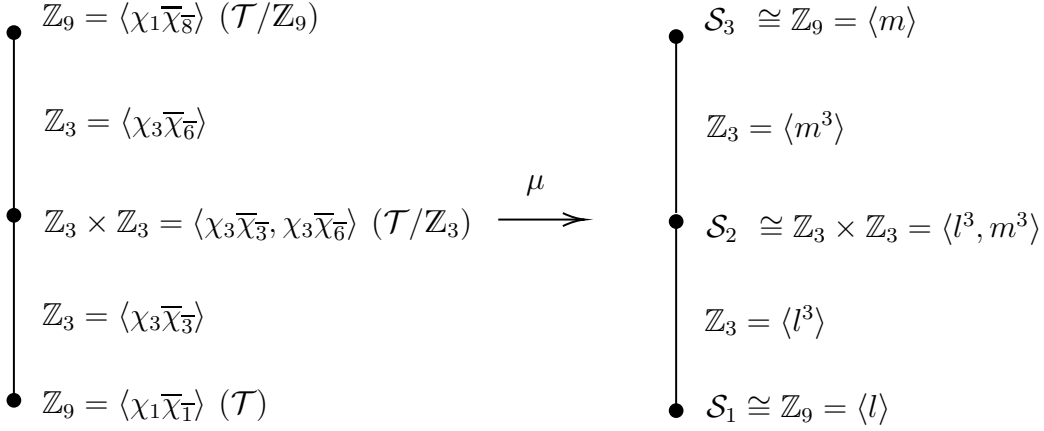

In this case, we have three CFTs in the orbifold family with partition functions
\bea 
Z_{\CT}:= \chi _0 \bar{\chi }_{\bar{0}}+\chi _1 \bar{\chi }_{\bar{1}}+\chi _2 \bar{\chi }_{\bar{2}}+\chi _3 \bar{\chi }_{\bar{3}}+\chi _4 \bar{\chi }_{\bar{4}}+\chi _5 \bar{\chi }_{\bar{5}}+\chi _6 \bar{\chi }_{\bar{6}}+\chi _7 \bar{\chi }_{\bar{7}}+\chi _8 \bar{\chi }_{\bar{8}}~, \nonumber \\
Z_{\CT/\DZ_3}:= \chi _0 \bar{\chi }_{\bar{0}}+\chi _3 \bar{\chi }_{\bar{3}}+\chi _6 \bar{\chi }_{\bar{6}}+\chi _0 \bar{\chi }_{\bar{3}}+\chi _3 \bar{\chi }_{\bar{6}}+\chi _6 \bar{\chi }_{\bar{0}}+\chi _0 \bar{\chi }_{\bar{6}}+\chi _3 \bar{\chi }_{\bar{0}}+\chi _6 \bar{\chi }_{\bar{3}}~,\\
Z_{\CT/\DZ_9}:= \chi _0 \bar{\chi }_{\bar{0}}+\chi _4 \bar{\chi }_{\bar{5}}+\chi _8 \bar{\chi }_{\bar{1}}+\chi _3 \bar{\chi }_{\bar{6}}+\chi _7 \bar{\chi }_{\bar{2}}+\chi _2 \bar{\chi }_{\bar{7}}+\chi _6 \bar{\chi }_{\bar{3}}+\chi _1 \bar{\chi }_{\bar{8}}+\chi _5 \bar{\chi }_{\bar{4}}~.\nonumber
\eea
These CFTs are mapped to three stabilizer groups in the generalized Pauli group acting on a qudit with $9$ states. The three CFTs, with partitions functions given above, can be depicted in the orbifold graph given in Fig. \ref{fig:orbgraph2}.
Note that the generators of the stabilizer group, $\CS_2$, are completely fixed by the choice of the generators for $\CS_1$ and $\CS_3$. This captures the fact the the primary operators of the CFT $\CT/\DZ_3$ are determined by a subset of primaries of $\CT$ and $\CT/\DZ_9$. 
Since the groups $\CS_1$, $\CS_2$, and $\CS_3$ must be distinct, we require the generators to satisfy
\be
l \neq m~, ~ l^3 \neq m^3~.
\ee
Moreover, we should make sure that the three groups $\CS_1$, $\CS_2$, and $\CS_3$ are indeed abelian. A consistent choice of generators is 
\be
l = X_{(9)}~, ~ m= Z_{(9)}~,
\ee
where $X$ and $Z$ are order-9 generalized Pauli group elements. 

\subsec{Is $\mu$ always an embedding?}
In the examples we have just discussed, the map between the orbifold and code graphs, $\mu$, is an embedding. More generally, depending on the choice of Rep$(\DV)$, we may get a complicated orbifold graph which leads to many constraints, via \eqref{eq:CFTcode condition 1} and \eqref{eq:CFTcode condition 2}, that the generators of the stabilizer groups in the image of the map $\mu$ must satisfy. One natural question that arises in this context is the following:
\vspace{0.2cm}

\noindent
{\it For an arbitrary $\DV$, can we always choose the graph homomorpism, $\mu$, to be an an embedding?} 
\vspace{0.2cm}

\noindent In Appendix \ref{Ap:1}, we answer this question in the negative. In particular, we study the constraints on stabilizer groups coming from Rep$(\DV)=E_{2^r}$ MTC (i.e., the bulk 3d theory is $\mathbb{Z}_{2^r}$ discrete gauge theory). Among prime CS theories, these QFTs have a particularly elaborate zoo of surface operators. We show that there are orbifold graphs for which $\mu$ cannot be chosen to be an embedding. Given this result, we can ask:
\vspace{0.2cm}

\noindent
{\it For an arbitrary $\DV$, is there a well-defined and universal subgraph of $\Gamma_{\DV}$ for which we can choose the graph homomorpishm, $\mu$, from the subgraph to $\CP_{\DV}$ to be an embedding? Does this $\mu$ naturally relate important physical properties of the CFT and the surface operator, $S(Q,[\sigma])$, with corresponding properties of the code?} 

\vspace{0.2cm}
\noindent In the next section, we show that such a subgraph and such a $\mu$ do indeed exist.

\section{From Abelian RCFTs to qudit stabilizer codes}

\label{sec:abelian RCFTS to qudit codes}
Given the results of the previous section and of Appendix \ref{Ap:1}, we know that $\mu$ cannot generally be an embedding of $\Gamma_{\mathds{V}}$ in $\Gamma_{\CP_{\mathds{V}}}$. As a result, we are, in some sense, free to choose $\mu$ and a universal subgraph of $\Gamma_{\mathds{V}}$ in a way that is physically interesting. In the discussion around \eqref{eq:CFTcode condition 1}, we saw that general considerations suggest that our quantum codes of interest, $\CS_{(Q,[\sigma])}$, are self-dual. Therefore, it is natural to choose a map, $\mu$, and a subgraph of $\Gamma_{\mathds{V}}$  such that the corresponding CS surface operators, $\CS(Q,[\sigma])$, are self-dual. In what follows, we will explicitly show that such a choice is possible.

To that end, recall that the abelianized generalized Pauli group, $V$, has a bilinear form defined in \eqref{eq:symplectic}. Also, from the explicit expression for the partition function of the CFT after orbifolding in equations \eqref{HZ} and \eqref{pconst2}, we know that the spectrum of primaries of the CFT after orbifolding depends crucially on how the Wilson lines of the bulk CS theory braid via the modular $S$ matrix.\footnote{Note that $S$ should not be confused with the stabilizer group, $\CS_{(Q,\sigma)}$, or the surface operator, $S(Q,[\sigma])$.} In fact, the modular $S$ matrix is a bilinear form on the abelian group of Wilson lines of the bulk CS theory, $K$. Therefore, it is natural to construct a map from operators of a CFT to operators in the generalized Pauli group such that the bilinear from on the latter is determined by the $S$ matrix. 

With this discussion in mind, let us describe the structure of the $S$ matrix explicitly. Since the bulk Chern-Simons theory admits a factorisation as in \eqref{pval}, the $S$ matrix of the full Chern-Simons theory is determined by the $S$ matrix of the individual factors. Let us denote the $S$-matrix of the theory as
\be
\label{Smatgen}
S_{\vec p, \vec q}=e^{2 \pi i \vec p^T M L \vec q}~,
\ee    
where $M$ and $L$ are block diagonal matrices, and the blocks are composed of the matrices
\be
\label{eq:L def}
L_{A_{2^r}}=1,~ L_{B_{2^r}}=-1,~ L_{C_{2^r}}=5,~ L_{D_{2^r}}=-5,~ L_{A_{p^r}}=4,~ L_{B_{p^r}}=2~,
\ee
for prime TQFTs corresponding to cyclic fusion groups,
\be
\label{eq:L def 2}
L_{E_{2^s}}=\begin{pmatrix}
0 & 1\\
1 & 0
\end{pmatrix}, ~ L_{F_{2^t}}=\begin{pmatrix}
2 & 1\\
1 & 2
\end{pmatrix}~,
\ee
and 
\bea
M_{A_{2^r}}&=&M_{B_{2^r}}=M_{C_{2^r}}=M_{D_{2^r}}=\frac{1}{2^r},~ M_{A_{p^r}}=M_{B_{p^r}}=\frac{1}{p^r}~,\nonumber\\
M_{E_{2^r}}&=&M_{F_{2^r}}=\begin{pmatrix}
\frac{1}{2^{r}} & 0\\
0 & \frac{1}{2^{r}} 
\end{pmatrix}~.
\eea
The $S$-matrix is then fixed by the decomposition in \eqref{pval}. We will use this form of the $S$-matrix to write down an explicit expression for the map between primary operators of a CFT and operators in the generalized Pauli group, $\CP_{\DV}$.

\subsection{Charge-conjugation modular invariant}

For a given chiral algebra, $\DV$, consider the factorisation of the bulk Chern-Simons theory as given in \eqref{pval}. The product of qudits which we use to construct our stabilizer codes is specified by this factorisation. For example, suppose the bulk Chern-Simons theory is $A_4 \times A_{3}$. Then, the fusion rules of chiral primaries is given by the group $\DZ_{4} \times \DZ_{3}$. In this case, we choose a system of one ``quadit" (a 4-state system) and a \lq\lq qutrit (a 3-state system). The CFTs specified by distinct surface operators in the $A_4 \times A_{3}$ Chern-Simons theory will be mapped to distinct stabilizer codes constructed from the generalized Pauli group acting on this system. An element of this generalized Pauli group is of the form 
\be
X_{(4)}^{\alpha_1} \otimes X_{(3)}^{\alpha_2} \circ Z_{(4)}^{\beta_1} \otimes Z_{(3)}^{\beta_2}~,
\ee
where $\alpha_1,\beta_1\in \DZ_{4}$ and $\alpha_2,\beta_2\in \DZ_{3}$. 

Let us return to the case of a general abelian MTC, Rep$(\DV)$. We will first describe a map from the CFT, $\CT$, with charge-conjugation parition function to quantum codes. The map from the orbifolds of $\CT$ to quantum codes will then follow. Consider the partition function of $\CT$
\begin{equation}
Z_{\CT}(q)=\sum_{\vec p}\chi_{\vec p}(q)\bar\chi_{\overline{\vec p}}(\bar q)~, \ \ \vec p+\overline{\vec p}=\vec 0~, \ \ N_{\overline{\vec p}},N_{\vec p},N_{\vec 0}\in{\rm Rep}(\DV)~.
\end{equation}
Let us choose the following simple map from primary operators to the quantum code that depends linearly on $\vec{p}$ (linearity is sufficient to preserve the additive structure of the RCFT fusion rules) 
\be
\mu: \CO_{\vec p, \overline{\vec p}} \to I \circ Z^{L \vec p}~.
\ee
Here, $L$ is the matrix that appears in \eqref{Smatgen} (we can further motivate its appearance as capturing the non-factorization of the $E_{2^r}$ and $F_{2^r}$ theories), and $I \circ Z^{L\vec p}$ is shorthand for the following expression
\bea
\label{ccqudit}
I \circ Z^{L\vec p}=  \bigotimes_r && \Bigg [\bigotimes_{i=1}^{n_{A_{2^r}}} (I_{(2^r)} \circ Z^{p_i^{(A_{2^r})}}_{(2^r)})  \bigotimes_{i=1}^{n_{B_{2^r}}} (I_{(2^r)} \circ Z^{-p_i^{(B_{2^r})}}_{(2^r)})  \bigotimes_{i=1}^{n_{C_{2^r}}} (I_{(2^r)} \circ Z^{5p_i^{(C_{2^r})}}_{(2^r)})  \bigotimes_{i=1}^{n_{D_{2^r}}} (I_{(2^r)} \circ Z^{-5p_i^{(D_{2^r})}}_{(2^r)}) \cr
&& \bigotimes_{i=1}^{n_{E_{2^r}}} (I_{(2^r)} \circ Z^{L_{E_{2^r} }\vec p_i^{(E_{2^r})}}_{(2^r)})  \bigotimes_{i=1}^{n_{F_{2^r}}} (I_{(2^r)} \circ Z^{L_{F_{2^r}} \vec p_i^{(F_{2^r})}}_{(2^r)}) \cr
&& \bigotimes_q \Bigg [\bigotimes_{i=1}^{n_{A_{q^r}}} (I_{(q^r)} \circ Z^{4p_i^{(A_{q^r})}}_{(q^r)})  \bigotimes_{i=1}^{n_{B_{q^r}}} (I_{(q^r)} \circ Z^{2p_i^{(B_{q^r})}}_{(q^r)}) \Bigg ] \Bigg ]~,
\eea
where $p_i^{(X)}$ denotes the $i^{\text{th}}$ prime $X$-type Chern-Simons factor in the vector $\vec p$. Since the resulting generalized Pauli group elements depend only on the generalized $Z$ Pauli matrices acting on various qudits, it is clear that these elements commute with each other and form a stabilizer group. In the CFT language, this statement corresponds to the fact that the resulting primary operators are mutually local.

As an example, let us write the general expression above in the specific case of Rep$(\DV)=A_4 \times A_3$. The charge conjugation partition function of the CFT $\CT$ is 
\be
Z_{\CT}(q)=\sum_{\vec p}\chi_{\vec p}(q)\bar\chi_{\overline{\vec p}}(\bar q)~, \ \ \vec p+\overline{\vec p}=\vec 0~, \ \ \vec p \in \DZ_4 \times \DZ_3~.
\ee
This CFT is mapped to the following stabilizer code
\be
\CO_{\vec p, \overline{\vec p}} \to I_{(4)} \otimes I_{(3)} \circ Z_{(4)}^{p_1} \otimes Z_{(3)}^{p_2} ~,  
\ee
where $\vec p= (p_1,p_2) \in \DZ_{4} \times \DZ_{3}$. 

\subsection{The general case}
Suppose we orbifold the CFT $\CT$ by a non-anomalous group, $Q$, to get the partition function
\begin{equation}
Z_{\CT/(Q,[\sigma])}=\sum_{\vec g\in Q}\sum_{\vec p\in B_{\vec g}}\chi_{\vec p}(q)\bar\chi_{\overline{\vec p+\vec g}}(\bar q)~,
\end{equation}
where $B_{\vec g}$ is defined in \eqref{pconst2}. Since orbifolding by a discrete group is an invertible operation, we should involve factors of the $X$ generalized Pauli matrices. A particularly simple choice that is linear in $\vec g$ is\footnote{We can motivate the absence of $L$ in the exponent of $X$ below as corresponding to the fact that higher-gauging does not depend on the braiding of the lines being higher gauged.}
\be
\label{genqudit}
\mu: \CO_{\vec p, \overline{\vec p + \vec g}} \to X^{\vec g} \circ Z^{L \vec p}~,
\ee
where $X^{\vec g} \circ Z^{L \vec p}$ is shorthand for an expression similar to \eqref{ccqudit}, with the $I$ factors replaced with appropriate factors of the generalized $X$ Pauli matrices
\bea
\label{genqudit2}
X^{\vec g} \circ Z^{L \vec p}=  \bigotimes_r && \Bigg [\bigotimes_{i=1}^{i=n_{A_{2^r}}} (X^{g_i^{(A_{2^r})}}_{(2^r)} \circ Z^{p_i^{(A_{2^r})}}_{(2^r)})  \bigotimes_{i=1}^{i=n_{B_{2^r}}} (X^{g_i^{(B_{2^r})}}_{(2^r)} \circ Z^{-p_i^{(B_{2^r})}}_{(2^r)}) \cr
&&  \bigotimes_{i=1}^{i=n_{C_{2^r}}} (X^{g_i^{(C_{2^r})}}_{(2^r)} \circ Z^{5p_i^{(C_{2^r})}}_{(2^r)})  \bigotimes_{i=1}^{i=n_{D_{2^r}}} (X^{g_i^{(D_{2^r})}}_{(2^r)} \circ Z^{-5p_i^{(D_{2^r})}}_{(2^r)}) \cr
&& \bigotimes_{i=1}^{i=n_{E_{2^r}}} (X^{\vec g_i^{(E_{2^r})}}_{(2^r)} \circ Z^{L_{E_{2^r} }\vec p_i^{(E_{2^r})}}_{(2^r)})  \bigotimes_{i=1}^{i=n_{F_{2^r}}} (X^{\vec g_i^{(F_{2^r})}}_{(2^r)} \circ Z^{L_{F_{2^r}} \vec p_i^{(F_{2^r})}}_{(2^r)}) \cr
&& \bigotimes_q \Bigg [\bigotimes_{i=1}^{i=n_{A_{q^r}}} (X^{g_i^{(A_{q^r})}}_{(q^r)} \circ Z^{4p_i^{(A_{q^r})}}_{(q^r)})  \bigotimes_{i=1}^{i=n_{B_{q^r}}} (X^{g_i^{(B_{q^r})}}_{(q^r)} \circ Z^{2p_i^{(B_{q^r})}}_{(q^r)}) \Bigg ] \Bigg ]~.
\eea
Note that the generalized Pauli group elements corresponding to $E_{2^r}$ and $F_{2^r}$ type theories are determined by vectors $\vec g_{i}$ and $L_{E_{2^r}/F_{2^r}} \vec p_i$, where $L_{E^{r}}$ and $L_{F_{2^r}}$ were defined in \eqref{eq:L def 2}. This is because for the $E_{2^r}$ and $F_{2^r}$ type theories the fusion group is $\DZ_{2^r} \times \DZ_{2^r}$. 

Let $G(\vec g,\vec p):=X^{\vec g} \circ Z^{L \vec p}$ be the generalized Pauli group element as defined in \eqref{genqudit}. In order to get a stabilizer group, we need to impose the constraint that corresponding Pauli group elements commute with each other. The commutation relations can be written as
\bea\label{stabCond}
G(\vec g_{1},\vec p_{1}) G(\vec g_{2},\vec p_{2})&=&
e^{2 \pi i [\vec g_{2}^{T}  ML \vec p_{1}- \vec g_{1}^{T} ML \vec p_{2}] }
  \ G(\vec g_{2},\vec p_{2}) G(\vec g_{1},\vec p_{1}) \nonumber \\
&=& S_{\vec g_{2}, \vec p_{1} } S^{-1}_{\vec g_{1}, \vec p_{2} }
  \ G(\vec g_{2},\vec p_{2}) G(\vec g_{1},\vec p_{1})\nonumber\\
&=& \Xi(\vec g_{2}, \vec g_{1} ) \Xi(\vec g_{1}, \vec g_{2} )^{-1} 
 \ G(\vec g_{2},\vec p_{2}) G(\vec g_{1},\vec p_{1})~,
\eea
where, in the third equality, we have used \eqref{pconst2}. We have also used the expression for the $S$ matrix, $S_{\vec p, \vec q}=e^{2 \pi i \vec p^T M L \vec q }$, as described in \eqref{Smatgen}. Therefore, $\CS_{\CT/H}$ is a stabilizer code if and only if 
\be
\label{stabcond2}
\Xi(\vec g_{1}, \vec g_{2} ) = \Xi(\vec g_{2}, \vec g_{1} )  ~ \forall \vec g_{1}, \vec g_{2} \in Q~.
\ee
If $\Xi$ is valued in $\pm 1$, then \eqref{stabcond2} is the same as 
\be
\Xi(\vec g_{1}, \vec g_{2} ) \Xi(\vec g_{2}, \vec g_{1} )= S_{\vec g_1,\vec g_2} =1 ~ \forall \vec g_{1}, \vec g_{2} \in Q~.
\ee
In other words, the 1-form symmetry group, $Q$, of the bulk Chern-Simons theory must be anomaly free. This is precisely the condition we derived in \cite{buican2021quantum} where $\Xi \in \{\pm 1\}$ was guaranteed because we only considered $Q$ with order-two elements.\footnote{More generally, even in the case considered here, we can relate this condition to the absence of certain 1-form 't Hooft anomalies. However, in our discussion below, we prefer to give a more physically immediate relation to properties of the surface operator.}

Consider a CFT $\CT/(Q,[\sigma])$ satisfying condition \eqref{stabcond2}. Then, the map $\mu$ in \eqref{genqudit} is an invertible map satisfying
\be
\mu(\CO_{\vec p_1, \overline{\vec p_1 + \vec g_1}}\CO_{\vec p_2, \overline{\vec p_2 + \vec g_2}})= \mu(\CO_{\vec p_1, \overline{\vec p_1 + \vec g_1}}) \mu(\CO_{\vec p_2, \overline{\vec p_2 + \vec g_2}})~.
\ee
Therefore, we find that the resulting stabilizer group $\CS_{(Q,[\sigma])}$ satisfies
\be
K_{(Q,[\sigma])} \cong \CS_{(Q,[\sigma])}~.
\ee
Moreover, since $\mu$ is a linear map, for two CFTs $\CT/(Q_1,[\sigma_1])$ and $\CT/(Q_2,[\sigma_2])$ satisfying condition \eqref{stabcond2}, under the map $\mu$, we get
\be
K_{(Q_1,[\sigma_1])} \cap K_{(Q_2,[\sigma_2])} \cong  \CS_{(Q_1,[\sigma_1])} \cap \CS_{(Q_2,[\sigma_2])}~.
\ee
Therefore, the two conditions \eqref{eq:CFTcode condition 1} and \eqref{eq:CFTcode condition 2} for a consistent CFT to quantum codes map are satisfied. As a result, $\mu$ defines an embedding of the subgraph of $\Gamma_{\DV}$ with CFTs satisfying condition \eqref{stabcond2} into the code graph $\Gamma_{\CP_{\DV}}$. 

The condition \eqref{stabcond2} is a necessary condition for the CFT $\CT/(Q,[\sigma])$ to admit a lift to an unoriented CFT which can be defined  on unoriented 2-manifolds \cite{fuchs2004tft}. More interestingly, in the next section, we will show a CFT, $\CT/(Q,[\sigma])$, which satisfies \eqref{stabcond2} corresponds to a surface operator, $S(Q,[\sigma])$, in the bulk CS theory which is insensitive to the orientation of the 2-manifold on which it is defined. 

\subsection{Self-dual stabilizer codes and self-dual surfaces}
\label{sec:selfdualsurfaces}

In the previous section, we derived a condition for the generalized Pauli group elements associated with the primary operators of a CFT to form a stabilizer group. This constraint can be translated into a constraint on the surface operator, $S(Q,[\sigma])$, in the bulk Chern-Simons theory corresponding to the partition function of the CFT, $\CT/(Q,[\sigma])$. 

To understand which surface operators lead to CFTs that can be mapped to quantum codes, consider the stabilizer elements $G(\vec g_{1},\vec p_{1})$ and $G(\vec g_{2},\vec p_{2})$ which, by definition, commute with each other. We will first show that the generalized Pauli group element, $G(\vec g_{1},\overline{\vec g_1 + \vec p_{1}})$, must be in the stabilizer group. Indeed, from \eqref{stabCond} we know this means that
\be
e^{2 \pi i [\vec g_{2}^T M L \vec p_{1}- \vec g_{1}^T M L \vec p_{2}] }=1~.
\ee
Now, consider the commutation relation of   $G(\vec g_{1},\overline{\vec g_1 + \vec p_{1}})$  with $G(\vec g_{2},\vec p_{2})$. We get
\bea
G(\vec g_{1},\overline{\vec g_1 + \vec p_{1}}) G(\vec g_{2},\vec p_{2})&=&
e^{2 \pi i [M \vec g_{2} \cdot L (\overline{\vec g_1 + \vec p_{1}})- M \vec g_{1} \cdot L \vec p_{2}] }
  \ G(\vec g_{2},\vec p_{2}) G(\vec g_{1},\overline{\vec g_1 + \vec p_{1}}) \nonumber \\
  &=&
S_{\vec g_2, \vec g_1}^{-1} S_{\vec g_2, \vec p_1}^{-1}S_{\vec g_1, \vec p_2}^{-1}
  \ G(\vec g_{2},\vec p_{2}) G(\vec g_{1},\overline{\vec g_1 + \vec p_{1}}) ~.\nonumber 
\eea  
Using the fact that $\vec p_1$ and $\vec p_2$ satisfies \eqref{pconst2} and the definition of $\Xi$ in \eqref{SXiZ2k}, we get
\be
S_{\vec g_2, \vec g_1}^{-1} S_{\vec g_2, \vec p_1}^{-1}S_{\vec g_1, \vec p_2}^{-1}= \Xi(\vec g_2, \vec g_1)^{-1}\Xi(\vec g_1, \vec g_2)^{-1} \Xi(\vec g_1, \vec g_2) \Xi(\vec g_2, \vec g_1)=1 ~.
\ee
Therefore, we find that if $G(\vec g_{1},\vec p_{1})$ and $G(\vec g_{2},\vec p_{2})$  commute then $G(\vec g_{1},\overline{\vec g_1 + \vec p_{1}})$ and $G(\vec g_{2},\vec p_{2})$ also commute. Note that  $G(\vec g_{2},\vec p_{2})$ is an arbitrary element of the stabilizer group, and therefore $G(\vec g_{1},\overline{\vec g_1 + \vec p_{1}})$ commutes with the full stabilizer group. Since we have a self-dual stabilizer group, $G(\vec g_{1},\overline{\vec g_1 + \vec p_{1}})$ must be in the stabilizer group. 

\begin{figure}[h!]
    \centering

\tikzset{every picture/.style={line width=0.75pt}} %set default line width to 0.75pt        

\begin{tikzpicture}[x=0.75pt,y=0.75pt,yscale=-1,xscale=1]

%Shape: Parallelogram [id:dp810447878811633] 
\draw  [color={rgb, 255:red, 245; green, 166; blue, 35 }  ,draw opacity=1 ][fill={rgb, 255:red, 248; green, 198; blue, 155 }  ,fill opacity=1 ] (123.32,233.99) -- (123.34,121.43) -- (206.59,70.06) -- (206.57,182.63) -- cycle ;
%Straight Lines [id:da6090870260961099] 
\draw [color={rgb, 255:red, 139; green, 6; blue, 24 }  ,draw opacity=1 ]   (164.95,152.03) -- (258.43,151.88) ;
%Straight Lines [id:da357988959458718] 
\draw [color={rgb, 255:red, 139; green, 6; blue, 24 }  ,draw opacity=1 ] [dash pattern={on 4.5pt off 4.5pt}]  (128.04,152.87) -- (164.95,152.03) ;
%Straight Lines [id:da015062114005278016] 
\draw [color={rgb, 255:red, 139; green, 6; blue, 24 }  ,draw opacity=1 ]   (65.97,152.33) -- (124.11,152.87) ;
\draw  [color={rgb, 255:red, 139; green, 6; blue, 24 }  ,draw opacity=1 ][fill={rgb, 255:red, 139; green, 6; blue, 24 }  ,fill opacity=1 ] (94.53,154.59) -- (91.89,152.5) -- (94.58,150.52) -- (93.22,152.53) -- cycle ;
%Shape: Ellipse [id:dp09900881215064994] 
\draw  [color={rgb, 255:red, 139; green, 6; blue, 24 }  ,draw opacity=1 ][fill={rgb, 255:red, 139; green, 6; blue, 24 }  ,fill opacity=1 ] (162.36,152.03) .. controls (162.36,151.13) and (162.94,150.4) .. (163.66,150.4) .. controls (164.37,150.4) and (164.95,151.13) .. (164.95,152.03) .. controls (164.95,152.93) and (164.37,153.66) .. (163.66,153.66) .. controls (162.94,153.66) and (162.36,152.93) .. (162.36,152.03) -- cycle ;
\draw  [color={rgb, 255:red, 139; green, 6; blue, 24 }  ,draw opacity=1 ][fill={rgb, 255:red, 139; green, 6; blue, 24 }  ,fill opacity=1 ] (233.04,153.93) -- (230.4,151.84) -- (233.09,149.86) -- (231.73,151.87) -- cycle ;
%Shape: Parallelogram [id:dp3578091522691098] 
\draw  [color={rgb, 255:red, 245; green, 166; blue, 35 }  ,draw opacity=1 ][fill={rgb, 255:red, 248; green, 198; blue, 155 }  ,fill opacity=1 ] (462.32,233.99) -- (462.34,121.43) -- (545.59,70.06) -- (545.57,182.63) -- cycle ;
%Straight Lines [id:da10193247304610997] 
\draw [color={rgb, 255:red, 139; green, 6; blue, 24 }  ,draw opacity=1 ]   (503.95,152.03) -- (597.43,151.88) ;
%Straight Lines [id:da5246158365653727] 
\draw [color={rgb, 255:red, 139; green, 6; blue, 24 }  ,draw opacity=1 ] [dash pattern={on 4.5pt off 4.5pt}]  (467.04,152.87) -- (503.95,152.03) ;
%Straight Lines [id:da9919732887253674] 
\draw [color={rgb, 255:red, 139; green, 6; blue, 24 }  ,draw opacity=1 ]   (404.97,152.33) -- (463.11,152.87) ;
\draw  [color={rgb, 255:red, 139; green, 6; blue, 24 }  ,draw opacity=1 ][fill={rgb, 255:red, 139; green, 6; blue, 24 }  ,fill opacity=1 ] (437.89,150.49) -- (440.55,152.53) -- (437.9,154.56) -- (439.22,152.53) -- cycle ;
%Shape: Ellipse [id:dp14990085672964493] 
\draw  [color={rgb, 255:red, 139; green, 6; blue, 24 }  ,draw opacity=1 ][fill={rgb, 255:red, 139; green, 6; blue, 24 }  ,fill opacity=1 ] (501.36,152.03) .. controls (501.36,151.13) and (501.94,150.4) .. (502.66,150.4) .. controls (503.37,150.4) and (503.95,151.13) .. (503.95,152.03) .. controls (503.95,152.93) and (503.37,153.66) .. (502.66,153.66) .. controls (501.94,153.66) and (501.36,152.93) .. (501.36,152.03) -- cycle ;
\draw  [color={rgb, 255:red, 139; green, 6; blue, 24 }  ,draw opacity=1 ][fill={rgb, 255:red, 139; green, 6; blue, 24 }  ,fill opacity=1 ] (569.34,149.86) -- (572.06,151.84) -- (569.45,153.93) -- (570.73,151.87) -- cycle ;

% Text Node
\draw (160.77,99.21) node [anchor=north west][inner sep=0.75pt]  [font=\scriptsize]  {$S( Q,[ \sigma ]){}$};
% Text Node
\draw (261.5,141.54) node [anchor=north west][inner sep=0.75pt]    {$\vec{p}$};
% Text Node
\draw (24.82,140.55) node [anchor=north west][inner sep=0.75pt]    {$\vec{g} +\vec{p}$};
% Text Node
\draw (602.5,139.54) node [anchor=north west][inner sep=0.75pt]    {$\vec{g} +\vec{p}$};
% Text Node
\draw (386.82,140.55) node [anchor=north west][inner sep=0.75pt]    {$\vec{p}$};
% Text Node
\draw (319,144.4) node [anchor=north west][inner sep=0.75pt]    {$\leftrightarrow $};
% Text Node
\draw (499.77,100.21) node [anchor=north west][inner sep=0.75pt]  [font=\scriptsize]  {$S( Q,[ \sigma ]){}$};

\end{tikzpicture}
    \caption{If the RCFT, $\CT/(Q,[\sigma])$, is mapped to a stabilizer code under the map \eqref{genqudit}, then the corresponding surface operator, $S(Q,[\sigma])$, is self-dual.}
    \label{fig:selfdualsurface}
\end{figure}
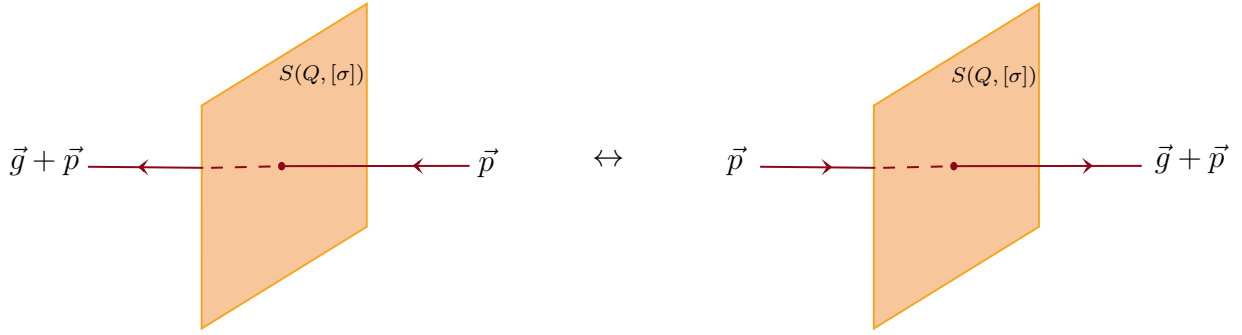

In terms of RCFT partition functions, the above discussion can be rephrased as follows. For a given RCFT, if the map \eqref{genqudit} defines a stabilizer code, then the RCFT must satisfy the following condition: if $\CO_{\vec p,\overline{\vec g + \vec p}}$ is a primary operator, then $\CO_{\overline{\vec g + \vec p},\vec p}$ must also be a primary operator. This statement implies that the surface operator in the bulk Chern-Simons theory which determines this partition function has an action on lines that satisfies Fig. \ref{fig:selfdualsurface}. In particular, the action of the surface operator on the lines must be the same irrespective of whether the action is from the left or from the right.

Which surface operators have this property? Consider the surface operator, $S(\Sigma,Q,[\sigma])$, where we include the 2-manifold, $\Sigma$, in the notation of the surface operator to emphasize the 2-manifold on which the surface operator is supported. The dual surface operator, $\bar S(\Sigma,Q,\sigma)$, is defined as 
\be
\bar S(\Sigma,Q,[\sigma]):= S(\bar \Sigma,Q,[\sigma])~,
\ee
where $\bar \Sigma$ is the orientation reversal of the 2-manfiold $\Sigma$. If the surface operator, $S(\Sigma,Q,[\sigma])$, acts on line operators as in the left diagram in Fig. \ref{fig:selfdualsurface}, then its dual, $\bar S(\Sigma,Q,\sigma)$, acts as in the right diagram in Fig. \ref{fig:selfdualsurface}. In abelian Chern-Simons theory, a surface operator is uniquely determined by its action on the line operators \cite{Kapustin:2010if,Buican:2023bzl}. Therefore, we find that for $\CT/(Q,[\sigma])$ to be mapped to a quantum stabilizer code under \eqref{genqudit}, the corresponding surface operator must satisfy the constraint
\be
\bar S(\Sigma,Q,[\sigma])= S(\Sigma, Q,[\sigma])~.
\ee
Such surface operators are {\it self-dual}.

In fact, the converse also holds. Suppose the surface operator, $S(Q,[\sigma])$, is self-dual. Then it satisfies Fig. \ref{fig:selfdualsurface}. That is, both $\CO_{\vec p,\overline{\vec g + \vec p}}$ and $\CO_{\overline{\vec g + \vec p},{\vec p}}$ are primary operator of the RCFT, $\CT/(Q,[\sigma])$. Then, from \eqref{pconst2}, we know that the following conditions are satisfied
\be
\label{eq:selfdualcond}
S_{\vec h,\vec  p} ~ \Xi(\vec h,\vec g)=1 ~ , ~ S_{\vec h,\overline{\vec g + \vec  p}} ~ \Xi(\vec h,\vec g)=1 \ \forall \vec h \in Q~.
\ee
The second equality can be simplified as follows.
\bea
1&=&S_{\vec h,\overline{\vec g + \vec  p}} ~ \Xi(\vec h,\vec g)= S_{\vec h, \overline{\vec g}} S_{\vec h, \overline{\vec p}} ~\Xi(\vec h,\vec g) \cr
&=& \Xi(\vec h,\vec g)^{-1} \Xi(\vec g,\vec h)^{-1} S_{\vec h, \overline{\vec p}} ~\Xi(\vec h,\vec g) \cr
&=&  \Xi(\vec g,\vec h)^{-1} S_{\vec h, \overline{\vec p}}~.
\eea
Comparing with the first equation in \eqref{eq:selfdualcond}, we find
\be
\Xi(\vec g,\vec h) = \Xi(\vec h,\vec g)~,
\ee
which is precisely the condition that we derived in equation \eqref{stabcond2} to get a stabilizer group. In summary, we find the following relation:
\be
\text{ Self-dual surface operator }\leftrightarrow \text{ Self-dual quantum stabilizer code~.}
\ee

\subsection{Examples}
Let us consider a few examples that illustrate the above discussion.
\subsubsection{$R=1$ and $R=2$ compact boson}\label{R12Ex}

The $R=1$ compact boson corresponds to the choice $n_{A_4}=1$ and all other $n_{*}=0$ in \eqref{pval}. In this case, the chiral algebra has the trivial, fundamental, spinor, and conjugate spinor representations, which we will denote by $N_{0}$, $N_{2}$, $N_{1}$, and $N_{3}$, respectively. These representations form the $K\cong\DZ_4$ group under fusion. The scaling dimensions of chiral primaries in these representations are
\be\label{chirscale}
h_{0}=0~,\ h_{2}=\frac{1}{2}~,\ h_{1}=h_{3}=\frac{1}{8}~.
\ee
The partition function is
\be
Z_{\CT}=\chi_0\bar\chi_0+\chi_{2}\bar\chi_{2}+\chi_1\bar\chi_3+\chi_{3}\bar\chi_{ 1}~,
\ee
which is the charge conjugation modular invariant. Using \eqref{genqudit}, these primaries can be mapped to the ``1-quadit" stabilizer code generated by the $Z_{(4)}$ Pauli matrix via 
\be
\label{eq:example code 1}
\mathcal{O}_{0,0} \xrightarrow{\mu} I_{(4)},~ 
\mathcal{O}_{2,2} \xrightarrow{\mu} Z_{(4)}^2,~ \mathcal{O}_{1,3} \xrightarrow{\mu} Z_{(4)},~ \mathcal{O}_{3,1} \xrightarrow{\mu} Z_{(4)}^3~.
\ee
Note that the subscript \lq\lq $4$" indicates that the order of the $Z_{(4)}$ matrix is $4$. 

In  our earlier work \cite{buican2021quantum}, the $R=1$ compact boson was mapped to a qubit code and the map was many-to-one: a set of primary operators were mapped to a single stabilizer group element. Here, we have a map to a quadit code, and the map is one-to-one. See Appendix \ref{ap:comparing with previous work} for further details on the relationships between the CFT-qudit codes map studied in this paper to the CFT-qubit codes map in \cite{buican2021quantum}. 

A topological line operator, $\CL_{2}$, labelled by $\vec p=2$ generates a non-anomalous $\DZ_2$ 0-form symmetry.  Taking the $\DZ_2$-orbifold, we get a CFT with partition function (using \eqref{HZ}, \eqref{pconst2})
\be
Z_{\CT/\DZ_2}=\chi_0\bar\chi_0+\chi_{2}\bar\chi_{2}+\chi_1\bar\chi_1+\chi_{3}\bar\chi_{3}~.
\ee
This is the partition function of the $R=2$ compact boson.  Using \eqref{genqudit}, these primaries can be mapped to a stabilizer code as follows
\be
\label{eq:example code 2}
\mathcal{O}_{0,0} \xrightarrow{\mu} I_{(4)},~ 
\mathcal{O}_{2,2} \xrightarrow{\mu} Z_{(4)}^2,~ \mathcal{O}_{1,1} \xrightarrow{\mu} X_{(4)}^2 \circ Z_{(4)},~ \mathcal{O}_{3,3} \xrightarrow{\mu} X_{(4)}^2 \circ Z_{(4)}^3~.
\ee
Therefore, we get a 1-quadit stabilizer code generated by $X_{(4)}^2 \otimes Z_{(4)}$.\footnote{Note that for the $Q=\DZ_2$ group generated by $\CL_2$, the $F$ matrix is trivial, and we have trivial discrete torsion. In order to get a stabilizer code, we should satisfy the constraint \eqref{stabcond2}. This condition is indeed satisfied because in the $A_4$ Chern-Simons theory, the $R$ matrix is symmetric in its arguments. This can be explicitly checked using the expression \eqref{eq:Rmatrixdef}.}

The $R=1$ and $R=2$ compact bosons are T-dual. This duality translates into the statement that the codes in \eqref{eq:example code 1} and \eqref{eq:example code 2} are code-equivalent. To understand this statement, note that code equivalences are generated by symmetries of the generalized Pauli group. That is, an outer automorphism of the generalized Pauli group. The codes in \eqref{eq:example code 1} and \eqref{eq:example code 2} are related by the operation
\be
Z_{(4)} \to X_{(4)}^2 \circ Z_{(4)}~,
\ee
which preserves the generalized Pauli group acting on a single quadit. Therefore, the two codes are indeed equivalent. More generally, two codes constructed from a set of qudits are equivalent if they are related by the Clifford group \cite{Gottesman:1997qd}. However, not all elements in the Clifford group correspond to symmetries/dualities of the corresponding CFTs \cite{dymarsky2020quantum}. We will return to this discussion in Sec. \ref{CodeEqSec}.

\subsubsection{$SU(3)_1$ WZW}\label{SU3ex}

The $SU(3)_1$ CFT corresponds to the choice $n_{B_3}=1$ and all others $n_{*}=0$ in \eqref{pval}. We denote the three representations of the chiral algebra by $N_0$, $N_1$, and $N_2$. They correspond to three primary fields with conformal weights $0$, $\frac{1}{3}$, and $\frac{1}{3}$, respectively. The partition function is
\be
Z_{\CT}=\chi_0\bar\chi_0+\chi_{1}\bar\chi_{2}+\chi_2\bar\chi_1~,
\ee
which is the charge conjugation modular invariant. Using \eqref{genqudit}, these primaries can be mapped to the 1-qutrit stabilizer code generated by the $Z_{(3)}$ Pauli matrix via 
\be\label{SU3code1}
\mathcal{O}_{0,0} \xrightarrow{\mu} I_{(3)},~ 
\mathcal{O}_{1,2} \xrightarrow{\mu} Z_{(3)},~ \mathcal{O}_{2,1} \xrightarrow{\mu} Z_{(3)}^2~.
\ee
Therefore, we get a 1-qutrit stabilizer code generated by $Z_{(3)}$. 

A topological line operator, denoted $\CL_{1}$, labelled by $\vec p=1$ generates a $\DZ_3$ 0-form symmetry.  Taking the $\DZ_3$-orbifold, we get a CFT with partition function (using \eqref{HZ}, \eqref{pconst2})
\be
Z_{\CT/\DZ_3}=\chi_0\bar\chi_0+\chi_{1}\bar\chi_{1}+\chi_2\bar\chi_2~.
\ee
  Using \eqref{genqudit}, these primaries can be mapped to the 1-qutrit stabilizer code as follows
\be\label{SU3code2}
\mathcal{O}_{0,0} \xrightarrow{\mu} I_{(3)},~ 
\mathcal{O}_{1,1} \xrightarrow{\mu} X_{(3)} \circ Z_{(3)},~ \mathcal{O}_{2,2} \xrightarrow{\mu} X_{(3)}^2 \circ Z_{(3)}^2~.
\ee
Therefore, we get a 1-qutrit stabilizer code generated by $X_{(3)} \circ Z_{(3)}$.\footnote{For the $Q=\DZ_3$ group generated by $\CL_2$, the $F$ matrix is trivial and we have trivial discrete torsion. Recall that, in order to get a stabilizer code, we should satisfy the constraint \eqref{stabcond2}. This condition is satisfied because in the $SU(3)_1$ CS theory, the $R$ matrix is symmetric in its arguments. This discussion can be explicitly checked using the expression \eqref{eq:Rmatrixdef}.} As in the previous example, the two codes presented in this subsection are code equivalent. We will explain this fact in Sec. \ref{CodeEqSec}.

\section{Generalized Pauli group from defects}

\label{sec:generalized pauli group}
For RCFTs which admit a quantum stabilizer group description, we explained how the stabilizer group elements are in one-to-one correspondence with primary operators of the CFT. In a self-dual quantum code, all generalized Pauli group elements which are not in the stabilizer group are errors that transform states in the code subspace to its complement.

In this section, we generalize the discussion in \cite{buican2021quantum} by studying symmetries implemented by certain line operators and arguing that the error operators correspond to the operators living at the end of these lines. These observables are called twisted-sector operators or non-genuine local operators. Let us write a general partition function of an RCFT with chiral algebra, Rep$(\DV)$, as
\be
Z_{\CT_{\CM}}:= \sum_{\vec p, \vec q } \CM_{\vec p, \vec q} ~ \chi_{\vec p}(q)\bar\chi_{\overline{\vec q}}(\bar q)~,
\ee
where $\CM$ is a modular-invariant matrix specifying the partition function. We first consider the case when $\CM$ is a permutation matrix (the corresponding surface in the bulk is invertible) before going into more general partition functions.

\subsection{Permutation modular invariants}

Suppose we have a permutation modular invariant, $\CM$ (this corresponds to having an invertible surface in the bulk). Then there is a permutation, $\sigma$, of the labels, $\vec p$, corresponding to the permutation matrix, $\CM$, and the partition function is
\be
Z_{\CT_{\CM}}:= \sum_{\vec p} \chi_{\vec p}(q)\bar\chi_{\overline{\sigma(\vec p)}}(\bar q)~.
\ee
Then, for every $\vec\ell \in \text{Rep}(\DV)$ we can define the Verlinde line 
\be
\CL_{\vec\ell}:= \sum_{\vec p} \frac{\bar S_{\vec\ell\vec p}}{\bar S_{\vec 0 \vec p}} \ket{\vec p, \sigma(\vec p)} \bra{\vec p, \sigma(\vec p)}~,
\ee
where $\ket{\vec p, \sigma(\vec p)} \bra{\vec p, \sigma(\vec p)}$ is a projector onto the conformal family corresponding to the primary state labelled by $(\vec p, \sigma(\vec p))$ \cite{Verlinde:1988sn,Gaiotto:2014lma,Buican:2017rxc}.

The spectrum of operators that live at the end of the line operator, $\CL_{\vec\ell}$, can be obtained by studying the partition function of the theory on the torus with the line operator $\CL_{\vec\ell}$ inserted along the time direction (e.g., see \cite{Petkova:2000ip,Buican:2017rxc,Chang:2018iay}). It is easy to compute this partition function, which we denote as $Z^{\vec\ell}_{\CT_{\CM}}$ (e.g., see \cite[Section II]{buican2021quantum} for details)
\be
Z^{\vec\ell}_{\CT_{\CM}}:= \sum_{\vec p, \vec q} \CM_{\vec p, \vec q} \chi_{\vec p+\vec\ell}(q)\bar\chi_{\overline{\vec q}}(\bar q)~.
\ee
Therefore, the point operators that live at the end of the line operator, $\CL_{\vec\ell}$, are labelled by 
\be
\CO^{\vec\ell}_{\vec p + \vec\ell, \overline{\vec p + \vec g}}~.
\ee
These can be mapped to elements of the generalized Pauli group as follows. First we write these operators in the form 
\be
\CO^{\vec\ell}_{\vec p + \vec\ell,~ \overline{\vec p + \vec\ell + \overline{\vec\ell} + \vec g}}~,
\ee
where $\overline{\vec\ell}$ is the conjugate representation of $\vec\ell$. Then, following \eqref{genqudit}, we can map these operators to elements of the generalized Pauli group as follows
\be
\label{defectqudit}
\mu: \CO_{\vec p + \vec\ell,~ \overline{\vec p + \vec\ell + \overline{\vec\ell} + \vec g}} \to X^{\vec g + \overline{\vec\ell}} \circ Z^{L (\vec p + \vec\ell)}~.
\ee
This map is valid for any Verlinde line. This is a generalization of our work in \cite{buican2021quantum}, where we only mapped twisted sector operators of order-two Verlinde lines to error operators in the corresponding qubit stabilizer code. Since $\vec\ell$ and $\vec p$ are arbitrary representations of the chiral algebra, the full set of defect end-point operators of the Verlinde lines maps to the  generalized Pauli group. Note that the map \eqref{defectqudit} does not capture the non-abelian nature of the generalized Pauli group, because the fusion of twisted-sector operators is abelian (since the fusion of Verlinde line operators is abelian). Indeed, this map reproduces the full abelianized generalized Pauli group. 

\subsection{General modular invariants}

\label{sec:gen modular inv}

In the case of a general modular invariant, we may not have enough Verlinde lines to recover the full (abelianized) generalized Pauli group. However, we can still define enough symmetries to construct the full (abelianized) generalized Pauli group. These symmetries (which we will denote as $\pi$ below) are defined by acting with phases on the primaries such that the action is compatible with fusion. 

Let $K_{(Q,[\sigma])}$ be the abelian group formed by the primaries of the theory, $\CT/(Q,[\sigma])$, under fusion. Let $\CM$ be the modular-invariant matrix defining the partition function. After inserting the topological defect, $D_{\pi}$, for the symmetry $\pi$ along a spatial cycle of the torus and computing the torus partition function some of the $1$ entries in $\CM$ become phases. We write the corresponding matrix as $\CM_{\pi}$.

The defect partition function can be obtained from performing an $S$ transformation to get $S^T\CM_{\pi}\bar S$. The characters arising from the defect partition functions for all possible symmetries, $\pi$, correspond to the non-zero entries of
\be\label{MSdef}
\sum_\pi S^T\CM_{\pi}\bar S= S^T \bigg (\sum_\pi \CM_\pi\bigg ) \bar S~,
\ee
where the sum is over all such $\pi$. Assigning phases to the primaries compatible with their fusion is the same as choosing an irreducible representation of $K_{(Q,[\sigma])}$. Therefore, for each $\pi$, we associate an irreducible representation, $R_{\pi}$ of $K_{(Q,[\sigma])}$. To identify the non-zero entries of $\sum_\pi \CM_\pi$ we should find when
\be\label{sumS}
\sigma(x):=\sum_{R_{\pi}} \chi_{R_{\pi}}(x)\ne0~.
\ee
In this expression, $\chi_{R_{\pi}}(x)$ is the character of $R_{\pi}$ evaluated on a given element, $x \in K_{(Q,[\sigma])}$.\footnote{Note that each element in $K_{(Q,\sigma)}$ represents a character combination $\chi_{\vec p}\bar\chi_{\overline{\vec g+\vec p}}\in Z_{\CT/(Q,[\sigma])}$. We will denote this combination as $(\vec p,\overline{\vec g+\vec p})$.} From the orthogonality of characters of groups, it follows that $\sigma(x)\ne0$ if and ony if $x$ is the identity element of $K_{(Q,[\sigma])}$. That is, $\sigma(x)\ne0$ if and only if $x=(\vec 0,\vec 0)$, where $\vec 0$ is the trivial representation of the chiral algebra. Therefore, the matrix, $\sum_\pi \CM_\pi$, has a non-zero entry only on the diagonal component corresponding to the trivial representation of the chiral algebra. Hence, we get
\be
\bigg (S^T \bigg (\sum_\pi \CM_\pi\bigg ) \bar S \bigg )_{ij} = S_{\vec 0, \vec i}S_{\vec 0, \vec j}~.
\ee
Therefore, the sum over defect parition functions for all $\pi$ is given by 
\bea\label{StransSumZ}
\sum_{\pi} Z_{\CT_{\CM}}^{\pi}& = &  \sum_{\vec i,\vec j} S_{ \vec 0, \vec i} \bar S_{\bar{\vec 0},\vec j} \chi_{\vec i} \bar \chi_{\overline{\vec j}} = \sum_{\vec i,\vec j} \chi_{\vec i} \bar \chi_{\overline{\vec j}}~,
\eea
where we have used the fact that the (unnormalized) S-matrix of an abelian Chern-Simons theory satisfies $S_{\vec 0,\vec i}=1$ for all $\vec i \in \text{Rep}(\DV)$. It is clear that we get characters for all $\vec i, \vec j \in \text{Rep}(\DV)$ in the expression above. In other words, the twisted-sector operators for all the lines $\pi$ together give us operators 
\be
\CO_{\vec i, \overline{\vec j}}~,
\ee
for all $\vec i, \vec j \in \text{Rep}(\DV)$. These operators can be mapped to generalized Pauli group elements as
\be
\mu: \CO_{\vec i, \overline{\vec j}} \to X^{\vec j - \vec i} \circ Z^{L \vec i}~.
\ee
Since the vectors $\vec i, \vec j$ exhaust all representations of the chiral algebra, we get the full abelianized generalized Pauli group from this map. 

\newsec{Quantum codes, gapped boundaries, and gapped interfaces}

\label{sec:gapped boundaries}

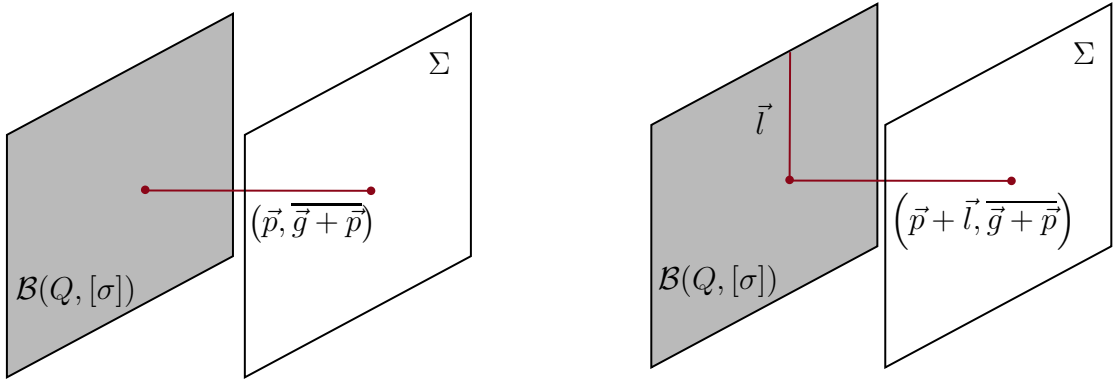
\begin{figure}[h!]
    \centering

\tikzset{every picture/.style={line width=0.75pt}} %set default line width to 0.75pt        

\begin{tikzpicture}[x=0.75pt,y=0.75pt,yscale=-1,xscale=1]
%uncomment if require: \path (0,330); %set diagram left start at 0, and has height of 330

%Shape: Parallelogram [id:dp3842962720757085] 
\draw  [color={rgb, 255:red, 0; green, 0; blue, 0 }  ,draw opacity=1 ][fill={rgb, 255:red, 74; green, 74; blue, 74 }  ,fill opacity=0.38 ] (53.99,262.16) -- (54.01,139.77) -- (168.02,78.6) -- (168,201) -- cycle ;
%Straight Lines [id:da9848515784535609] 
\draw  [color={rgb, 255:red, 139; green, 6; blue, 24 }  ,draw opacity=1 ]  (121.84,167.84) -- (236,168) ;
%Shape: Circle [id:dp3377344092080524] 
\draw  [color={rgb, 255:red, 139; green, 6; blue, 24 }  ,draw opacity=1 ][fill={rgb, 255:red, 139; green, 6; blue, 24 }  ,fill opacity=1 ] (123.75,169.15) .. controls (122.84,169.15) and (122.1,168.41) .. (122.1,167.5) .. controls (122.1,166.59) and (122.84,165.85) .. (123.75,165.85) .. controls (124.66,165.85) and (125.4,166.59) .. (125.4,167.5) .. controls (125.4,168.41) and (124.66,169.15) .. (123.75,169.15) -- cycle ;
%Shape: Parallelogram [id:dp9396233592520278] 
\draw  [color={rgb, 255:red, 0; green, 0; blue, 0 }  ,draw opacity=1 ][fill={rgb, 255:red, 74; green, 74; blue, 74 }  ,fill opacity=0.38 ] (378.99,257.16) -- (379.01,134.77) -- (493.02,73.6) -- (493,196) -- cycle ;
%Straight Lines [id:da9164932012039712] 
\draw  [color={rgb, 255:red, 139; green, 6; blue, 24 }  ,draw opacity=1 ]  (446.84,162.84) -- (561,163) ;
%Shape: Circle [id:dp020735238221246455] 
\draw  [color={rgb, 255:red, 139; green, 6; blue, 24 }  ,draw opacity=1 ][fill={rgb, 255:red, 139; green, 6; blue, 24 }  ,fill opacity=1 ] (448.75,164.15) .. controls (447.84,164.15) and (447.1,163.41) .. (447.1,162.5) .. controls (447.1,161.59) and (447.84,160.85) .. (448.75,160.85) .. controls (449.66,160.85) and (450.4,161.59) .. (450.4,162.5) .. controls (450.4,163.41) and (449.66,164.15) .. (448.75,164.15) -- cycle ;
%Straight Lines [id:da4150144169100386] 
\draw [color={rgb, 255:red, 139; green, 6; blue, 24 }  ,draw opacity=1 ]   (449,98) -- (448.75,160.85) ;
%Shape: Parallelogram [id:dp7125683037277415] 
\draw  [color={rgb, 255:red, 0; green, 0; blue, 0 }  ,draw opacity=1 ] (173.99,262.16) -- (174.01,139.77) -- (288.02,78.6) -- (288,201) -- cycle ;
%Shape: Circle [id:dp4513780538937081] 
\draw  [color={rgb, 255:red, 139; green, 6; blue, 24 }  ,draw opacity=1 ][fill={rgb, 255:red, 139; green, 6; blue, 24 }  ,fill opacity=1 ] (237.65,169.65) .. controls (236.74,169.65) and (236,168.91) .. (236,168) .. controls (236,167.09) and (236.74,166.35) .. (237.65,166.35) .. controls (238.56,166.35) and (239.3,167.09) .. (239.3,168) .. controls (239.3,168.91) and (238.56,169.65) .. (237.65,169.65) -- cycle ;
%Shape: Parallelogram [id:dp6748984568638553] 
\draw  [color={rgb, 255:red, 0; green, 0; blue, 0 }  ,draw opacity=1 ] (496.99,257.16) -- (497.01,134.77) -- (611.02,73.6) -- (611,196) -- cycle ;
%Shape: Circle [id:dp09663433192125925] 
\draw  [color={rgb, 255:red, 139; green, 6; blue, 24 }  ,draw opacity=1 ][fill={rgb, 255:red, 139; green, 6; blue, 24 }  ,fill opacity=1 ] (560.65,164.65) .. controls (559.74,164.65) and (559,163.91) .. (559,163) .. controls (559,162.09) and (559.74,161.35) .. (560.65,161.35) .. controls (561.56,161.35) and (562.3,162.09) .. (562.3,163) .. controls (562.3,163.91) and (561.56,164.65) .. (560.65,164.65) -- cycle ;

% Text Node
\draw (57.01,209.17) node [anchor=north west][inner sep=0.75pt]    {$\CB( Q,[ \sigma ])$};
% Text Node
\draw (174,172.4) node [anchor=north west][inner sep=0.75pt]    {$\left(\vec{p} ,\overline{\vec{g} +\vec{p}}\right)$};
% Text Node
\draw (381.01,204.17) node [anchor=north west][inner sep=0.75pt]    {$\CB( Q,[ \sigma ])$};
% Text Node
\draw (498,167.4) node [anchor=north west][inner sep=0.75pt]    {$\left(\vec{p} +\vec{l} ,\overline{\vec{g} +\vec{p}}\right)$};
% Text Node
\draw (265,96.4) node [anchor=north west][inner sep=0.75pt]    {$\Sigma $};
% Text Node
\draw (590,91.4) node [anchor=north west][inner sep=0.75pt]    {$\Sigma $};
% Text Node
\draw (430,122.4) node [anchor=north west][inner sep=0.75pt]    {$\vec{l}$};

\end{tikzpicture}
    \caption{Left: Folding Fig. \ref{fig:TQFT-CFT non-trivial surface} turns the surface operator, $S(Q,[\sigma])$, into a gapped boundary condition, $\CB(Q,[\sigma])$. The Wilson lines in the $\CI \times \bar \CI$ Chern-Simons theory corresponding to local CFT primaries can end on the boundary. Right: Line operators that correspond to error operators in the code end on the gapped boundary and form a junction with a non-trivial line operator on the gapped boundary. By shrinking this diagram, it is clear that error operators correspond to point operators at the end of the Verlinde line labelled by $\vec\ell \in \text{Rep}(\DV)$. This picture can also be generalized to the $\pi$ symmetries discussed in section \ref{sec:gen modular inv}.}
    \label{fig:codes and lagrangian subgroups}
\end{figure}

In section \ref{sec:selfdualsurfaces}, we showed that, under the map $\mu$, an RCFT, $\CT/(Q,[\sigma])$, is mapped to a stabilizer code if and only if the surface operator, $S(Q,[\sigma])$, in the corresponding bulk Chern-Simons theory is self-dual. By folding Fig. \ref{fig:TQFT-CFT non-trivial surface}, we can convert the surface operator, $S(Q,[\sigma])$, into a gapped boundary, $\CB(Q,[\sigma])$, to get the left diagram in Fig. \ref{fig:codes and lagrangian subgroups}.

The Wilson lines of the bulk Chern-Simons theory that can end on this gapped boundary are precisely those determined by the CFT partition function. Let us show this statement algebraically. To that end, recall that the representations, $\vec p$, of the chiral algebra, $\mathds{V},$ define the Wilson lines of a TQFT which we will denote by $\CI$. Consider the product TQFT, $\CI \times \overline \CI$, which has Wilson lines labelled by
\be
(\vec p, \vec q)~,
\ee
where $\vec p$ and $\vec q$ are Wilson lines in $\CI$ and $\bar \CI$, respectively. The bar on $\overline\CI$ indicates that it has Wilson lines with topological spins that are complex conjugates of the topological spins of the lines in $\CI$. Note that the number of line operators in $\CI \times \bar \CI$ is $|K|^2$ and that these operators form the group $K \times K$ under fusion. Recall that the $Q$-orbifold torus partition function is
\begin{equation}
Z_{(\CT/Q,[\sigma])}=\sum_{\vec g\in Q}\sum_{\vec p\in B_{\vec g}}\chi_{\vec p}(q)\bar\chi_{\overline{\vec p+\vec g}}(\bar q)~,
\end{equation}
where  
\be
\label{pconstgapped}
B_{\vec g}:=\left\{\vec p\ \Big|\ S_{\vec h,\vec  p} ~ \Xi(\vec h,\vec g)=1 ~,\ \forall \vec h \in Q\right\}~.
\ee
The CFT primaries of the form $\CO_{(\vec p, \overline{\vec g + \vec p})}$ define a subgroup of $|K|$ Wilson lines labelled by $(\vec p, \overline{\vec g + \vec p})$ in $\CI \times \bar \CI$. In fact, these Wilson lines are all bosons. To see this, consider
\be
\theta_{(p, \overline{\vec p + \vec g})}= \frac{\theta_{p}}{\theta_{\vec p + \vec g}} =  \frac{\theta_{p}\theta_{g}}{\theta_{\vec p + \vec g}} \frac{1}{\theta_{\vec g}} = \frac{1}{S_{\vec g, \vec p}\theta_{\vec g}}=1~,
\ee
where the last equality follows form setting $\vec h= \vec g$ in \eqref{pconstgapped}.

Therefore, the $\CO_{(\vec p, \overline{\vec g + \vec p})}$ operators define a Lagrangian subgroup of $|K|$ bosons in $\CI \boxtimes \bar \CI$. This subgroup is isomorphic to the group formed by the CFT primaries under fusion, and so we will denote it as $K_{(Q,[\sigma])}$. The corresponding set of Wilson lines consists of the lines that can end on the gapped boundary, $\CB(Q,[\sigma])$. 

The relationship between CFT partition functions and gapped boundaries of  $\CI \boxtimes \bar \CI$ provides a new interpretation for our quantum codes. First of all, note that the self-duality of the quantum code is directly related to the Lagrangian property of $K_{(Q,[\sigma])}$. The Wilson lines labelled by $(\vec p, \overline{\vec g + \vec p})$ in the subgroup $K_{(Q,[\sigma])}$ can end on the gapped boundary $\CB(Q,[\sigma])$. Any other Wilson line braids non-trivially with at least one of the Wilson lines in $K_{(Q,[\sigma])}$. Therefore, the stabilizer group corresponds to Wilson lines that can end on the boundary, and the rest of the elements of the generalized Pauli group correspond to the other Wilson lines in $\CI \boxtimes \bar \CI$. The line operators that correspond to errors end on $\CB(Q,[\sigma])$ with a tail on the boundary (see Fig. \ref{fig:codes and lagrangian subgroups}). 

\begin{figure}[h!]
    \centering

\tikzset{every picture/.style={line width=0.75pt}} %set default line width to 0.75pt        

\begin{tikzpicture}[x=0.75pt,y=0.75pt,yscale=-1,xscale=1]
%uncomment if require: \path (0,330); %set diagram left start at 0, and has height of 330

%Shape: Parallelogram [id:dp3842962720757085] 
\draw  [color={rgb, 255:red, 0; green, 0; blue, 0 }  ,draw opacity=1 ][fill={rgb, 255:red, 74; green, 74; blue, 74 }  ,fill opacity=0.38 ] (219.99,258.16) -- (220.01,135.77) -- (334.02,74.6) -- (334,197) -- cycle ;
%Straight Lines [id:da9848515784535609] 
\draw  [color={rgb, 255:red, 139; green, 6; blue, 24 }  ,draw opacity=1 ]  (287.84,163.84) -- (402,164) ;
%Shape: Circle [id:dp3377344092080524] 
\draw  [color={rgb, 255:red, 139; green, 6; blue, 24 }  ,draw opacity=1 ][fill={rgb, 255:red, 139; green, 6; blue, 24 }  ,fill opacity=1 ] (289.75,165.15) .. controls (288.84,165.15) and (288.1,164.41) .. (288.1,163.5) .. controls (288.1,162.59) and (288.84,161.85) .. (289.75,161.85) .. controls (290.66,161.85) and (291.4,162.59) .. (291.4,163.5) .. controls (291.4,164.41) and (290.66,165.15) .. (289.75,165.15) -- cycle ;
%Shape: Parallelogram [id:dp7125683037277415] 
\draw  [color={rgb, 255:red, 0; green, 0; blue, 0 }  ,draw opacity=1 ] (339.99,258.16) -- (340.01,135.77) -- (454.02,74.6) -- (454,197) -- cycle ;
%Shape: Circle [id:dp4513780538937081] 
\draw  [color={rgb, 255:red, 139; green, 6; blue, 24 }  ,draw opacity=1 ][fill={rgb, 255:red, 139; green, 6; blue, 24 }  ,fill opacity=1 ] (403.65,165.65) .. controls (402.74,165.65) and (402,164.91) .. (402,164) .. controls (402,163.09) and (402.74,162.35) .. (403.65,162.35) .. controls (404.56,162.35) and (405.3,163.09) .. (405.3,164) .. controls (405.3,164.91) and (404.56,165.65) .. (403.65,165.65) -- cycle ;

% Text Node
\draw (227.01,219.17) node [anchor=north west][inner sep=0.75pt]    {$\mathcal{W}$};
% Text Node
\draw (340,170.4) node [anchor=north west][inner sep=0.75pt]  [font=\scriptsize]  {$\left(\vec{p} ,\overline{\vec{g} +\vec{p}}\right) \in \mu ^{-1}( A)$};
% Text Node
\draw (431,92.4) node [anchor=north west][inner sep=0.75pt]    {$\Sigma $};
% Text Node
\draw (338,89.4) node [anchor=north west][inner sep=0.75pt]    {$\mathcal{I} \times \overline{\mathcal{I}}$};
% Text Node
\draw (143,91.4) node [anchor=north west][inner sep=0.75pt]    {$\mathcal{I} \times \overline{\mathcal{I}} /\mu ^{-1}( A)$};

\end{tikzpicture}
    \caption{A non-self-dual stabilizer code with stabilizer group given by a subgroup, $A \subset \CS$, corresponds to a set of Wilson lines, $\mu^{-1}(A)$, which can end on the gapped interface $\CW$ separating the TQFT $\CI \times \bar\CI$ and $\CI \times \bar\CI / \mu^{-1}(A)$ where $\CI \times \bar\CI / \mu^{-1}(A)$ is obtained from condensing the Wilson lines in $\mu^{-1}(A)$.}
    \label{fig:gapped interface}
\end{figure}
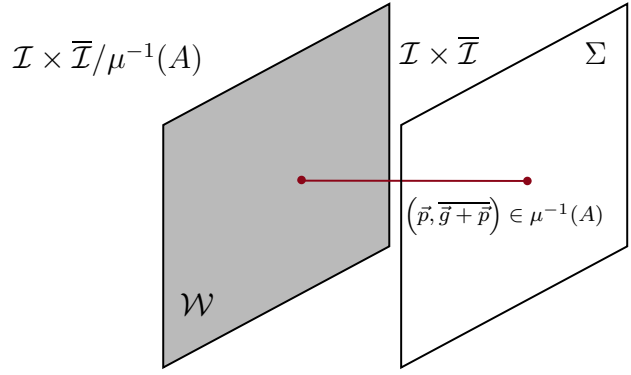

More generally, a proper subgroup, $A<\CS_{(Q,[\sigma])}$, is a stabilizer group for a non-self-dual code. Corresponding to $A$, we have a subgroup, $\mu^{-1}(A) \subset K_{(Q,[\sigma])}$, of bosonic Wilson lines in $\CI \times \bar \CI$. $\mu^{-1}(A)$ is a set of bosonic Wilson lines that can be condensed to define a gapped interface. Therefore, non-self-dual sub-codes determined by proper subgroups of $\CS_{(Q,[\sigma]])}$ correspond to gapped interfaces obtained from condensing a subset of Wilson lines in $K_{(Q,[\sigma])}$ in the bulk CS theory (see Fig. \ref{fig:gapped interface}).\footnote{This relation between Lagrangian subgroups and self-dual codes and between non-Lagrangian isotropic subgroups (gapped interfaces) and non-self-dual codes were first annonced in two talks by the authors at King's College London during the workshop \href{https://nms.kcl.ac.uk/gerard.watts/DefSym22/}{Defects and Symmetry 2022}.} 

\subsection{SymTFTs, code equivalences, and invertible surfaces}\label{CodeEqSec}
In this section, we relate the above discussion to SymTFTs (e.g., see \cite{Freed:2012bs,Freed:2022qnc,Kaidi:2022cpf,Kaidi:2023maf} for an introduction to these TQFTs) and shed light on CFT dualities and code equivalences.

To that end, note that the $\CT$ theory with charge conjugation modular invariant has an Abelian $K$ 0-form symmetry. Since the theory is modular, the SymTFT for this symmetry is a (twisted) $K$ Dijkgraaf-Witten (DW) theory (i.e., a discrete gauge theory with gauge group $K$ and a possible DW twist, $\omega\in H^3(K,U(1))$), which we denote as $D(K)_{\omega}$. More succinctly, we have
\begin{equation}\label{symTFTId}
D(K)_{\omega}\cong\CI\times\bar\CI~.
\end{equation}
This TQFT is precisely a theory of the type described in Fig. \ref{fig:codes and lagrangian subgroups} (with $\CB(Q,[\sigma])=\CB(1,[1])$).

\begin{figure}[h!]
    \centering

\tikzset{every picture/.style={line width=0.75pt}} %set default line width to 0.75pt        

\begin{tikzpicture}[x=0.75pt,y=0.75pt,yscale=-1,xscale=1]
%uncomment if require: \path (0,330); %set diagram left start at 0, and has height of 330

%Shape: Parallelogram [id:dp3842962720757085] 
\draw  [color={rgb, 255:red, 0; green, 0; blue, 0 }  ,draw opacity=1 ][fill={rgb, 255:red, 74; green, 74; blue, 74 }  ,fill opacity=0.38 ] (21.99,257.03) -- (22.01,138.66) -- (116.42,79.5) -- (116.4,197.88) -- cycle ;
%Shape: Parallelogram [id:dp20738060683314363] 
\draw  [color={rgb, 255:red, 245; green, 166; blue, 35 }  ,draw opacity=1 ][fill={rgb, 255:red, 248; green, 198; blue, 155 }  ,fill opacity=1 ] (82.96,258) -- (82.98,139.63) -- (177.39,80.47) -- (177.37,198.84) -- cycle ;
%Shape: Parallelogram [id:dp9396233592520278] 
\draw  [color={rgb, 255:red, 0; green, 0; blue, 0 }  ,draw opacity=1 ][fill={rgb, 255:red, 74; green, 74; blue, 74 }  ,fill opacity=0.38 ] (454.85,256.13) -- (454.87,137.76) -- (549.28,78.6) -- (549.26,196.97) -- cycle ;
%Shape: Parallelogram [id:dp7125683037277415] 
\draw  [color={rgb, 255:red, 0; green, 0; blue, 0 }  ,draw opacity=1 ] (147.28,257.03) -- (147.3,138.66) -- (241.71,79.5) -- (241.69,197.88) -- cycle ;
%Shape: Parallelogram [id:dp6748984568638553] 
\draw  [color={rgb, 255:red, 0; green, 0; blue, 0 }  ,draw opacity=1 ] (552.57,257.13) -- (552.59,138.76) -- (647,79.6) -- (646.98,197.97) -- cycle ;

% Text Node
\draw (26.42,207.43) node [anchor=north west][inner sep=0.75pt]  [font=\scriptsize]  {$\CB( 1,[ 1])$};
% Text Node
\draw (457.96,209.46) node [anchor=north west][inner sep=0.75pt]  [font=\scriptsize]  {$\CB( Q,[ \sigma ])$};
% Text Node
\draw (216.47,96.47) node [anchor=north west][inner sep=0.75pt]    {$\Sigma $};
% Text Node
\draw (617.73,102.37) node [anchor=north west][inner sep=0.75pt]    {$\Sigma $};
% Text Node
\draw (85.39,205.5) node [anchor=north west][inner sep=0.75pt]  [font=\scriptsize]  {$S( E(Q),[ \sigma ])$};
% Text Node
\draw (259,169.4) node [anchor=north west][inner sep=0.75pt]    {$\xrightarrow{\text{fuse}\ S( E(Q),[ \sigma ]) \ \text{with}\ \CB( 1,[ 1])}$};

\end{tikzpicture}
    \caption{Fusing the surface operator $S(E(Q),[\sigma])$ with the gapped boundary $\CB(1,[1])$ changes the gapped boundary to $\CB(Q,[\sigma])$.}
    \label{fig:surface fusing with gapped boundary}
\end{figure}
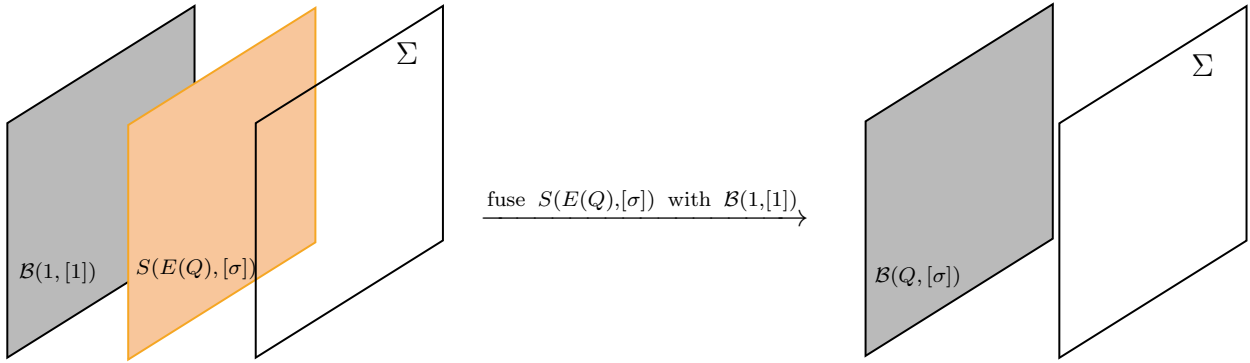

Different orbifold theories, $\CT/(Q,[\sigma])$, correspond to the same bulk SymTFT but have different gapped boundary conditions gotten by fusing the embedding of the surface $S(Q,[\sigma])$ in the folded theory, which we will call $S(E(Q),[\sigma])$, with the gapped boundary, $\CB(1,[1])$. This maneuver results in the transformation $\CB(1,[1])\to\CB(Q,[\sigma])$ (i.e., a transformation from a Dirichlet boundary condition to a partial Neumann boundary condition specified by $Q\le K$. See Fig. \ref{fig:surface fusing with gapped boundary}). In general, such new boundary conditions are allowed whenever the restriction of the DW twist to $Q$, $\omega|_Q$, is trivial in $H^3(Q,U(1))$. 

Now, when the surface, $S(E(Q),[\sigma]))$, is invertible, it generates an $H\cong\mathbb{Z}_n$ bulk 0-form symmetry, where $n$ is the order of $E(S(Q,[\sigma]))$ under fusion.\footnote{Note that, in general, $\mathbb{Z}_n$ is not isomorphic to $Q$ as a group (see the discussion of $c=1$ theories below).} Since the corresponding 't Hooft anomaly vanishes ($H^4(\mathbb{Z}_n,U(1))$ is trivial for all $n$), we can gauge the $H$ symmetry. Such gauging has been described in the condensed matter literature in \cite{barkeshli2019symmetry} and in the SymTFT literature in \cite{Kaidi:2022cpf}.

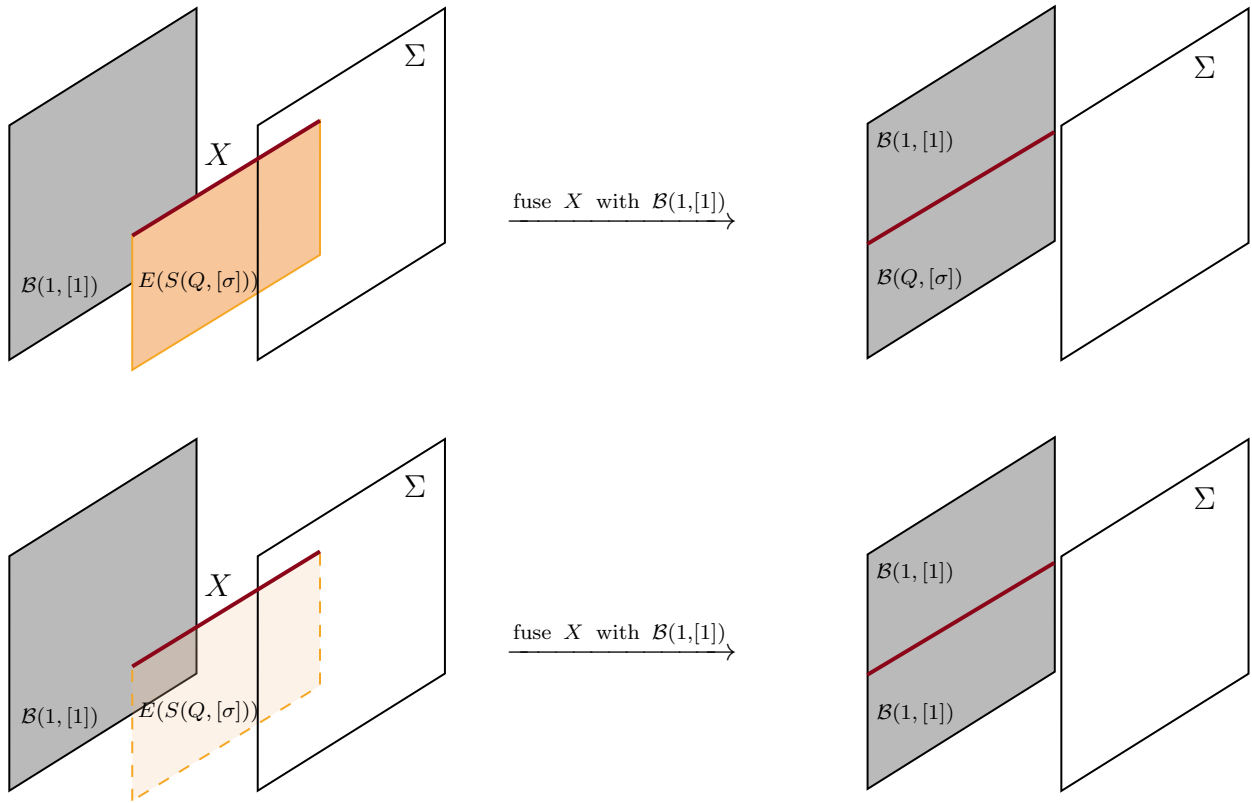
\begin{figure}[h!]
    \centering

\tikzset{every picture/.style={line width=0.75pt}} %set default line width to 0.75pt        

\begin{tikzpicture}[x=0.75pt,y=0.75pt,yscale=-1,xscale=1]
%uncomment if require: \path (0,330); %set diagram left start at 0, and has height of 330

%Shape: Parallelogram [id:dp3842962720757085] 
\draw  [color={rgb, 255:red, 0; green, 0; blue, 0 }  ,draw opacity=1 ][fill={rgb, 255:red, 74; green, 74; blue, 74 }  ,fill opacity=0.38 ] (21.99,257.03) -- (22.01,138.66) -- (116.42,79.5) -- (116.4,197.88) -- cycle ;
%Shape: Parallelogram [id:dp20738060683314363] 
\draw  [color={rgb, 255:red, 245; green, 166; blue, 35 }  ,draw opacity=1 ][fill={rgb, 255:red, 248; green, 198; blue, 155 }  ,fill opacity=1 ] (83.99,262.08) -- (84.01,194.33) -- (178.7,136.25) -- (178.69,204) -- cycle ;
%Shape: Parallelogram [id:dp9396233592520278] 
\draw  [color={rgb, 255:red, 0; green, 0; blue, 0 }  ,draw opacity=1 ][fill={rgb, 255:red, 74; green, 74; blue, 74 }  ,fill opacity=0.38 ] (454.85,256.13) -- (454.87,137.76) -- (549.28,78.6) -- (549.26,196.97) -- cycle ;
%Shape: Parallelogram [id:dp7125683037277415] 
\draw  [color={rgb, 255:red, 0; green, 0; blue, 0 }  ,draw opacity=1 ] (147.28,257.03) -- (147.3,138.66) -- (241.71,79.5) -- (241.69,197.88) -- cycle ;
%Shape: Parallelogram [id:dp6748984568638553] 
\draw  [color={rgb, 255:red, 0; green, 0; blue, 0 }  ,draw opacity=1 ] (552.57,257.13) -- (552.59,138.76) -- (647,79.6) -- (646.98,197.97) -- cycle ;
%Straight Lines [id:da41466650118075077] 
\draw [color={rgb, 255:red, 139; green, 6; blue, 24 }  ,draw opacity=1 ][line width=1.5]    (178.7,136.25) -- (84.01,194.33) ;
%Straight Lines [id:da7621114342280537] 
\draw [color={rgb, 255:red, 139; green, 6; blue, 24 }  ,draw opacity=1 ][line width=1.5]    (549,142) -- (454.71,198.52) ;

% Text Node
\draw (26.42,213.43) node [anchor=north west][inner sep=0.75pt]  [font=\scriptsize]  {$\CB( 1,[ 1])$};
% Text Node
\draw (457.96,209.46) node [anchor=north west][inner sep=0.75pt]  [font=\scriptsize]  {$\CB( Q,[ \sigma ])$};
% Text Node
\draw (219.47,96.47) node [anchor=north west][inner sep=0.75pt]    {$\Sigma $};
% Text Node
\draw (617.73,102.37) node [anchor=north west][inner sep=0.75pt]    {$\Sigma $};
% Text Node
\draw (85.39,210.5) node [anchor=north west][inner sep=0.75pt]  [font=\scriptsize]  {$E( S( Q,[ \sigma ]))$};
% Text Node
\draw (271,169.4) node [anchor=north west][inner sep=0.75pt]    {$\xrightarrow{\text{fuse}\ X\ \text{with}\ \CB( 1,[ 1])}$};
% Text Node
\draw (119,146.4) node [anchor=north west][inner sep=0.75pt]    {$X$};
% Text Node
\draw (457.87,140.16) node [anchor=north west][inner sep=0.75pt]  [font=\scriptsize]  {$\CB( 1,[ 1])$};

\end{tikzpicture}
\vspace{0.2cm}

\begin{tikzpicture}[x=0.75pt,y=0.75pt,yscale=-1,xscale=1]
%uncomment if require: \path (0,330); %set diagram left start at 0, and has height of 330

%Shape: Parallelogram [id:dp3842962720757085] 
\draw  [color={rgb, 255:red, 0; green, 0; blue, 0 }  ,draw opacity=1 ][fill={rgb, 255:red, 74; green, 74; blue, 74 }  ,fill opacity=0.38 ] (21.99,257.03) -- (22.01,138.66) -- (116.42,79.5) -- (116.4,197.88) -- cycle ;
%Shape: Parallelogram [id:dp20738060683314363] 
\draw  [color={rgb, 255:red, 245; green, 166; blue, 35 }  ,draw opacity=1 ][fill={rgb, 255:red, 248; green, 198; blue, 155 }  ,fill opacity=0.23 ][dash pattern={on 4.5pt off 4.5pt}] (83.99,262.08) -- (84.01,194.33) -- (178.7,136.25) -- (178.69,204) -- cycle ;
%Shape: Parallelogram [id:dp9396233592520278] 
\draw  [color={rgb, 255:red, 0; green, 0; blue, 0 }  ,draw opacity=1 ][fill={rgb, 255:red, 74; green, 74; blue, 74 }  ,fill opacity=0.38 ] (454.85,256.13) -- (454.87,137.76) -- (549.28,78.6) -- (549.26,196.97) -- cycle ;
%Shape: Parallelogram [id:dp7125683037277415] 
\draw  [color={rgb, 255:red, 0; green, 0; blue, 0 }  ,draw opacity=1 ] (147.28,257.03) -- (147.3,138.66) -- (241.71,79.5) -- (241.69,197.88) -- cycle ;
%Shape: Parallelogram [id:dp6748984568638553] 
\draw  [color={rgb, 255:red, 0; green, 0; blue, 0 }  ,draw opacity=1 ] (552.57,257.13) -- (552.59,138.76) -- (647,79.6) -- (646.98,197.97) -- cycle ;
%Straight Lines [id:da41466650118075077] 
\draw [color={rgb, 255:red, 139; green, 6; blue, 24 }  ,draw opacity=1 ][line width=1.5]    (178.7,136.25) -- (84.01,194.33) ;
%Straight Lines [id:da7621114342280537] 
\draw [color={rgb, 255:red, 139; green, 6; blue, 24 }  ,draw opacity=1 ][line width=1.5]    (549,142) -- (454.71,198.52) ;

% Text Node
\draw (26.42,213.43) node [anchor=north west][inner sep=0.75pt]  [font=\scriptsize]  {$\CB( 1,[ 1])$};
% Text Node
\draw (457.96,211.46) node [anchor=north west][inner sep=0.75pt]  [font=\scriptsize]  {$\CB( 1,[ 1])$};
% Text Node
\draw (219.47,96.47) node [anchor=north west][inner sep=0.75pt]    {$\Sigma $};
% Text Node
\draw (617.73,102.37) node [anchor=north west][inner sep=0.75pt]    {$\Sigma $};
% Text Node
\draw (85.39,210.5) node [anchor=north west][inner sep=0.75pt]  [font=\scriptsize]  {$E( S( Q,[ \sigma ]))$};
% Text Node
\draw (271,169.4) node [anchor=north west][inner sep=0.75pt]    {$\xrightarrow{\text{fuse}\ X\ \text{with}\ \CB( 1,[ 1])}$};
% Text Node
\draw (119,146.4) node [anchor=north west][inner sep=0.75pt]    {$X$};
% Text Node
\draw (457.87,140.16) node [anchor=north west][inner sep=0.75pt]  [font=\scriptsize]  {$\CB( 1,[ 1])$};

\end{tikzpicture}
    \caption{Top: Fusing the twisted sector line $X$ with the gapped-boundary $\CB(1,[1])$ gives a 1-dimensional interface between the $\CB(1,[1])$ and $\CB(Q,[\sigma])$ gapped-boundary conditions. Bottom: Gauging the surface operator $E(S(Q,[\sigma]))$ makes it trivial and liberates the line operator $X$. In this case, fusing the line with the gapped-boudary gives a line operator of the $\CB(1,[1])$ gapped boundary \cite{Sun:2023xxv}.}
    \label{fig:fusing X with gapped boundary}
\end{figure}

The upshot is that, by 0-form gauging, we trivialize the $S(E(Q),[\sigma])$ surface and liberate non-genuine lines that bound it (gauging effectively \lq\lq erases" the surfaces attached to these non-genuine lines). As a result, we produce a bulk theory that includes a non-invertible line, $X$, such that
\begin{equation}\label{liberatedTwist}
X\times\bar X=\sum_{\ell\in \widetilde{E(Q)}}\ell~,
\end{equation}
where the lines on the RHS of the above equation are images, under 0-form gauging, of the lines in $\CI\times\bar\CI$ that are condensed to produce $E(S(Q,[\sigma]))$. 
If the boundary theory is invariant under $Q$ gauging, then, bringing the lines in \eqref{liberatedTwist} to the boundary, we obtain a fusion rule for the non-invertible line in a $Q$ Tambara-Yamagami (TY) fusion category (see Fig. \ref{fig:fusing X with gapped boundary}). This is the CFT manifestation of duality.

As a well-known class of examples, consider certain RCFT points on the $c=1$ compact boson conformal manifold with $R^2=2k$ and $k\in\mathbb{Z}$. This set of theories is related by T-duality to points with $R^2=2/k$. It is easy to check that, in the unfolded CS bulk picture, these theories differ by the surface operator that has been inserted: in the $R^2=2k$ case we have theories with a trivial surface operator inserted, while, in the case of $R^2=2/k$, the corresponding bulk CS theory has an insertion of an order-two charge conjugation surface, $S(\mathbb{Z}_k,[1])$.\footnote{This statement corresponds to the well-known fact that the $R^2=2k$ and $R^2=2/k$ theories correspond to charge conjugation and diagonal modular invariants, respectively (e.g., see \cite{Gukov:2002nw,Thorngren:2021yso,Fuchs:2001am,Fuchs:2007tx}).} In the folded picture of this section, the bulk theories differ by their boundary conditions: $\CB(1,[1])$ versus $\CB(\mathbb{Z}_k,[1])$. In \eqref{liberatedTwist} we find lines organized in orbits of charge conjugation. Taking these lines to the boundary gives us $\mathbb{Z}_k$ Tambara-Yamagami fusion rules \cite{Thorngren:2021yso}. These non-invertible lines capture the T-duality of the CFT.

Our assignment of codes to QFTs involves associating a code with a particular surface operator. In our discussion, distinct $(Q,[\sigma])$ led to distinct surface operators and hence to distinct codes. On the other hand, in the $c=1$ RCFTs reviewed in the previous paragraph, the $S(\mathbb{Z}_k,[1])$ charge conjugation surfaces led to T-duality in the RCFT, it is natural to expect that the corresponding codes generated by $\mu$ are (in a sense we will make precise momentarily) equivalent.\footnote{Note that the example discussed in Fig. \ref{fig:orbgraph2} shows that this logic cannot extend in general to non-invertible surfaces since boundary CFTs corresponding to non-invertible bulk surfaces generally have different fusion rules from those with invertible surfaces.} More generally, we can also think at the level of the 3d bulk and note that we can always gauge an invertible symmetry corresponding to $S(Q,[\sigma])$. In the corresponding CS theory, this gauging trivializes the surface. Since gauging is an invertible procedure, it is natural to imagine that codes generated by $\mu$ should be equivalent whenever two CFTs, $\CT(Q_1,[\sigma_1])$ and $\CT(Q_2,[\sigma_2])$, correspond to surfaces related as follows
\begin{equation}
S(Q_1,[\sigma_1])=S(Q_2,[\sigma_2])\times S(Q_3, [\sigma_3])~,
\end{equation}
where $S(Q_3,[\sigma_3])$ is an invertible surface.

In fact, we already alluded to such code equivalences in our discussion of the case of the $SU(3)_1$ theory in Sec. \ref{SU3ex}, where $Q\cong\mathbb{Z}_3$ and $H\cong\mathbb{Z}_2$. There we commented that the codes in \eqref{SU3code1} and \eqref{SU3code2} were in fact code equivalent (i.e., that there is an element of the Clifford group that transforms one code into the other).

We are therefore motivated to study more general conditions under which $\mu$ produces code equivalences via orbifolding by a symmetry that corresponds to a bulk invertible surface. To that end, consider a single qudit of dimension given by an integer $d\geq 2$ with Pauli group, $\CP$, generated by the $X$ and $Z$ operators, both of order $d$. A general element of the Pauli group (up to overall phases) can be written as
\be
X^{\alpha} \circ Z^{\beta}~.
\ee
In this case, the abelianized Pauli group is isomorphic to the finite group $\mathds{Z}_{d} \times \mathds{Z}_{d}$. The automorphisms of this Pauli group are unitaries, $U$, which act as
\be
U (X^{\alpha} \circ Z^{\beta}) U^{-1} \in \CP~.
\ee
Such unitaries form the single qudit Clifford group, $\CJ$.

To get a handle on $\CJ$, we wish to build up a generating set of elements. Without loss of generality, we can consider the following unitary
\be
U X^{\alpha} U^{-1} = Z^{-\alpha} ~, ~~ U Z^{\alpha} U^{-1} = X^{\alpha} ~.
\ee
It is clear that $U$ sends Pauli group elements to themselves and is therefore an automorphism of the Pauli group (i.e., $U\in\CJ$). The action of this unitary on the vector $(\alpha, \beta)$ specifying a general element of the Pauli group is given by the matrix
\be
F:= \begin{pmatrix}
    0 & 1\\
    -1 & 0
\end{pmatrix}~.
\ee

To get a generating set for $\CJ$, let us also consider the following actions on the Pauli group
\be
X^{\alpha} \circ Z^{\beta} \to X^{\gamma^{-1} \alpha} \circ Z^{\gamma \beta}~,
\ee
and 
\be
X^{\alpha} \circ Z^{\beta} \to X^{\alpha} \circ Z^{\gamma \alpha + \beta}~,
\ee
where $\gamma \in \mathds{Z}^{\times}_d$, and the multiplicative inverse is denoted by $\gamma^{-1}$. Note that for $\gamma$ to be invertible, it must be co-prime to $d$. As matrices acting on the vector $(\alpha, \beta)$, the above operations can be written as
\be
 M_{\gamma}=\begin{pmatrix}
    \gamma^{-1} & 0\\
    0 & \gamma
\end{pmatrix}~,~~P_{\gamma}=\begin{pmatrix}
    1 & 0\\
    \gamma & 1
\end{pmatrix}~,
\ee
respectively. In fact, the unitaries $F$, $M_{\gamma}$, and $P_{\gamma}$ generate the full Clifford group $\CJ$ \cite{Gottesman:1998se,Farinholt_2014}. For two stabilizer groups $\CS_1$ and $\CS_2$, if there exists a unitary $U\in \CJ$ such that 
\be
U \CS_1 U^{-1}= \CS_2~,
\ee
then we say that the stabilizer codes determined by $\CS_1$ and $\CS_2$ are equivalent.

Let us now use the above operations in the context of our CFT map. To that end, suppose that $\CT/(Q,[\sigma])$ has a permutation modular invariant (i.e., $S(Q,[\sigma])$ is invertible), and, under the map $\mu$, this CFT has a corresponding stabilizer group, $\CS_{(Q,[\sigma])}$, with elements of the form
\be
X^{g} \circ Z^{L p}~.
\ee
In particular, recall that $\CT$ (with charge-conjugation partition function) gives the stabilizer group with elements of the form $Z^{L p}$. Note that in this case $L$ is an integer which is co-prime to $d$ (see equation \eqref{eq:L def}). Since we have a single qudit, the stabilizer group has a single generator. Without loss of generality, we can choose the generator to be 
\be
X^{g_0} \circ Z^{L p_0}~,
\ee
for some $g_0$ and $p_0$.
Now, we will show that there exists a unitary, $U$, such that 
\be
U(X^{g_0} \circ Z^{L p_0})U^{-1}= Z~.
\ee
In fact, this statement is a direct consequence of the Pauli-Euclid-Gottesman (PEG) Lemma:
\vspace{0.2cm}

\noindent \textbf{Lemma (PEG) \cite{Nielsen_2002}:} For integers $0\leq \alpha,\beta < d$, there exists a unitary operator, $U(\alpha,\beta) \in \CJ$, such that 
\be
U(\alpha,\beta) (X^{\alpha} \circ Z^{\beta}) U(\alpha,\beta)^{-1} = Z^{\text{gcd}(\alpha,\beta)}~,
\ee
where we define gcd$(0,\alpha)=\alpha$. 
\vspace{0.2cm}

\noindent
Applying the PEG lemma to the generator, $X^{g_0} \circ Z^{L p_0}$, we find that there exists a unitary, $U(g_0,p_0)$, such that 
\be
U(X^{g_0} \circ Z^{L p_0})U^{-1}= Z^{\text{gcd}(g_0,Lp_0)}~.
\ee
Since we started with a self-dual code, the generator $X^{g_0} \circ Z^{L p_0}$ has order $d$. Of course, the conjugation action by the unitary $U$ does not change the order. Therefore, $Z^{\text{gcd}(g_0,L p_0)}$ also has order $d$. Moreover, the resulting stabilizer group only depends on the $Z$ generalized Pauli matrix. Such a self-dual stabilizer group is unique, and so the stabilizer group generated by $Z^{\text{gcd}(g_0,Lp_0)}$ is the same as the stabilizer group generated by $Z$. In other words, the stabilizer groups $\CS_{(Q,[\sigma])}$ and $\CS_{(\DZ_1,[1])}$ are related by a Clifford group element. We have shown that the codes associated with the charge-conjugation CFT $\CT$ and its orbifold, $\CT/(Q,[\sigma])$, corresponding to an invertible surface operator, $S(Q,[\sigma])$, are equivalent. 

So far, we considered the case of a single qudit of dimension $d$. Let us generalize this discussion to a system of $n$ qudits, each of dimension $d$. Let $\CP_n$ be the Pauli group acting on these $n$ qudits. In this case, the abelianized generalized Pauli group is the group $\mathds{F}^{n}_d \times \mathds{F}^n_d$. The set of unitaries which preserve the Pauli group is the Clifford group, $\CJ_n$. Note that the unitaries $F$, $P_{\gamma}$, and $M_{\gamma}$ act on a single qudit. Even though they generate the Clifford group acting on single qudits, they do not generate $\CJ_{n}$. In order to generate $\CJ_{n}$, we need one more generator that acts on two qudits. This 2-qudit unitary, $D_{i,j}$, acts on the $i^{\text{th}}$ and $j^{\text{th}}$ qudits as
\be
X^{\alpha_1} \otimes X^{\alpha_2} \circ Z^{\beta_1} \otimes Z^{\beta_2} \to X^{\alpha_1} \otimes X^{\alpha_2-\alpha_1} \circ Z^{\beta_1+\beta_2} \otimes Z^{\beta_2}~.
\ee
As a matrix acting on the vector $(\alpha_1,\beta_1,\alpha_2,\beta_2)$, we have
\be
D_{i,j}:=\begin{pmatrix}
    1 & 0 & 0 & 0 \\
     0 & 1 & 0 & 1 \\
      0 & 0 & 1 & 0 \\
       0 & 1 & 0 & 1 
\end{pmatrix}~.
\ee
The single qudit unitaries $F_i$, $P_{\gamma,i}$, and $M_{\gamma,i}$ acting on the $i^{\text{th}}$ qudit\footnote{We have added the subscript $i$ to the notation for the single qudit unitaries to emphasize which qudit they act on.} and the 2-qudit unitary $D_{i,j}$ together generate the full Clifford group, $\CJ_n$ \cite{Farinholt_2014}.  For two stabilizer groups $\CS_1$ and $\CS_2$, if there exists a unitary, $U\in \CJ_n$, such that 
\be
U \CS_1 U^{-1}= \CS_2~,
\ee
then we say that the stabilizer codes determined by $\CS_1$ and $\CS_2$ are equivalent. 

Let us now relate the above discussion to the codes appearing via $\mu$. To that end, consider the CFT, $\CT/(Q,[\sigma])$, determined by an invertible self-dual surface operator, $S(Q,[\sigma])$, in the bulk CS theory. Now, consider the corresponding stabilizer code, $\CS_{(Q,[\sigma])}$, with elements of the form
\be
X^{\vec g} \circ Z^{L \vec p}~.
\ee
Without loss of generality, we can choose the generators of the stabilizer group to be
\be
\label{eq:stabilizer gens}
\{X^{\vec g_i} \circ Z^{L \vec p_i}\}, ~~ i \in {1,\dots,n} ~,
\ee
The explicit values of the integer vectors $\vec g_i,\vec p_i$ will not play a role in our discussion below. Since $S(Q,[\sigma])$ is invertible, $\CT/(Q,[\sigma])$ has a permutation modular invariant. Therefore, we have
\be
\CS_{(Q,[\sigma])}\cong K \cong \DZ_{d}^n~,
\ee
where $K$ is the group formed by the chiral primaries under fusion. As a result, the generators in \eqref{eq:stabilizer gens} are all of order $d$.  

We will show that the stabilizer code, $S(Q,[\sigma])$, is equivalent to the code, $S(\DZ_1,[1])=\mu(\CT)$. That is, there exists a unitary operator, $U \in \CJ_n$, such that 
\be
U S(Q,[\sigma]) U^{-1} = S(\DZ_1,[1])  ~.
\ee
In order to prove this assertion, we will use the Generalized PEG Lemma:
\vspace{0.2cm}

\noindent \textbf{Lemma (Generalized PEG) \cite{Nielsen_2002}:} For integers $0\leq \alpha_1,\dots, \alpha_n,\beta_1,\dots , \beta_n< d$, there exists a unitary operator $\tilde U_n(\vec \alpha,\vec \beta) \in \CJ_n$ such that 
\be
\tilde U_n(\vec \alpha,\vec \beta) (X^{\vec \alpha} \circ Z^{\vec \beta}) \tilde U_n(\vec \alpha,\vec \beta)^{-1} = I^{\otimes n-1} \otimes Z^{\text{gcd}(\alpha_1,\dots,\alpha_n,\beta_1,\dots \beta_n)}~,
\ee
where we define gcd$(0,a)=\text{gcd}(a,0)=a$.\footnote{For $n=1$, we get the PEG lemma discussed above.}
\vspace{0.2cm}

\noindent
If $X^{\vec \alpha} \circ Z^{\beta}$ is order $d$, then $I^{\otimes n-1} \otimes Z^{\text{gcd}(\alpha_1,\dots,\alpha_n,\beta_1,\dots \beta_n)}$ is also order $d$. This logic implies that the integer, $\text{gcd}(\alpha_1,\dots,\alpha_n,\beta_1,\dots \beta_n)$, is coprime to $d$ and is an element of $\DZ_{d}^{\times}$. Therefore, we have
\be
\label{eq:U actio}
U_n(\alpha,\beta) (X^{\vec\alpha}\circ Z^{\vec\beta}) U_n^{-1}(\alpha,\beta)= I^{\otimes n-1}  \otimes Z~,
\ee
where 
\be
\label{eq: U def}
U_n(\alpha,\beta):= M_{\text{gcd}(\alpha_1,\dots,\alpha_n,\beta_1,\dots \beta_n)^{-1},n} \tilde U_n(\vec \alpha,\vec \beta)~.
\ee
Note that in \eqref{eq:U actio}, the $Z$ matrix acts on the last qudit. But this choice is arbitrary. We can choose the $U_n(\alpha,\beta)$ such that the $Z$ action of the resulting generalized Pauli group element is on any of the $n$ qudits. Moreover, if we have a set of qudits, we can
apply the Generalized PEG to any subset of qudits. 

Let us first apply the Generalized PEG to the case of two qudits of dimension $d$. It is useful to write the generators \eqref{eq:stabilizer gens} as an augmented matrix as follows
\be
\label{eq:stabilzer gen mat0}
\left(\begin{array}{cc|cc}  
  a_{11} & a_{12} & b_{11} & b_{12}\\  
  a_{21} & a_{22} & b_{21} & b_{22} 
\end{array}\right)~,
\ee
where $\vec g_i=(a_{i1},a_{i2})$, and $L \vec p_i=(b_{i1},b_{i2})$. Note that columns 1 and 3 specify the generalized Pauli matrices acting on the first qudit, while columns 2 and 4 specify the generalized Pauli matrices acting on the second qudit. We will show that there exists a combination of the unitaries $F_i$, $P_{\gamma,i}$, $M_{\gamma,i}$, and $D_{i,j}$ which converts the matrix \eqref{eq:stabilzer gen mat0} to the matrix
\be
\left(\begin{array}{cc|cc}  
  0 & 0 & i & j\\  
  0 & 0 & k & l 
\end{array}\right)~,
\ee
for some $i$, $j$, $k$, $l \in \mathds{\DZ}_d$. 

To show this statement, let us act with the unitary \eqref{eq: U def} on the two qudits. Then the matrix \eqref{eq:stabilzer gen mat} changes as follows
\be
\label{eq:stabilzer gen mat}
\left(\begin{array}{cc|cc}  
  a_{11} & a_{12} & b_{11} & b_{12}\\  
  a_{21} & a_{22} & b_{21} & b_{22} 
\end{array}\right) \xrightarrow{U_2(\vec g_1,L \vec p_1)} 
\left(\begin{array}{cc|cc}  
  0 & 0 & 0 & 1\\  
  a_{21}^{'} & a_{22}^{'} & b_{21}^{'} & b_{22}^{'} 
\end{array}\right)~,
\ee
for some integers $a_{21}^{'}$, $a_{22}^{'}$, $b_{21}^{'}$, and $b_{22}^{'}$. Since this new generator matrix also gives a stabilizer group, the generators determined by the first and second rows of this matrix must commute with each other. This logic implies that $a_{22}^{'}=0$. Also, we can subtract $b_{22}'$ copies of the first row from the second row. This does not change the stabilizer group. Therefore, we get the generator matrix 
\be
\left(\begin{array}{cc|cc}  
  0 & 0 & 0 & 1\\  
  a_{21}^{'} & 0 & b_{21}^{'} & 0 
\end{array}\right)~.
\ee
Now, note that $\text{gcd}(a_{21}^{'},b_{21}^{'})$ is co-prime to $d$ as the second row must give an order $d$ stabilizer element. Therefore, we can act with the unitary \eqref{eq: U def} on the first qudit to transform the generator matrix as
\be
\left(\begin{array}{cc|cc}  
  0 & 0 & 0 & 1\\  
  a_{21}^{'} & 0 & b_{21}^{'} & 0
\end{array}\right)\xrightarrow{U_1(a_{21}^{'}, b_{21}^{'})} 
\left(\begin{array}{cc|cc}  
  0 & 0 & 0 & 1\\  
  0 & 0 & 1 & 0 
\end{array}\right)= \CS_{(\DZ_1,[1])} ~,
\ee
In summary, we found a sequence of Clifford group operations that shows our map, $\mu$, assigns equivalent self-dual codes to the CFT, $\CT$, and the dual CFT, $\CT/(Q,[\sigma])$, corresponding to a self-dual invertible surface operator $S(Q,[\sigma])$.  

Next, let us consider a system of three qudits and show that a simple generalization of the argument above holds. To that end, consider the stabilizer group, $\CS_{(Q,[\sigma])}$, with generator matrix
\be
\label{eq:stabilzer gen mat 2}
\left(\begin{array}{ccc|ccc}  
  a_{11} & a_{12} & a_{13} & b_{11} & b_{12} & b_{13}\\  
  a_{21} & a_{22} & a_{23} & b_{21} & b_{22} & b_{23}\\
  a_{31} & a_{32} & a_{33} & b_{31} & b_{32} & b_{33}
\end{array}\right)~,
\ee
where $\vec g_i=(a_{i1},a_{i2},a_{i3})$ and $L \vec p_i=(b_{i1},b_{i2},b_{i3})$. Let us act with the unitary \eqref{eq: U def} on all three qudits to get
\be
\left(\begin{array}{ccc|ccc}  
  a_{11} & a_{12} & a_{13} & b_{11} & b_{12} & b_{13}\\  
  a_{21} & a_{22} & a_{23} & b_{21} & b_{22} & b_{23}\\
  a_{31} & a_{32} & a_{33} & b_{31} & b_{32} & b_{33}
\end{array}\right) \xrightarrow{U_{3}(\vec g_1,L \vec p_1)}
\left(\begin{array}{ccc|ccc}  
  0 & 0 & 0 & 0 & 0 & 1\\  
  a_{21}^{'} & a_{22}^{'} & a_{23}^{'} & b_{21}^{'} & b_{22}^{'} & b_{23}^{'}\\
  a_{31}^{'} & a_{32}^{'} & a_{33}^{'} & b_{31}^{'} & b_{32}^{'} & b_{33}^{'}
\end{array}\right)~.
\ee
We know that the three generators of the stabilizer group determined by the three rows of the matrix above must commute with each other. This logic implies that $a_{23}'=a_{33}'=0$. Moreover, by subtracting multiples of the first row from the second and third rows, we can set $b_{23}^{'}=b_{33}^{'}=0$ and obtain
\be
\left(\begin{array}{ccc|ccc}  
  0 & 0 & 0 & 0 & 0 & 1\\  
  a_{21}^{'} & a_{22}^{'} & 0 & b_{21}^{'} & b_{22}^{'} & 0\\
  a_{31}^{'} & a_{32}^{'} & 0 & b_{31}^{'} & b_{32}^{'} & 0
\end{array}\right)~.
\ee
Now, note that $\text{gcd}(a_{21}^{'},a_{22}^{'},b_{21}^{'},b_{22}^{'})$ is co-prime to $d$ as the second row must give an order $d$ stabilizer element. Therefore, we can act with the unitary \eqref{eq: U def} on the first two qudits to transform the generator matrix as
\be
\left(\begin{array}{ccc|ccc}  
  0 & 0 & 0 & 0 & 0 & 1\\  
  a_{21}^{'} & a_{22}^{'} & 0 & b_{21}^{'} & b_{22}^{'} & 0\\
  a_{31}^{'} & a_{32}^{'} & 0 & b_{31}^{'} & b_{32}^{'} & 0
\end{array}\right) \xrightarrow{U_{2}(a_{21}',a_{22}',b_{21}',b_{22}')}
\left(\begin{array}{ccc|ccc}  
  0 & 0 & 0 & 0 & 0 & 1\\  
  0 & 0 & 0 & 0 & 1 & 0\\
  a_{31}^{''} & a_{32}^{''} & 0 & b_{31}^{''} & b_{32}^{''} & 0
\end{array}\right)~,
\ee
for some integers $a_{31}^{''},a_{32}^{''},b_{31}^{''},b_{32}^{''},b_{33}^{''}$. Once again, using the constraint that the three stabilizers must commute, we find that $a_{32}^{''}=0$. Also, by subtracting multiples of the second row from the third row, we can set $b_{32}^{'}=0$ to get
\be
\left(\begin{array}{ccc|ccc}  
  0 & 0 & 0 & 0 & 0 & 1\\  
  0 & 0 & 0 & 0 & 1 & 0\\
  a_{31}^{''} & 0 & 0 & b_{31}^{''} & 0 & 0
\end{array}\right)~.
\ee
Finally, acting with the unitary \eqref{eq: U def} on the first qudit, we obtain
\be
\left(\begin{array}{ccc|ccc}  
  0 & 0 & 0 & 0 & 0 & 1\\  
  0 & 0 & 0 & 0 & 1 & 0\\
  0 & 0 & 0 & 1 & 0 & 0
\end{array}\right)=\CS_{(\DZ_1,[1])}~.
\ee

The argument above can be generalized to a system of $n$ qudits of the same dimension $d$. In this case, we will have an $n \times 2n$ generator matrix which can be turned into a matrix with zeroes on the left $n$ columns through a sequence of unitaries using the Generalized PEG lemma. In fact, this statement generalizes if we have a system of $n_1$ qudits of dimension $d_1$, $n_2$ qudits of dimension $d_2$ etc., such that the dimensions $d_1,d_2,\dots$ are co-prime to each other. This statement holds because an invertible surface operator of the bulk CS theory in this case will not mix Wilson line operators corresponding to qudits of different dimensions. Also, the Clifford group of this system does not mix Pauli matrices acting on qudits of different dimensions. Therefore, the above argument can be applied to each subset of qudits of the same dimension separately.

Let us summarize what we have seen above:
\begin{itemize}
\item Consider a charge conjugation abelian RCFT, $\CT$, having $\prod_i\mathbb{Z}_{d_i}^{n_i}$ fusion rules with each $d_i$, $n_i\in\mathbb{N}$ any (composite) natural numbers such that the $d_i$ are coprime. The corresponding quantum code is equivalent to the quantum code assigned by $\mu$ to any orbifold theory, $\CT/(Q,[\sigma])$, described by an invertible bulk surface operator, $\CS(Q,[\sigma])$. This is the code version of the QFT equivalences we have described above. In particular, we have code equivalence in this case when
\begin{equation}\label{codeEq}
\CS(Q_1,[\sigma_1])=\CS(Q_2,[\sigma_2])\times\CS(Q_3,[\sigma_3])~,
\end{equation}
and all surfaces are invertible.
\item For non-invertible surface operators, the statement in the previous bullet does not hold in general (e.g., see the example discussed in Fig. \ref{fig:orbgraph2}). However, we conjecture that two theories, $\CT(Q_1,[\sigma_1])$ and $\CT(Q_2,[\sigma_2])$, described by bulk surfaces, $\CS(Q_i,[\sigma_i])$, satisfying \eqref{codeEq} with general $\CS(Q_{1,2},[\sigma_{1,2}])$ and invertible $\CS(Q_3,[\sigma_3])$ are assigned equivalent codes by $\mu$ (at least if the fusion rules involve groups of the form $\prod_i\mathbb{Z}_{d_i}^{n_i}$ for co-prime $d_i$).\footnote{It can be explicitly checked that this conjecture is true for CFTs corresponding to a bulk CS theory with a cyclic fusion group in \eqref{MTCclass}. In this case, we get a single qudit quantum code, and codes corresponding to non-invertible surfaces related by fusion with an invertible surface are equivalent.}
\item Clearly, it would be interesting to understand if we can relax the co-primeness condition on the qudit dimensions in the previous two bullets.
\end{itemize}

In fact, the idea of code equivalences arising from invertible surfaces in the relevant CS theories is more general. Indeed, it can happen that the bulk theory itself has a dual description. As an example, consider the equivalence of the following 3d TQFTs: $D(\mathbb{Z}_2)\times SU(2)_1$ and $SU(2)_1\times SU(2)_{-1}\times SU(2)_1$, where $D(\mathbb{Z}_2)$ is the untwisted $\mathbb{Z}_2$ discrete gauge theory. In the language of $K$ matrices,
\begin{equation}
\CL={1\over4\pi}\vec a^tKd\vec a~,
\end{equation}
where $\vec a$ is a vector of gauge fields, we have
\begin{equation}
K_1=\left(\begin{array}{ccc}  
  0 & 2 & 0 \\  
  2 & 0 & 0 \\
  0 & 0 & 2 
\end{array}\right) \ \leftrightarrow\ \left(\begin{array}{ccc}  
  2 & 0 & 0 \\  
  0 & -2 & 0 \\
  0 & 0 & 2 
\end{array}\right)= K_2~.
\end{equation}
It is straightforward to check that these theories, although dual, give different codes. In particular, even if we choose the same physical surface operator in these different descriptions, we obtain distinct codes. However, there is an invertible map between these theories (which we can imagine being implemented by an invertible surface), and the codes generated by the dual descriptions are, in fact, equivalent.

\subsection{1+1d TQFTs and stabilizer codes}

\begin{figure}[h!]
    \centering

\tikzset{every picture/.style={line width=0.75pt}} %set default line width to 0.75pt        

\begin{tikzpicture}[x=0.75pt,y=0.75pt,yscale=-1,xscale=1]
%uncomment if require: \path (0,330); %set diagram left start at 0, and has height of 330

%Shape: Parallelogram [id:dp9581542666770164] 
\draw  [color={rgb, 255:red, 0; green, 0; blue, 0 }  ,draw opacity=1 ][fill={rgb, 255:red, 74; green, 74; blue, 74 }  ,fill opacity=0.38 ] (171.4,252) -- (171.42,138.37) -- (282,81.61) -- (281.98,195.24) -- cycle ;
%Shape: Parallelogram [id:dp3842962720757085] 
\draw  [color={rgb, 255:red, 0; green, 0; blue, 0 }  ,draw opacity=1 ][fill={rgb, 255:red, 74; green, 74; blue, 74 }  ,fill opacity=0.38 ] (39.99,252) -- (40.01,138.37) -- (150.59,81.61) -- (150.57,195.24) -- cycle ;
%Straight Lines [id:da9848515784535609] 
\draw  [color={rgb, 255:red, 139; green, 6; blue, 24 }  ,draw opacity=1 ]  (105.8,164.44) -- (172.32,164.59) ;
%Shape: Ellipse [id:dp3377344092080524] 
\draw  [color={rgb, 255:red, 139; green, 6; blue, 24 }  ,draw opacity=1 ][fill={rgb, 255:red, 139; green, 6; blue, 24 }  ,fill opacity=1 ] (107.65,165.66) .. controls (106.77,165.66) and (106.05,164.97) .. (106.05,164.13) .. controls (106.05,163.28) and (106.77,162.6) .. (107.65,162.6) .. controls (108.53,162.6) and (109.25,163.28) .. (109.25,164.13) .. controls (109.25,164.97) and (108.53,165.66) .. (107.65,165.66) -- cycle ;
%Shape: Ellipse [id:dp4513780538937081] 
\draw  [color={rgb, 255:red, 139; green, 6; blue, 24 }  ,draw opacity=1 ][fill={rgb, 255:red, 139; green, 6; blue, 24 }  ,fill opacity=1 ] (218.12,166.12) .. controls (217.24,166.12) and (216.52,165.44) .. (216.52,164.59) .. controls (216.52,163.74) and (217.24,163.06) .. (218.12,163.06) .. controls (219.01,163.06) and (219.72,163.74) .. (219.72,164.59) .. controls (219.72,165.44) and (219.01,166.12) .. (218.12,166.12) -- cycle ;
%Straight Lines [id:da8108754008873513] 
\draw [color={rgb, 255:red, 139; green, 6; blue, 24 }  ,draw opacity=1 ] [dash pattern={on 4.5pt off 4.5pt}]  (172.32,164.59) -- (216.52,164.59) ;
%Shape: Parallelogram [id:dp04416957728004234] 
\draw  [color={rgb, 255:red, 0; green, 0; blue, 0 }  ,draw opacity=1 ][fill={rgb, 255:red, 74; green, 74; blue, 74 }  ,fill opacity=0.38 ] (509.4,250) -- (509.42,136.37) -- (620,79.61) -- (619.98,193.24) -- cycle ;
%Shape: Parallelogram [id:dp5701359309724269] 
\draw  [color={rgb, 255:red, 0; green, 0; blue, 0 }  ,draw opacity=1 ][fill={rgb, 255:red, 74; green, 74; blue, 74 }  ,fill opacity=0.38 ] (377.99,250) -- (378.01,136.37) -- (488.59,79.61) -- (488.57,193.24) -- cycle ;
%Straight Lines [id:da25369802000404407] 
\draw  [color={rgb, 255:red, 139; green, 6; blue, 24 }  ,draw opacity=1 ]  (443.8,162.44) -- (510.32,162.59) ;
%Shape: Ellipse [id:dp3515039623469257] 
\draw  [color={rgb, 255:red, 139; green, 6; blue, 24 }  ,draw opacity=1 ][fill={rgb, 255:red, 139; green, 6; blue, 24 }  ,fill opacity=1 ] (445.65,163.66) .. controls (444.77,163.66) and (444.05,162.97) .. (444.05,162.13) .. controls (444.05,161.28) and (444.77,160.6) .. (445.65,160.6) .. controls (446.53,160.6) and (447.25,161.28) .. (447.25,162.13) .. controls (447.25,162.97) and (446.53,163.66) .. (445.65,163.66) -- cycle ;
%Shape: Ellipse [id:dp36983034359287115] 
\draw  [color={rgb, 255:red, 139; green, 6; blue, 24 }  ,draw opacity=1 ][fill={rgb, 255:red, 139; green, 6; blue, 24 }  ,fill opacity=1 ] (556.12,164.12) .. controls (555.24,164.12) and (554.52,163.44) .. (554.52,162.59) .. controls (554.52,161.74) and (555.24,161.06) .. (556.12,161.06) .. controls (557.01,161.06) and (557.72,161.74) .. (557.72,162.59) .. controls (557.72,163.44) and (557.01,164.12) .. (556.12,164.12) -- cycle ;
%Straight Lines [id:da255811672943542] 
\draw [color={rgb, 255:red, 139; green, 6; blue, 24 }  ,draw opacity=1 ] [dash pattern={on 4.5pt off 4.5pt}]  (510.32,162.59) -- (554.52,162.59) ;
%Straight Lines [id:da30071958211134053] 
\draw [color={rgb, 255:red, 139; green, 6; blue, 24 }  ,draw opacity=1 ]   (446,102) -- (445.79,137) -- (445.65,160.6) ;

% Text Node
\draw (48.6,190.21) node [anchor=north west][inner sep=0.75pt]  [font=\small]  {$\CB( Q_{1} ,[ \sigma _{1}])$};
% Text Node
\draw (178.81,192.62) node [anchor=north west][inner sep=0.75pt]  [font=\small]  {$\CB( Q_{2} ,[ \sigma _{2}])$};
% Text Node
\draw (386.6,188.21) node [anchor=north west][inner sep=0.75pt]  [font=\small]  {$\CB( Q_{1} ,[ \sigma _{1}])$};
% Text Node
\draw (516.81,190.62) node [anchor=north west][inner sep=0.75pt]  [font=\small]  {$\CB( Q_{2} ,[ \sigma _{2}])$};

\end{tikzpicture}
    \caption{Left: Line operators that can end on both boundaries become local operators in the 2d TQFT upon shrinking the slab. Right: Line operators that can only end on one boundary become non-genuine local operators upon shrinking the slab.}
    \label{fig:two gapped boundaries}
\end{figure}
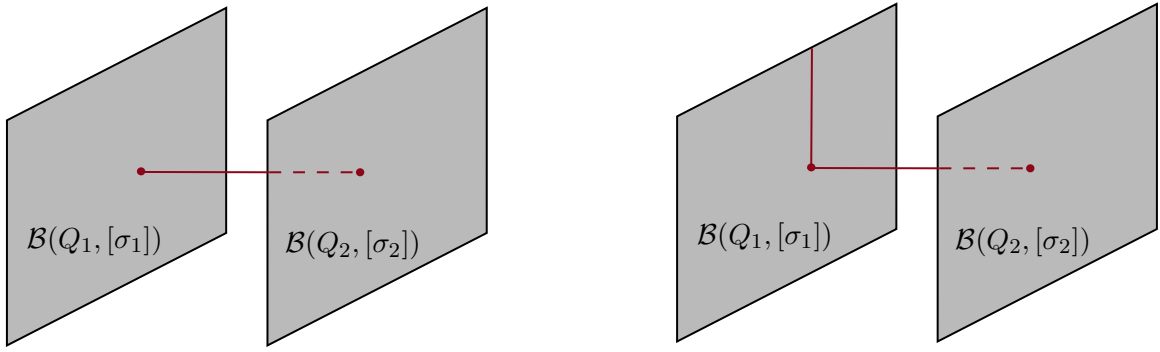

In Fig. \ref{fig:codes and lagrangian subgroups}, we understood how gapped boundary conditions describing the partition function of the CFT on $\Sigma$ are related to self-dual stabilizer codes. This discussion also applies if the theory on $\Sigma$ is gapped. Indeed, consider the situation where we have a 3d abelian CS theory of the form $\CI \times \bar \CI$ with two gapped boundaries $\CB(Q_1,[\sigma_1])$ and $\CB(Q_2,[\sigma_2])$ forming a slab. If the gapped boundary $\CB(Q_1,[\sigma_1])$ realizes a symmetry $Y$ on the boundary, then shrinking this slab results in a 2d TQFT with $Y$ symmetry. The line operators of the bulk TQFT which can end on both gapped boundaries become local operators of the 2d TQFT (see Fig. \ref{fig:two gapped boundaries}) \cite{Zhang:2023wlu,Sun:2023xxv,Bhardwaj:2023idu,Bhardwaj:2023fca} (see also related results in 4d \cite{Antinucci:2023ezl,Cordova:2023bja}). If a line operator can end on the boundary $\CB(Q_2,[\sigma_2])$ but not on $\CB(Q_1,[\sigma_1])$, then shrinking the slab gives us twisted sector operators of the 2d TQFT (see Fig. \ref{fig:two gapped boundaries}). 

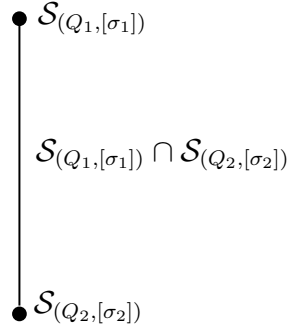
\begin{figure}
    \centering

\tikzset{every picture/.style={line width=0.75pt}} %set default line width to 0.75pt        

\begin{tikzpicture}[x=0.75pt,y=0.75pt,yscale=-1,xscale=1]

%Shape: Circle [id:dp48759178582002205] 
\draw  [fill={rgb, 255:red, 0; green, 0; blue, 0 }  ,fill opacity=1 ] (303,77.21) .. controls (303,75.44) and (304.44,74) .. (306.21,74) .. controls (307.98,74) and (309.42,75.44) .. (309.42,77.21) .. controls (309.42,78.98) and (307.98,80.42) .. (306.21,80.42) .. controls (304.44,80.42) and (303,78.98) .. (303,77.21) -- cycle ;
%Shape: Circle [id:dp6351142949072132] 
\draw  [fill={rgb, 255:red, 0; green, 0; blue, 0 }  ,fill opacity=1 ] (303,226.21) .. controls (303,224.44) and (304.44,223) .. (306.21,223) .. controls (307.98,223) and (309.42,224.44) .. (309.42,226.21) .. controls (309.42,227.98) and (307.98,229.42) .. (306.21,229.42) .. controls (304.44,229.42) and (303,227.98) .. (303,226.21) -- cycle ;
%Straight Lines [id:da38752251405447835] 
\draw    (306.21,80.42) -- (306.21,222) ;

% Text Node
\draw (314,67.4) node [anchor=north west][inner sep=0.75pt]    {$\CS_{( Q_1,[\sigma_1] )}$};
% Text Node
\draw (312,213.4) node [anchor=north west][inner sep=0.75pt]    {$\CS_{( Q_2,[\sigma_2] )}$};
% Text Node
\draw (313,136.4) node [anchor=north west][inner sep=0.75pt]    {$\CS_{( Q_1,[\sigma_1] )}\cap \CS_{( Q_2,[\sigma_2] )}$};

\end{tikzpicture}
    \caption{The label of the edge of the vertices connecting the stabilizer codes $\CS_{( Q_1,[\sigma_1] )}$ and $\CS_{( Q_1,[\sigma_1] )}$ gives us the group formed by the local operators of the 2D TQFT under fusion.}
    \label{fig:Code Graph gapped boundaries}
\end{figure}

Let us assume that these gapped boundaries correspond to self-dual surface operators $S(Q_1,[\sigma_1])$ and $S(Q_2,[\sigma_2])$ of $\CI$. Then, we have corresponding stabilizer codes $\CS_{(Q_1,[\sigma_1])}$ and $\CS_{(Q_2,[\sigma_2])}$ that form two vertices of the code graph (see Fig. \ref{fig:Code Graph gapped boundaries}). The group labelling the edge between the two vertices is the group formed by the local operators of the 2d TQFT under fusion and corresponds to stabilizer elements in the intersection, $\CS_{(Q_1,[\sigma_1])}\cap\CS_{(Q_2,\sigma_2)}$. In particular, the size of the group is the number of topological local operators. Moreover, the line operators that can end on the boundary $\CB(Q_2,[\sigma_2])$ but not on $\CB(Q_1,[\sigma_1])$ are in one-to-one correspondence with stabilizer elements in $\CS_{(Q_2,[\sigma_2])}$ which are not in $\CS_{(Q_2,\sigma_2)}$. 

Finally, note that, if the 0-form symmetry group, $Y$, realized on the gapped boundary, $\CB(Q_1,[\sigma_1])$, has an anomaly, then we know that a 2d TQFT with $Y$ symmetry cannot be trivially gapped. In other words, it must necessarily have non-trivial local operators. This is reflected in the code graph in the following way. The vertex corresponding to the gapped boundary, $\CB(Q_1,[\sigma_1])$, is necessarily connected to all other vertices. If this was not the case, then there will be another gapped boundary, say $\CB(Q_2,[\sigma_2])$, such that there are no non-trivial line operators that can end on both $\CB(Q_1,[\sigma_1])$ and $\CB(Q_2,[\sigma_2])$. Then, shrinking a slab with these two gapped boundaries will result in a trivially gapped theory with $Y$ symmetry, which contradicts the assumption that $K$ is anomalous. This discussion also implies that if there is a vertex of the code graph that is connected to all other vertices, then the symmetry realized by the gapped boundary corresponding to that vertex is necessarily anomalous.

\section{Conclusion}

In this paper, we studied various aspects of a map from abelian RCFTs and associated CS theories to quantum stabilizer codes. We considered several natural conditions that such a map should satisfy. For a fixed chiral algebra, we defined the orbifold graph, $\Gamma_{\DV}$, of RCFTs and showed that a map from abelian RCFTs to quatum codes satisfying some natural constraints is a graph homomorphism from $\Gamma_{\DV}$ to the code graph. We then showed that this map cannot be an embedding of the full orbifold graph into the code graph for all chiral algebras. 

However, motivated by the relation between RCFTs through discrete orbifolds, we defined an explicit map from abelian RCFTs to qudit stabilizer codes which preserves a universal subgraph of the orbifold graph and embeds it in the code graph. Using self-duality of the resulting quantum codes, we showed that this subgraph contains RCFTs whose partition functions are determined by self-dual surface operators in the bulk Chern-Simons theory. We also showed that the full abelianized generalized Pauli group acting on the qudits can be obtained from the twisted sector operators of symmetries of the RCFT.

Finally, we related self-dual quantum codes obtained from RCFTs to gapped boundaries of the corresponding bulk CS theories. This discussion interfaced with SymTFTs and 0-form gauging of invertible symmetries in the bulk. Motivated by this gauging, we explained certain code equivalences as arising in theories defined by surface operators that are related by fusion with invertible surfaces as in \eqref{codeEq}. We also related certain gapped interfaces to non-self-dual codes. 

The analysis in this paper leads to several natural questions:

\begin{itemize}
    \item In Appendix \ref{ap:non-self-dual}, we showed that the full Orbifold Graph can be mapped to stabilizer codes if we forego self-duality of the codes. It would be interesting to understand if there is a canonical construction of non-self-dual codes from abelian RCFTs.
    \item It would also be interesting to understand if our code / CFT map can be related to other code / CFT maps by certain natural transition functions. Perhaps quantum codes can then be understood as belonging to some \lq\lq code manifold" corresponding to a larger subspace of CFTs (where our present construction only captures a particular patch of this space). 
    \item It is interesting to extend the map constructed in this paper to non-abelian RCFTs. A related question is to understand how non-invertible symmetries of RCFTs are captured by the quantum codes. We have seen that some of these symmetries correspond to code equivalences and CFT dualities, but it would be nice to understand a more general picture (and perhaps also to understand the role played by non-invertible surfaces in the bulk).
    \item Closely related to the previous bullet point, it would be interesting to understand whether the discussion around \eqref{codeEq} can be extended to an equivalence relation among codes corresponding to general (self-dual) surfaces that are related by the fusion of invertible surfaces.
        \item We showed that the Orbifold Graph captures several properties of 1+1 TQFTs defined using Lagrangian subgroups of a SymTFT. It will be interesting to explore this direction further. 
    \item In this paper, we focused on quantum stabilizer codes. In particular, the fact that the stabilizer group is abelian puts strong constraints on the CFTs which admit a code description. It will be interesting to understand maps from CFTs to more general codes.\footnote{Indeed, if we focus on general additive codes whose inner product is determined by the modular S-matrix of an abelian CS theory, then all gapped boundaries of such a TQFT correspond to self-dual additive codes \cite{Barbar:2023ncl}.} A natural starting point is to look at subsystem codes in which the check operators are not required to commute \cite{eczoo_oecc}. It will be interesting to explore relations to the lattice stabilizer and subsystem codes obtained from arbitrary abelian CS theories in \cite{Ellison:2022web}. Another interesting study would be to include Floquet codes which have received a lot of attention in the quantum information theory community \cite{eczoo_floquet,Hastings:2021ptn,Davydova:2023mnz}. 
\end{itemize}

\ack{We thank M.~Balasubramanian, A.~Banerjee, Z.~Duan, A.~Dymarsky, I.~Runkel, and A.~Sharon for discussions and / or collaboration on related work. M.~B. thanks the Simons Center for Geometry and the Institute for Advanced Study for hospitality during extended visits in 2022, where this work was undertaken. Some of the results in this paper were first announced at King's College London during the workshop \href{https://nms.kcl.ac.uk/gerard.watts/DefSym22/}{Defects and Symmetry 2022}. M. B. was partially supported by the grant “Relations, Transformations, and Emergence in Quantum Field Theory” from the Royal Society and the grant “Amplitudes, Strings and Duality” from STFC. No new data were generated or analyzed in this study.}

\newpage

\begin{appendices}

\section{A proof there is no universal $\Gamma_{\mathds{V}}\hookrightarrow\Gamma_{\CP}$ embedding}

\label{Ap:1}

In this appendix, we show there is no universal embedding of the orbifold graph, $\Gamma_{\mathds{V}}$, in the code graph, $\Gamma_{\CP}$. In other words, there are MTCs and corresponding CFTs where such an embedding is impossible. We do this by finding a particular set of counterexamples to such an embedding in the case of 3d $\mathbb{Z}_2^r$ discrete gauge theories with $r>1$. These are $E_{2^r}$ MTCs in the nomenclature of \cite{wang2020and} with fusion rules given by the abelian group $\DZ_{2^r} \times \DZ_{2^r}$. The $F$ matrices are all trivial. The $R$ matrix, the topological spins and the $S$ matrix are given by
\bea
R(\vec g, \vec h)&=& e^{\frac{2\pi i}{2^r} g_1 h_2 } ~,\nonumber\\
\theta(\vec g)&=&R(\vec g,\vec g)= e^{\frac{2\pi i}{2^r} g_1 g_2 }~, \nonumber\\
S_{\vec g,\vec h}&=& R(\vec g, \vec h)R(\vec h, \vec g) = e^{\frac{2\pi i}{2^r} (g_1 h_2+g_2 h_1) }~.
\eea
Given this data, let us consider the constraint 
\be
\label{orbiconst}
S_{\vec h,\vec  p} ~ \Xi(\vec h,\vec g)=1 ~, \ \forall \vec h \in Q~.
\ee
When the discrete torsion is trivial, we can write this constraint explicitly in terms of the MTC data above to get
\be
\label{constsimp}
e^{\frac{2\pi i}{2^r} (h_1 p_2+h_2 p_1)}  e^{\frac{2\pi i}{2^r} g_1 h_2 } = e^{\frac{2\pi i}{2^r} (h_1 p_2+h_2 p_1 + g_1 h_2)}=1  ~ \forall ~ \vec h \in Q~.
\ee
The operator $\CO_{\vec p, \overline{\vec g + \vec p}}$ is a primary in $\CT/(Q,[1])$ precisely when the vectors $\vec g$ and $\vec p$ satisfy the above constraint. 

Let us consider specific orbifolds and look at the constraints coming from the CFT to code map. Since the fusion of chiral primaries form the group $\DZ_{2^r} \times \DZ_{2^r}$, we choose a 2-qudit system where each qudit has a Hilbert space of dimension $2^r$. We want to map distinct CFTs with the $E_{2^r}$ MTC to distinct stabilizer codes in the Pauli group acting on the 2-qudit Hilbert space. 

Consider the charge-conjugation CFT whose primaries from a $\DZ_{2^r} \times \DZ_{2^r}$ group under fusion. In particular, we have the primaries $\CO_{(0,1),\overline{(0,1)}}$ and  $\CO_{(1,0),\overline{(1,0)}}$. Note that these primaries are both order $2^r$ under fusion. The CFT to code map, $\mu$, assigns 
\be
\mu(\CO_{(0,1),\overline{(0,1)}})= \vec v_1, ~ \mu(\CO_{(1,0),\overline{(1,0)}})= \vec v_2~,
\ee
for some vectors, $\vec v_1$ and $\vec v_2$, in the vector space, $\DZ_{2^r}^4$. We require $\mu(\CT)$ to be a Lagrangian subspace. Therefore, we get the constraint
\be
\label{c1}
\omega(\vec v_1,\vec v_2)=0~.
\ee
We also require $\mu(\CT) \cong \DZ_{2^r} \times \DZ_{2^r}$ as groups. Therefore, $\vec v_1$ and $\vec v_2$ are order $2^r$ elements. 

Now, let us consider the subgroup $Q=\langle (0,1) \rangle $. It is easy to verify that the CFT $\CT/(Q,[1])$ has primaries $\CO_{(0,1),\overline{(0,1)}}$ and $\CO_{(0,0),\overline{(0,1)}}$. The map $\mu$ assigns 
\be
\mu(\CO_{(0,1),\overline{(0,1)}})= \vec v_1~, ~ \mu(\CO_{(0,0),\overline{(0,1)}})= \vec v_3~,
\ee
for some vector, $\vec v_3$, in the vector space $\DZ_{2^r}^4$. We require $\mu(\CT/(Q,[1]))$ to be a Lagrangian subspace and so
\be
\label{c2}
\omega(\vec v_1,\vec v_3)=0~.
\ee

Similarly, consider the subgroup  $Q=\langle (1,0) \rangle $. it is easy to verify that the CFT $\CT/(Q,[1])$ has primaries $\CO_{(1,0),\overline{(1,0)}}$ and $\CO_{(0,0),\overline{(1,0)}}$. The map $\mu$ assigns 
\be
\mu(\CO_{(1,0),\overline{(1,0)}})= \vec v_2~, ~ \mu(\CO_{(0,0),\overline{(1,0)}})= \vec v_4~,
\ee
for some vector $\vec v_4$ in the vector space $(\DZ_{2^r})^4$. We require $\mu(\CT/(Q,[1]))$ to be a Lagrangian subspace. Therefore, we get the constraint
\be
\label{c3}
\omega(\vec v_2,\vec v_4)=0~.
\ee

Let us now specialize to the case $r\geq 2$ and consider the subgroup $Q=\langle (0,2^{r-1}), (2^{r-1},0) \rangle $. It is easy to verify that the CFT $\CT/(Q,[1])$ has primaries $\CO_{(0,2),\overline{(0,2)}}$, $\CO_{(2,0),\overline{(2,0)}}$, $\CO_{(0,0),\overline{(0,2^{r-1})}}$, and $ \CO_{(0,0),\overline{(2^{r-1},0)}}$. The map $\mu$ assigns 
\bea
\label{mainorb}
\mu(\CO_{(0,2),\overline{(0,2)}})=2 \vec v_1~, ~  \mu(\CO_{(2,0),\overline{(2,0)}})=2 \vec v_2 ~,\cr
 \mu(\CO_{(0,0),\overline{(0,2^{r-1})}})=2^{r-1}\vec v_3~, ~  \mu(\CO_{(0,0),\overline{(2^{r-1},0)}})=2^{r-1} v_4~.
\eea
We require $\mu(\CT/(Q,[1]))$ to be a Lagrangian subspace, and so
\be
\omega(i,j)=0~,
\ee
for all $i,j \in \{ 2\vec v_1,2 \vec v_2,2^{r-1}\vec v_3,2^{r-1}\vec v_4 \}$.

Now, let us consider $Q=\langle (1,2^{r}-1) \rangle $. It is easy to verify that the CFT $\CT/(Q,[1])$ has primaries $\CO_{(1,1),\overline{(1,1)}}$ and $\CO_{(0,1),\overline{(1,0)}}$. The map $\mu$ assigns 
\be
\mu(\CO_{(1,1),\overline{(1,1)}})= \vec v_1 + \vec v_2~, ~ \mu(\CO_{(0,1),\overline{(1,0)}})= \vec v_1 - \vec v_3 + \vec v_4~.
\ee
We require $\mu(\CT/(Q,[1]))$ to be a Lagrangian subspace. Therefore, we get the constraint
\be
\label{c4}
\omega(\vec v_1 + \vec v_2, \vec v_1 - \vec v_3 + \vec v_4)=0~.
\ee

Finally, let us consider the subgroup $Q=\langle (0,1), (1,0)\rangle$. It is easy to verify that the CFT $\CT/(Q,[1])$ has primaries $\CO_{(0,0),\overline{(0,1)}}$ and $\CO_{(1,0),\overline{(0,0)}}$. The map $\mu$ assigns 
\be
\mu(\CO_{(0,0),\overline{(0,1)}})= \vec v_3~, ~ \mu(\CO_{(1,0),\overline{(0,0)}})= \vec v_2 -\vec v_4~.
\ee
We require $\mu(\CT/(Q,[1]))$ to be a Lagrangian subspace. Therefore, we get the constraint
\be
\label{c5}
\omega(\vec v_3, \vec v_2 - \vec v_4)=0~.
\ee

Now let us consider the CFT $\CT/(Q,[\sigma])$, where $\sigma$ is the 2-cocycle given by
\be
\sigma(\vec g,\vec h)= e^{\frac{2 \pi i}{2^r} g_1 h_2}~. 
\ee
The constraint \eqref{orbiconst} can be explicitly written as
\be
\label{constsimp2}
e^{\frac{2\pi i}{2^r} (h_1 p_2+h_2 p_1)}  e^{\frac{2\pi i}{2^r} g_1 h_2 } e^{\frac{2 \pi i}{2^r} g_1 h_2} e^{-\frac{2 \pi i}{2^r} h_1 g_2} = e^{\frac{2\pi i}{2^r} (h_1 p_2+h_2 p_1 + 2 g_1 h_2-g_2h_1)}=1~,\   \forall \vec h \in Q~.
\ee
We can verify that the CFT $\CT/(Q,[\sigma])$  has primaries $\CO_{(0,1),\overline{(0,2)}}$ and $\CO_{(2^r-2,0),\overline{(2^r-1,0)}}$. The map $\mu$ then assigns 
\be
\mu(\CO_{(0,1),\overline{(0,2)}})= \vec v_1 + \vec v_3, ~ \mu(\CO_{(2^r-2,0),\overline{(2^r-1,0)}})= -2 \vec  v_2 + \vec v_4~.
\ee
We require $\mu(\CT/(Q,[1]))$ to be a Lagrangian subspace, and so
\be
\label{c6}
\omega( \vec v_1 + \vec v_3,-2\vec  v_2 + \vec v_4)=0~.
\ee

Let us summarize all the constraints we have so far
\bea
&& \omega(\vec v_1,\vec v_2)=0~,\nonumber\\
&& \omega(\vec v_1,\vec v_3)=0~, \nonumber\\
&& \omega(\vec v_2,\vec v_4)=0~, \nonumber\\
&& \omega(i,j)=0 ~ \forall i,j ~ \in \{ 2\vec v_1,2 \vec v_2,2^{r-1}\vec v_3,2^{r-1}\vec v_4 \}~, \nonumber\\
&& \omega(\vec v_1 + \vec v_2, \vec v_1 -  v_3 + v_4)=0 
\implies  \omega(\vec v_1,\vec v_4) - \omega(\vec v_2, \vec v_3)=0~,   \nonumber\\
&& \omega(\vec v_3, \vec v_2 -\vec v_4)=0 \implies \omega(\vec v_3, \vec v_2) - \omega(\vec v_3, \vec v_4)=0~, \nonumber\\
&&\omega( \vec v_1 + \vec v_3, -2\vec  v_2 + \vec v_4)=0 \implies  \omega(\vec v_1, \vec v_4) -2 \omega(\vec v_3,\vec v_2) + \omega(\vec v_3, \vec v_4)=0~, 
\eea
where we have used the first three constraints to simplify the last three. Noting that $\omega$ is anti-symmetric in its arguments and adding the last three constraints together, we get
\be
2 \omega(\vec v_1, \vec v_4)=0~.
\ee
This result, along with the first two constraints above, implies
\be
\omega(\vec v_1, 2\vec v_2)=0, ~ \omega(\vec v_1, 2 \vec v_3)=0, ~ \omega(\vec v_1, 2 \vec v_4)=0~.
\ee
Therefore, from the discussion around \eqref{mainorb}, we find that the vector $\vec v_1$ commutes with all generators of the stabilizer code associated with the CFT $\CT/(Q,[1])$ for $Q=\langle (0,2^{r-1}),(2^{r-1},0)\rangle$. Since this stabilizer code is self-dual, $\vec v_1$ should in fact be an element of this code. This statement is a contradiction, because this stabilizer code is isomorphic to $\DZ_{2^{r-1}}\times \DZ_{2^{r-1}}\times \DZ_2 \times \DZ_2$ as a group while $\vec v_1$ has order $2^r$. 

This argument shows that the full orbifold graph of CFTs with $E_{2^r}$ MTC and $r\geq 2$ cannot be embedded in the code graph. Therefore, there is no universal $\Gamma_{\mathds{V}}\hookrightarrow\Gamma_{\CP}$ embedding.

\section{Comparing with the CFT-Qubit code map in \cite{buican2021quantum}}

\label{ap:comparing with previous work}

Let us compare the CFT to qudit codes map defined in the main text to the map in \cite{buican2021quantum}, where we map CFTs to qubit codes. Recall that in \cite{buican2021quantum}, the general map from a CFT to a qubit code is given by
\be 
\label{decmap}
\CO_{\vec p, \overline{\vec p + \vec g}} \leftrightarrow X^{M \vec g} \circ Z^{A \vec p}~,
\ee
where $X$ and $Z$ are Pauli matrices acting on a qubit Hilbert space. Note that the matrix $M$ is diagonal, with diagonal entries of the form $M_{ii}=2^{r_i-1}$ is the order of the $i^{th}$ cyclic factor in the group $K$. 

The map in \eqref{decmap} differs from the map \eqref{genqudit}, because we don't have the matrix $M$ in \eqref{genqudit}. This difference arises because in \cite{buican2021quantum} we were focusing on $Q$ with order-two elements. Therefore, $\vec g \in Q$ is an order-two element. For such an element to contribute non-trivially to a qubit code we need to renormalize $\vec g$ so that its components are valued in $\pm 1$. That is precisely what $M \vec g$ in \eqref{decmap} achieves. On the other hand, we do not need such a factor in \eqref{genqudit} since we are mapping the CFT to qudit codes.

Also, note that the matrix, $A$, multiplying $\vec p$ in \eqref{decmap} differs from the matrix, $L$, multiplfying $\vec p$ in \eqref{genqudit}. In particular, we have 
\be
L_{E_{2^s}}:=\begin{pmatrix}
0 & 1\\
1 & 0
\end{pmatrix}~,
\ee
\be
L_{F_{2^t}}:=\begin{pmatrix}
2 & 1\\
1 & 2
\end{pmatrix}~,
\ee
while 
\be
A_{E_{2^s}}:=\begin{pmatrix}
0 & 1\\
1 & 0
\end{pmatrix}~,
\ee
\be
A_{F_{2^t}}:=\begin{pmatrix}
0 & 1\\
1 & 0
\end{pmatrix}~.
\ee
This difference arises because, in the map to qubit codes, the $X$ and $Z$ Pauli matrices are order two. Therefore, the factors of $2$ in $L_{F_{2^t}}$ do not contribute to the map. On the other hand, since in this note we are mapping CFTs to qudit codes, factors of 2 are important. 
\bigskip

\noindent \textit{When do both constructions agree?}

\noindent Both maps \eqref{decmap} and \eqref{genqudit} agree when the the fusion group, $K$, of the chiral primaries of the CFT is of the form $\DZ_2^{\ell}$ for some integer $\ell$. Indeed, in this case, the map \eqref{genqudit} is
\be
\CO_{\vec p, \overline{\vec p + \vec g}} \leftrightarrow X^{\vec g} \circ Z^{L \vec p}~,
\ee
where $X$ and $Z$ are Pauli matrices acting on the Hilbert space of $\ell$-qubits. If we restrict to the case of $K \simeq \DZ_2^{\ell}$, the map \eqref{decmap} also reduces to
\be 
\CO_{\vec p, \overline{\vec p + \vec g}} \leftrightarrow X^{\vec g} \circ Z^{A \vec p}~,
\ee
since, in this case, $M$ is the identity matrix. Moreover, $A_{F_{2^t}} = L_{F_{2^t}} \text{ mod } 2$. Therefore, the two maps agree in this case. 

\bigskip

\noindent \textbf{Remark:} When $K \simeq \DZ_2^k$, the insensitivity of the CFT to qubit code map to the distinction between $L_{E_{2^t}}$ and $L_{F^{2^t}}$
does not mean that $E$ and $F$ type theories cannot be distinguished at the level of the code. Indeed, consider toric code versus the 3-fermion model (i.e., ${\rm Spin}(16)_1$ versus ${\rm Spin}(8)_1$ Chern-Simons theory). They have the same $S$-matrix but different topological spins. For a CFT with either of these bulk TQFTs and the charge conjugation modular invariant, we obtain the same quantum code. However, due to the difference in topological spins, the quantum codes obtained for other modular invariants are different. For example, in the Toric code case, consider the partition functions
\bea
Z_{\CT/Q_1}= \chi_{(0,0)}\bar\chi_{(0,0)} &+& \chi_{(0,0)}\bar\chi_{(0,1)} + \chi_{(0,1)}\bar\chi_{(0,0)}\cr &&\ + \chi_{(0,1)}\bar\chi_{(0,1)} ~,\cr
Z_{\CT/Q_2}= \chi_{(0,0)}\bar\chi_{(0,0)} &+& \chi_{(0,0)}\bar\chi_{(1,0)} + \chi_{(1,0)}\bar\chi_{(0,0)}\cr &&\ + \chi_{(1,0)}\bar\chi_{(1,0)} ~,\cr  
Z_{\CT/Q_3}= \chi_{(0,0)}\bar\chi_{(0,0)} &+& \chi_{(1,1)}\bar\chi_{(1,1)} + \chi_{(0,1)}\bar\chi_{(1,0)}\cr&&\ + \chi_{(1,0)}\bar\chi_{(0,1)} ~,
\eea
obtained from gauging the symmetries $Q_1=\langle(0,1)\rangle$, $Q_2=\langle(1,0)\rangle$, and $Q_3=\langle(1,1)\rangle$, respectively. Using \eqref{genqudit}, these CFTs can be mapped to the stabilizer codes $\langle Z \otimes I, I \otimes X\rangle$, $\langle I \otimes Z, X \otimes I\rangle$, and $\langle Z \otimes Z, Y \otimes X\rangle$ respectively. On the other hand, for the 3-fermion case, we get
\bea
Z_{\CT/Q_1}= \chi_{(0,0)}\bar\chi_{(0,0)} &+& \chi_{(0,1)}\bar\chi_{(0,1)} + \chi_{(1,1)}\bar\chi_{(1,0)}\cr &&\ + \chi_{(1,0)}\bar\chi_{(1,1)} ~,\cr
Z_{\CT/Q_2}= \chi_{(0,0)}\bar\chi_{(0,0)} &+& \chi_{(1,0)}\bar\chi_{(1,0)} + \chi_{(1,1)}\bar\chi_{(0,1)}\cr &&\ + \chi_{(0,1)}\bar\chi_{(1,1)} ~,\cr  
Z_{\CT/Q_3}= \chi_{(0,0)}\bar\chi_{(0,0)} &+& \chi_{(1,1)}\bar\chi_{(1,1)} + \chi_{(0,1)}\bar\chi_{(1,0)}\cr&&\ + \chi_{(1,0)}\bar\chi_{(0,1)} ~,
\eea
obtained from gauging the symmetries $Q_1=\langle(0,1)\rangle$, $Q_2=\langle(1,0)\rangle$, and $Q_3=\langle(1,1)\rangle$ respectively. Using \eqref{genqudit}, these CFTs can be mapped to the stabilizer codes $\langle Z \otimes Y, I \otimes Y\rangle$, $\langle(Y \otimes Z, Y \otimes I\rangle$, and $\langle Z \otimes Z, Y \otimes X\rangle$ respectively.

\section{Non-self-dual codes from CFTs}

\label{ap:non-self-dual}

In Appendix \ref{Ap:1}, we found that for a class of chiral algebras, there is no embedding of the orbifold graph in the code graph. In other words, for some chiral algebras, not all CFTs admit a universal description in terms of qudit stabilizer codes. In this section, we show that maps from CFTs to codes for arbitrary chiral algebras exist if we forego the self-duality condition.

To understand this statement, suppose the group $K$, formed by the chiral primaries under fusion, factorises into cyclic groups as
\be
K \simeq \DZ_{n_1} \times \DZ_{n_2} \times \dots \times \DZ_{n_k} ~.
\ee
Then, we can choose a product of qudit codes of the form
\be
\CH_{1} \otimes \dots \otimes \CH_{k}~,
\ee
with dimensions $n_1^2,\dots,n_k^2$, respectively. Consider the generalized Pauli group acting on this system of qudits. We can define the map from abelian RCFTs to the generalized Pauli group as
\be
\label{eq:non-self-dual code map}
\CO_{\vec p, \overline{\vec g+ \vec p}} \to X^{\vec n * \vec g} \circ Z^{\vec n * L \vec p}~,
\ee
where $\vec n$ is the vector with components $(n_1,n_2,\dots,n_k)$ and the product \lq\lq$*$" (sometimes called the Hadamard product) is defined as $\vec n * \vec g=(n_1 g_1, n_2 g_2, \dots , n_k g_k)$. The commutation relations for the generalized Pauli group elements so obtained are given by
\bea
&& X^{\vec n * \vec g^{(1)}} \circ Z^{\vec n * L \vec p^{(1)}} \cdot X^{\vec n * \vec g^{(2)}} \circ Z^{\vec n * L \vec p^{(2)}} = e^{2 \pi i [(\vec n * \vec g^{(1)}) M^2 (\vec n * L \vec p^{(2)})-(\vec n * \vec g^{(2)}) M^2 (\vec n * L \vec p^{(1)})] }\\
&& \hspace{7.0cm}  X^{\vec n * \vec g^{(2)}} \circ Z^{\vec n * L \vec p^{(2)}} \cdot X^{\vec n * \vec g^{(1)}} \circ Z^{\vec n * L \vec p^{(1)}}~, \\
&& = X^{\vec n * \vec g^{(2)}} \circ Z^{\vec n * L \vec p^{(2)}} \cdot X^{\vec n * \vec g^{(1)}} \circ Z^{\vec n * L \vec p^{(1)}}~.
\eea
Recall that the matrix $M$ is diagonal with components $M_{ii}=\frac{1}{n_i}$, and the contribution of the two $\vec n$ cancels with $M^2$ to yield a trivial phase. Therefore, the map \eqref{eq:non-self-dual code map} defines a non-self-dual qudit code for any abelian RCFT.

\end{appendices}

\newpage

\bibliography{chetdocbib}

\end{document}